\documentclass[11pt]{article}
\RequirePackage{amsmath,amsfonts,amssymb,amsthm}
\usepackage{graphicx,psfrag,epsf}
\usepackage{enumerate}
\usepackage{natbib}
\usepackage[dvipsnames]{xcolor}
\usepackage[hypertexnames=false,colorlinks,citecolor=OliveGreen,linkcolor=blue,backref=true]{hyperref}
\usepackage{booktabs}
\usepackage{dsfont}
\usepackage{here}
\usepackage{rotating,multirow}

\newcommand{\eps}{\varepsilon}

\def\T{{\intercal}}
\newcommand{\bpsi}{\boldsymbol{\psi}}
\DeclareMathOperator{\sign}{sign}
\DeclareMathOperator{\pr}{\mathrm{P}}
\newcommand{\D}{\mathcal{D}}

\newcommand{\W}{\mathcal{W}}
\newtheorem{theorem}{Theorem}[]
\newtheorem{remark}{Remark}[]
\newcommand{\blind}{0}

\begin{document}

\def\spacingset#1{\renewcommand{\baselinestretch}%
{#1}\small\normalsize} \spacingset{1}


\if0\blind
{
 \title{\bf Testing for threshold regulation in presence of measurement error with an application to the PPP hypothesis.}
  \author{Kung-Sik Chan
  \hspace{.2cm}\\
    {Department of Statistics and Actuarial Science, The University of Iowa, USA}\\
    and \\
    Simone Giannerini \\
    Department of Statistical Sciences, University of Bologna, Italy\\
    and \\
    Greta Goracci \\
    Department of Statistical Sciences, University of Bologna, Italy\\
    Faculty of Economics and Management, Free University of Bozen/Bolzano, Italy\\
    and \\
    Howell Tong\\
    University of Electronic Science and Technology of China, Chengdu, China;\\
    Tsinghua University, China; 
    London School of Economics and Political Science, U.K.
    }
  \maketitle
} \fi

\if1\blind
{
  \bigskip
  \bigskip
  \bigskip
  \begin{center}
    {\LARGE\bf  Testing for threshold regulation in presence of measurement error with an application to the PPP hypothesis.}
\end{center}
  \medskip
} \fi
\vspace*{-0.95cm}
\begin{abstract}
Regulation is an important feature characterising many dynamical phenomena and can be tested within the threshold autoregressive setting, with the null hypothesis being a global non-stationary process. Nonetheless, this setting is debatable since data are often corrupted by measurement errors. Thus, it is more appropriate to consider a threshold autoregressive moving-average model as the general hypothesis. We implement this new setting with the integrated moving-average model of order one as the null hypothesis. We derive a Lagrange multiplier test which has an asymptotically similar null distribution and provide the first rigorous proof of tightness pertaining to testing for threshold nonlinearity against difference stationarity, which is of independent interest. Simulation studies show that the proposed approach enjoys less bias and higher power in detecting threshold regulation than existing tests when there are measurement errors. We apply the new approach to the daily real exchange rates of Eurozone countries. It lends support to the purchasing power parity hypothesis, via a nonlinear mean-reversion mechanism triggered upon crossing a threshold located in the extreme upper tail. Furthermore, we analyse the Eurozone series and propose a threshold autoregressive moving-average specification, which sheds new light on the purchasing power parity debate.
\end{abstract}

\noindent%
{\it Keywords:} Lagrange multiplier test; Random walk hypothesis;  Threshold autoregressive moving-average model;   Purchasing power parity; Volatility.
\vfill

\spacingset{1.1} 

\section{Introduction} \label{sec:intro}
%
Regulation plays a fundamental role in various fields including economics, finance, biological growth and population fluctuations, etc. Growth  processes are generally regulation-free until they enter into extreme phases. For instance, there are strong theoretical economic arguments to support the view that the price gap for goods of a particular kind (or a basket of goods) in different countries should rapidly converge to zero. In other words, currencies should have the same long-run purchasing power implying that the real exchange rates would be regulated, and that  there exists a threshold that triggers a mean‐reversion \citep{Tay01b}. However, empirical evidence points to a strong persistence of the price gap and the inability of existing unit-root tests to reject the null hypothesis of a random walk. This is also known as the Purchasing Power Parity (PPP) puzzle, which has been a subject of debate ever since its first formulation. See, e.g., \cite{Tay04} for a discussion.
\par
 The random walk is a simple model for regulation-free dynamics. On the other hand, regulation from above (below) may be captured by a first-order threshold autoregressive model (TAR) which follows a random walk until the process crosses a certain threshold above (below) which mean-reversion takes place, while the process as a whole is \emph{stationary}. A nonlinear stationary process generally renders nonlinear and state-dependent the impulse response to a random shock, which is consequential and could be leveraged in economic regulation. Thus, an approach to test for dynamic regulation is to adopt the preceding threshold model as the general model and test whether it reduces to a \emph{global} random walk. This approach has received much attention in the literature \citep{End98,Can01,Bec04,Kap06,Bec08,Seo08,Par16,deJ07,Gio17}. However, data are almost always corrupted by measurement error. In this case, the TAR model is not appropriate and the null hypothesis should be an exponential smoothing model instead, i.e., the integrated moving-average IMA(1,1) model.  Then, the general hypothesis may be taken as the first-order threshold autoregressive moving-average model, i.e., TARMA(1,1),  which is driven by an IMA(1,1) model in one of its two regimes. See Section~\ref{sec:Merr} of the Supplementary Material for further justification. Above all, we cannot over-emphasize the critical importance of the role of the moving average term for practical applications.
\par
Just as ARMA models provide a parsimonious approximation to some long AR models, so may TARMA models well approximate some high-order TAR models parsimoniously \cite{Gor20,Gor21}. Thus, the TARMA model holds substantial promise as a class of nonlinear time series models for exploring nonlinear dynamics in economics and other fields.  Yet, the TARMA model has been under-explored, partly because of a lack of progress in obtaining conditions on stationarity and ergodicity. Unlike the AR-ARMA analogy, the incorporation of a moving-average part in a nonlinear framework poses major theoretical challenges and has non-trivial implications on the probabilistic structure of the process. Recent work by \cite{Cha19} provides, for the first time, a breakthrough in deriving a set of necessary and sufficient conditions for the (multi-regime) TARMA(1,1) model to admit an irreducible and invertible state-space representation. Moreover, they derived a set of necessary and sufficient conditions for stationarity and ergodicity of the TARMA(1,1) model.
\par
By leveraging on the recent results of \cite{Cha19}, we deploy a supremum Lagrange Multiplier test (supLM) for threshold regulation, with the TARMA(1,1) model as the general framework. We specify an IMA$(1,1)$ model as the null hypothesis and a TARMA$(1,1)$ with a unit-root regime as the alternative. A difficulty  arising from testing for a unit-root against a TARMA model is that the threshold parameters are absent under the null hypothesis.
 This non-standard situation, in the nonlinear time series context, is well recognized both in the TAR setting \citep{Cha90,Han96} and in the TARMA setting \cite{Li11,Gor21c}. Fortunately, the supLM framework overcomes this problem. We derive its asymptotic distribution both under the null hypothesis and local alternatives. We prove that the test is consistent and asymptotically similar in that its asymptotic null distribution does not depend on the value of the MA parameter. Moreover, we provide the first rigorous proof of tightness pertaining to testing for threshold nonlinearity against difference stationarity. The tightness result is of independent interest. It constitutes a general theoretical framework for ARIMA versus TARMA testing. We also introduce a wild bootstrap version of the supLM statistic that, for finite samples, possesses good properties and robustness against heteroskedasticity. We perform a large scale simulation study to compare our tests with existing tests, in which the alternative hypothesis is that of a threshold model. In general, the size of the latter tests is severely biased in a number of cases to the extent that their use in practical applications remains questionable unless additional information on the data generating process is available. The surprisingly good size property of our tests may be owing to the versatility of an IMA(1,1) model in approximating general non-seasonal difference stationary processes. In addition, the comparison includes some of the best performing unit-root tests to date, where the alternative hypothesis does not specify explicitly a nonlinear process.
\par
The paper is structured as follows. In Section~\ref{sec:TARMA} we present some fundamentals of the first-order TARMA model and a parametrization  that reduces to the IMA(1,1) process under the null hypothesis. In Section~\ref{sec:LM} we present a supremum Lagrange Multiplier test, which we denote by  supLM, including the theoretical framework based on Brownian local time. Section~\ref{sec:null} is devoted to the derivation of the asymptotic distribution of the supLM test statistic under the null hypothesis and we show that  it is nuisance-parameter-free and depends only on the search range of the threshold. The results concerning the local power of the proposed test are summarized in Section~\ref{sec:LP}. In Section~\ref{sec:sim} we perform a large scale simulation study to show the performance of the supLM test and a wild bootstrap version of it and compare them with numerous existing tests in the recent literature. Section~\ref{sec:real} presents the application of the tests to the daily real exchange rates of the panel of Eurozone countries, plus the TARMA-GARCH modelling for the global Eurozone series. All the proofs are collected in the Supplementary Material, which contains further results from the Monte Carlo study and from the real data application.

\section{Threshold autoregressive moving-average model}\label{sec:TARMA}
Consider the following first-order threshold autoregressive moving-average (TARMA) model:
\begin{equation}
X_t=\begin{cases}
\phi_{1,0}+\phi_{1,1}X_{t-1} +\eps_t- \theta \eps_{t-1}, & \text{if } X_{t-d} \le r \\
 \phi_{2,0}+\phi_{2,1}X_{t-1} +\eps_t- \theta \eps_{t-1}           & \text{otherwise},
\end{cases}
\label{tarma.r}
\end{equation}
where $\phi_{2,1}$ is fixed at 1 unless stated otherwise,
the innovations $\{\eps_t\}$ are independent and identically distributed random variables with zero mean and  variance $\sigma^2$, $\eps_t$ is independent of $X_{t-j}, j\ge 1$, the delay $d$ is a positive integer which, for simplicity, is taken to be 1 henceforth,  $r$ is the real-valued threshold parameter, and the $\phi$'s and $\theta$'s are unknown coefficients. The assumption of independence and identical distribution of the innovations will be relaxed later to a martingale difference sequence. The preceding (constrained) TARMA model assumes that the sub-model in the upper regime is a first-order IMA model while the lower regime specifies a general first-order ARMA model. Statistical inference with a TARMA model hinges on whether the model is invertible.  We assume $|\theta|<1$ since it is a necessary and sufficient condition for the invertibility of Model~(\ref{tarma.r}) \citep{Cha10}. By assuming that the innovations admit a positive, continuous probability density function with finite absolute first moment, \cite{Cha19} showed that Model~(\ref{tarma.r})  is an ergodic Markov chain if and only if $\phi_{2,0}<0$ and either $(i)$ $\phi_{1,1}<1$,  or $(ii)$ $\phi_{1,1}=1$, $\phi_{1,0}>0$; ergodicity then implies that the first-order TARMA model admits  a unique stationary distribution. Furthermore, under the stronger condition that the innovations admit a finite absolute $k$th moment for some $k>2$,  \cite{Cha19} provides a complete classification of  the parametric regions of Model~(\ref{tarma.r})  into sub-regions of ergodicity, null recurrence and transience. In particular, the (constrained) first-order TARMA model defined by Model~(\ref{tarma.r}) is null-recurrent if any of the following holds:
$(iii) \,\,  \phi_{1,1}=1,   \phi_{2,0}=0,  \phi_{1,0}\geq0; \quad
(iv)  \,\, \phi_{1,1}=1,   \phi_{2,0}<0,  \phi_{1,0}=0; \quad
(v)   \,\, \phi_{1,1}<1,   \phi_{2,0}=0.$
If none of the conditions $(i)$--$(v)$ holds, then the model is transient. Therefore, Model~(\ref{tarma.r}) is a rich model that encompasses both linear and nonlinear processes spanning  a wide spectrum of long-run behaviors including ergodicity, null recurrence and transience.
%
\section{Lagrange multiplier test for threshold regulation}\label{sec:LM}
We first formulate a framework for testing for threshold regulation from below.
Let $\{X_t, t=0,1,\ldots\}$, be a time series and assume that, for  $t\ge 1$, $X_t$ satisfies the equation
\begin{equation}
H:\quad X_t=\phi_0+X_{t-1} + \eps_{t}-\theta\eps_{t-1} + (\phi_{1,0}+\phi_{1,1} X_{t-1})\times I(X_{t-1}\le r),
\label{reparm.model}
\end{equation}
which is a re-parameterization of  Model~(\ref{tarma.r})
 with
 $\phi_0=\phi_{2,0}$ and by an abuse of notation, $\phi_{1,0}$ and $\phi_{1,1}$ represent, respectively, the difference of intercept and slope of the lower regime relative to their upper-regime counterparts;
 the initial value $X_0$  can be fixed at, say, 0. Our interest is in testing whether $\phi_{1,0}=\phi_{1,1}=0$, in which case the data are generated by
the IMA(1,1) model
\begin{equation}
 H_0:\quad X_t=\phi_0+X_{t-1}+\eps_t-\theta\eps_{t-1},
\label{reparm.modelH0}
\end{equation}
where $|\theta|<1$.
If the intercept $\phi_0\not = 0$, then the IMA(1,1) process has a linear trend.  If no such linear trend is apparent in the data, it is reasonable to omit the intercept. Henceforth, we assume that $\phi_0=0$ under $H_0$. The case for $\phi_0\not= 0$ will be studied elsewhere.  However, the intercept terms on the two regimes of any competing stationary first-order TARMA model will be required to model the mean of the data. Indeed, even for mean-deleted data, the intercept terms of the first-order TARMA model are not necessarily zero. Thus, the intercept terms are essential and retained in the constrained TARMA model under $H$. Note that testing for threshold regulation from above can be conducted by applying the test to $\{-X_t\}$.
\par
Under the null hypothesis, the threshold parameter is absent thereby complicating the test \citep{Cha90, Han96, Li11, Gor21c}. Our approach is to develop a Lagrange multiplier test statistic for $H_0$ initially with the threshold parameter fixed at some $r$. Denote the test statistic as $T_{n}(r)$. Since $r$ is unknown and indeed absent under $H_0$, we shall compute $T_{n}(r)$ for all $r$ over some data-driven interval, say, $[a,b]$ with the end points being some percentiles  of the observed data. For instance, $a$ could be the 20-th percentile and $b$ the 80-th percentile.  Then the overall test statistic results in $T_n=\sup_{r\in [a,b]} T_{n}(r)$. Besides taking the supremum, other approaches including integration can be employed to derive an overall test statistic.
\par
For fixed $r$, the Lagrange multiplier test is developed based on the Gaussian likelihood conditional on $X_0$:
\begin{equation}
\ell = -\log (2\pi\sigma^2)\times n/2 -\sum_{t=1}^n \eps_t^2/(2\sigma^2),
\label{log-like}
\end{equation}
where, by an abuse of notation,
\begin{equation}
\eps_{t}=X_t-\{\phi_0+X_{t-1} + (\phi_{1,0}+\phi_{1,1} X_{t-1})\times I(X_{t-1}\le r)\}+\theta\eps_{t-1},\quad \forall t\ge 1, \label{residual}
\end{equation}
with the unknown $\eps_0$ set to be zero;  $\eps_t$ in the preceding formula  is  a function of $\phi_0,\phi_{1,0}, \phi_{1,1}, \theta$ and  $r$, but the  arguments are generally suppressed for simplicity.
Let $\bpsi=(\phi_0,\theta, \sigma^2,$ $ \phi_{1,0}, \phi_{1,1})^\T$, with its components denoted by $\psi_j, j=1,2,\ldots,5$, and let it be partitioned into $\bpsi_1=(\phi_0,\theta, \sigma^2)^\T$   and $\bpsi_2=(\phi_{1,0}, \phi_{1,1})^\T$. The null hypothesis can be succinctly expressed as $H_0: \bpsi_2=0$.
The score vector is
\begin{align*}
\frac{\partial \ell}{\partial \psi_j}&= -\sum_{t=1}^n \frac{\eps_t}{\sigma^2} \frac{\partial \eps_t}{\partial \psi_j},\quad
1\le j \le 5, j \not = 3, \\
\frac{\partial \ell}{\partial \psi_3}&= \frac{\partial \ell}{\partial \sigma^2} =\sum_{t=1}^n \frac{\eps_t^2-\sigma^2}{2\sigma^4}
\end{align*}
where for $t> 1$,
\begin{align}
\frac{\partial \eps_t}{\partial \phi_0} &=
-1+\theta \frac{\partial \eps_{t-1}}{\partial \phi_0} = -\sum_{j=0}^{t-1}\theta^j,\label{eq:ppsi0}\\
\frac{\partial \eps_t}{\partial \theta} &= \eps_{t-1}+\theta \frac{\partial \eps_{t-1}}{\partial \theta}
=\sum_{j=0}^{t-1}\theta^j\eps_{t-1-j}, \label{eq:ptheta}\\
\frac{\partial \eps_t}{\partial \phi_{1,0}} &= -I(X_{t-1}\le r)+ \theta \frac{\partial \eps_{t-1}}{\partial \phi_{1,0}}
=-\sum_{j=0}^{t-1}\theta^jI\left(X_{t-1-j}\le r\right), \label{eq:pphi10}\\
\frac{\partial \eps_t}{\partial \phi_{1,1}} &= -X_{t-1}I(X_{t-1}\le r)+ \theta \frac{\partial \eps_{t-1}}{\partial \phi_{1,1}}
=-\sum_{j=0}^{t-1}\theta^jX_{t-1-j}I\left(X_{t-1-j} \le r\right), \label{eq:pphi11}
\end{align}
with  initial values given by
$\partial \eps_1/\partial \phi_0=-1, \partial \eps_1/\partial \theta=0,
\partial \eps_1/\partial \phi_{1,0}=-I(X_0\le r)$ and $\partial \eps_1/\partial \phi_{1,1}=-X_0I(X_0\le r)$. Below, we sometimes write, as a typical example,
$\partial \eps_t/\partial \phi_{1,1}=-(1-\theta B)^{-1} \left\{ X_{t-1}I(X_{t-1} \le r)\right\}$, where $B$ is the backshift operator that shifts the indices backward by 1 unit. The IMA(1,1) model under the null hypothesis can be estimated by solving the score equation
$
\partial \ell/\partial \bpsi_1=0,
$
yielding $\hat{\bpsi}_1=\hat{\bpsi}_{1,n}=(\hat{\phi}_{0,n}, \hat{\theta}_n, \hat{\sigma}^2_n)^\T$. Thus, the overall estimator of $\bpsi$ under $H_0$ is $\hat{\bpsi}=(\hat{\phi}_{0,n}, \hat{\theta}_n, \hat{\sigma}^2_n, 0,0)^\T$, with the residuals given by
\begin{equation}
\hat{\eps}_{t}=X_t-X_{t-1}-\hat{\phi}_0 +\hat{\theta}\hat{\eps}_{t-1},\quad \forall t\ge 1, \label{residualunderH0}
\end{equation}
where $\hat{\eps}_0=0$. The observed Fisher information (excluding the threshold parameter) is given by
$$I_n = -\frac{\partial^2 \ell}{\partial \bpsi \partial \bpsi^\T}, $$
whose $(i,j)$-th element  with $i,j\not= 3$ is given by
\begin{equation}
\sum_{t=1}^n
\frac{1}{\sigma^2}\frac{\partial \eps_t}{\partial \psi_i}\frac{\partial \eps_t}{\partial \psi_j} + \sum_{t=1}^n
\frac{\eps_t}{\sigma^2} \frac{\partial^2 \eps_t}{\partial \psi_i \partial\psi_j} = (1+o_p(1))\times\sum_{t=1}^n
\frac{1}{\sigma^2}\frac{\partial \eps_t}{\partial \psi_i}\frac{\partial \eps_t}{\partial \psi_j},
\label{info1}
\end{equation}
its  $(3,i)$-th element with $i\not = 3$  equal to
\begin{equation}
\sum_{t=1}^n  \frac{\eps_t}{\sigma^4}
\frac{\partial \eps_t}{\partial \psi_i}=o_p(n)
\label{info2}
\end{equation}
and the $(3,3)$-th element  equal to
$$
\sum_{t=1}^n  \left(\frac{1}{2\sigma^4}-\frac{\eps_t^2}{\sigma^6}\right),
$$
where the $o_p(1)$ and $o_p(n)$ terms hold uniformly in $r$, when the expressions are evaluated at the true parameter value under the null hypothesis; hence they are asymptotically negligible (via arguments similar to those in the proof of Theorem~\ref{thm1}), and omitted in all numerical work reported herein. Partition the Fisher information matrix according to $\bpsi_i, i=1,2$ into
\begin{equation}\label{eq:In}
I_n =
\begin{pmatrix}
I_{1,1, n} & I_{1,2,n}\\
I_{2,1,n}  & I_{2,2,n}
\end{pmatrix}.
\end{equation}
Note $\partial \ell/\partial \psi_j, \partial \eps_t/\partial \psi_i, I_n,
$  depend on $\bpsi$ and $r$ implicitly. Below, we sometimes  write, e.g., $\partial \ell/\partial \psi_j(\bpsi;r)$,  to highlight the role of the arguments; we further simplify the notation, for example, from $\partial \ell/\partial \psi_j(\bpsi_0;r)$ to $\partial \ell/\partial \psi_j(r)$,  with $\bpsi_0$ denoting the true value under $H_0$. Moreover, $I_{1,1,n}(\bpsi_0;r)$ and $\partial \ell/\partial \bpsi_1(\bpsi_0;r)$ are further simplified as $I_{1,1,n}$ and $\partial \ell/\partial \bpsi_1$ as they do not depend on $r$. By an abuse of notation, the true values of the moving-average coefficient and the innovation variance under $H_0$ are simply denoted by $\theta$ and $\sigma^2$; no confusion should arise as the context will make  clear whether  they stand for the generic parameters or their true  values.
\par
The Lagrange multiplier test statistic is an asymptotic approximation of twice the Gaussian likelihood ratio statistic, based on a second-order Taylor expansion. For fixed $r$, it equals
\begin{equation}
T_{n}(r)=\frac{\partial \hat{\ell} }{\partial \bpsi_2^\T}(r) \left\{\hat{I}_{2,2,n}(r)-\hat{I}_{2,1,n}(r)\hat{I}_{1,1,n}^{-1}(r) \hat{I}_{1,2,n}(r)\right\}^{-1}\frac{\partial \hat{\ell} }{\partial \bpsi_2}(r) \label{LM-stat}
\end{equation}
where $\partial \hat{\ell}/\partial \bpsi_2(r)$ is equal to $\partial \ell/\partial \bpsi_2$ evaluated at $\bpsi_1=\hat{\bpsi}_1$, $\bpsi_2=0$ and  the threshold parameter  $r$. Similarly defined are $\hat{I}_{i,j,n}(r), 1\le i,j,\le 2$. Because  the threshold $r$ is unknown, the overall supLM   statistic is   $T_n=\sup_{r\in [a, b]} T_{n}(r)$ with $a$ and $b$, for instance, being some pre-specified percentiles  of the observed data. For theoretical analysis, the threshold range is specified as $R_n=(n^{1/2}(1-\theta) \sigma \times r_L, n^{1/2}(1-\theta) \sigma \times r_U)$ where $r_L<r_U$ are two fixed finite numbers.  We now justify this choice of the threshold range. First, some heuristics will be employed.
Under the null hypothesis (with $\phi_0=0$),
\begin{equation*}
X_t =
\eps_t+(1-\theta)\sum_{s=1}^{t-1} \eps_s-\theta\eps_0+  X_0.
\end{equation*}
Hence, $\{n^{-1/2}\,X_{[sn]}, 0\le s\le 1\}$, where $X_{[sn]} = \sum_{t=1}^{[sn]} X_t $ and $[sn]$ is the largest integer less than or equal to $sn$, converges in distribution to $\{(1-\theta)\sigma W_s\}$  where $\{W_s\}$ is the standard Brownian motion. It is well known \citep[Theorems 3.1 and 3.2]{Bjo19} that the Brownian local time $\{L_t^x, t\ge 0,  -\infty<x<\infty\}$ defined as follows:
$$
L_t^x=|W_t-x|-|x|-\int_0^t \sign(W_s-x)ds,
$$
where $\sign(x)$ denotes the sign of $x$, is essentially the probability density function of the Brownian realization in the sense that for any bounded real-valued Borel function $f$,
$$
\int_0^1 f(W_s)ds= \int_{-\infty}^\infty f(x) L_1^x dx.
$$
Thus, any quantile of $\{X_t, t=0,\ldots,n\}$ is asymptotically equal to $n^{1/2}(1-\theta)\sigma$ times the corresponding quantile of $\{W_s, 0\le s\le 1\}$. Since the Brownian local time process is a random process, so the quantiles are realization specific! This motivates us to set the threshold to be of the form $r_n=(1-\theta)\tau\sigma n^{1/2}$ for some fixed $\tau$, in which case
\begin{equation}
 n^{-1/2} \frac{\partial \ell}{\partial \phi_{1,0} }(r_n)=
n^{-1/2}\sum_{t=1}^n   \frac{\eps_t}{\sigma^2} \frac{1}{1-\theta B}\left\{ I\left(\frac{X_{t-1}}{n^{1/2}(1-\theta)\sigma}\le \tau\right)\right\}.
\label{score10}
\end{equation}
The right side of (\ref{score10})  is a Riemann-Stieltjes sum over $[0,1]$, with a step integrator jumping at $t/n$ with jump size $ (n\sigma^{2})^{-1/2} \eps_t$ and the integrand is a piecewise constant function which equals $\sum_{j=0}^{t-1}\theta^j I\left(\{n^{1/2}(1-\theta)\sigma\}^{-1}X_{t-1-j} \le \tau \right)$ over the interval $[n^{-1}(t-1), n^{-1}t]$, for $t=1,2,\ldots,n$. The integrator converges weakly to the standard Brownian motion whereas the integrand to $(1-\theta)^{-1} I(W_s\le \tau)$ as $t, n\to \infty$ such that $t/n\to s$ in $[0,1]$. Thus, heuristically, $ n^{-1/2} \partial \ell/\partial \phi_{1,0}(r_n)$ converges in distribution to $(1-\theta)^{-1} \sigma\int_0^1 I(W_s\le \tau )dW_s $
under $H_0$ and as $n\to\infty$, or in symbol,
\begin{equation}
 n^{-1/2} \frac{\partial \ell}{\partial \phi_{1,0} }(r_n) \rightsquigarrow \frac{1}{(1-\theta) \sigma}\int_0^1 I(W_s\le \tau )dW_s.
\label{wconv1}
\end{equation}
This  asymptotic result and other heuristic results stated below can be essentially justified using Theorem 7.10 in \cite{Kur96}.
Similarly,
\begin{align}
 n^{-1}  \frac{\partial \ell}{\partial \phi_{1,1} }(r_n)&= n^{-1/2}\sum_{t=1}^n   \frac{\eps_t}{\sigma} \frac{1}{1-\theta B}\left[\frac{X_{t-1}}{n^{1/2}\sigma} I\left\{\frac{X_{t-1}}{n^{1/2}(1-\theta)\sigma}\le \tau\right\}\right]\nonumber \\
&\rightsquigarrow \int_0^1 W_sI(W_s\le \tau )dW_s \label{wconv2} \\
 n^{-1/2} \frac{\partial \ell}{\partial \phi_{0} }&=
n^{-1/2} \sum_{t=1}^n   \frac{\eps_t}{\sigma^2} \frac{1}{1-\theta B} (1) \rightsquigarrow \frac{1}{(1-\theta)\sigma}\int_0^1 dW_s=\frac{W_1}{(1-\theta)\sigma}.
\label{wconv3}
\end{align}
%
 Note the different rates of normalization. Let $K_n$ be the $5\times 5$ diagonal matrix with the last diagonal elements being $n$ and  other diagonal elements all being $n^{1/2}$.  We can also show that
$
K_n^{-1} I_n(r_n) K_n^{-1}$ converges in probability to a matrix denoted by $\mathcal{I}(\tau)$ which can be blocked as $I_n$ (see Eq.~\ref{eq:In}). In particular, $\mathcal{I}_{1,1}$ is a diagonal matrix comprising $(1-\theta)^{-2}\sigma^{-2}, (1-\theta^2)^{-1}, (4\sigma^4)^{-1}$ as its diagonal elements,
\begin{equation*}
\mathcal{I}_{2,2}(\tau)=
\begin{pmatrix}
\frac{1}{(1-\theta)^2 \sigma^2}\int_0^1 I(W_s\le \tau) ds &
\frac{1}{(1-\theta) \sigma}\int_0^1 W_sI(W_s\le \tau) ds\\
\frac{1}{(1-\theta) \sigma}\int_0^1 W_sI(W_s\le \tau) ds & \int_0^1 W_s^2 I(W_s\le \tau) ds
\end{pmatrix};
\label{I22}
\end{equation*}
\begin{equation*}
\mathcal{I}_{2,1}(\tau)=
\begin{pmatrix}
\frac{1}{(1-\theta)^2 \sigma^2}\int_0^1 I(W_s\le \tau) ds &
0 & 0\\
\frac{1}{(1-\theta) \sigma}\int_0^1 W_sI(W_s\le \tau) ds & 0 & 0
\end{pmatrix}.
\label{I21}
\end{equation*}
 Note that $\mathcal{I}_{1,1}$ does not depend on $\tau$.
Thus, $\theta$ and $\sigma^2$ are locally orthogonal to the other parameters around the true parametric value under $H_0$ . Hence, their estimates are expected to be asymptotically independent of the proposed test statistic, as will be shown below  to be  the case.
\section{The null distribution}\label{sec:null}
We now derive the asymptotic  distribution of $T_{n}(r)$ under  the null hypothesis of an IMA(1,1) model with zero intercept.
Using second-order Taylor expansion and after some routine algebra, it holds that
\begin{equation}
\frac{\partial \hat{\ell} }{\partial \bpsi_2} (r_n)
\approx\frac{\partial \ell}{\partial \bpsi_2}(r_n)-I_{2,1,n}(r_n)I_{1,1,n}^{-1}\frac{\partial \ell}{\partial \bpsi_1}. \label{eq:scoreapprox}
\end{equation}
More rigorously, letting
$$Q_n=\begin{pmatrix}
n^{-1/2} & 0 \\
0 & n^{-1}
\end{pmatrix}, \quad
P_n=n^{-1/2}\begin{pmatrix}
1 & 0 &0 \\
0 & 1 & 0\\
0 & 0 & 1
\end{pmatrix},
$$
we shall prove below that uniformly for $ r_n=n^{1/2}(1-\theta) \sigma\tau\in R_n=(n^{1/2}(1-\theta) \sigma \times r_L, n^{1/2}(1-\theta) \sigma \times r_U)$, where $r_L< r_U$ are fixed numbers,
\begin{align}
Q_n\frac{\partial \hat{\ell} }{\partial \bpsi_2}(r_n)
&=Q_n\frac{\partial \ell}{\partial \bpsi_2}(r_n)
-I_{2,1}(\tau)I_{1,1}^{-1}
P_n\frac{\partial \ell}{\partial \bpsi_1}+o_P(1) \nonumber \\
&= Q_n\frac{\partial \ell}{\partial \bpsi_2}(r_n)
-\tilde{I}_{2,1}(\tau)\tilde{I}_{1,1}^{-1}
P_n\frac{\partial \ell}{\partial \phi_0}+o_P(1),
\label{eq:exactscoreapprox}
\end{align}
where,
owing to the form of $I_{2,1}(\tau)$,
$\tilde{I}_{1,1}= (1-\theta)^{-2}\sigma^{-2}$ and
$$
\tilde{I}_{2,1}=
\begin{pmatrix}
\frac{1}{(1-\theta)^2 \sigma^2}\int_0^1 I(W_s\le \tau) ds \\
\frac{1}{(1-\theta) \sigma}\int_0^1 W_sI(W_s\le \tau) ds
\end{pmatrix}.
$$
The intercept $\hat{\phi}_{0,n}$ admits the asymptotic representation under $H_0$ \citep[c.f. Eqn.(8.11.5)]{Bro01}
$$P_n^{-1}(\hat{\phi}_{0,n}-\phi_0)= (\tilde{I}_{1,1})^{-1} P_n \frac{\partial \ell}{\partial \phi_0}+o_P(1).$$
A key step in deriving the limiting null distribution of the proposed test is then to demonstrate that uniformly for $r_n=n^{1/2}(1-\theta) \sigma\tau\in R_n$
\begin{equation}
Q_n\frac{\partial \hat{\ell} }{\partial \bpsi_2}(r_n)
= Q_n\frac{\partial \ell}{\partial \bpsi_2}(r_n)-\tilde{I}_{2,1}(\tau)P_n^{-1}(\hat{\phi}_{0,n} -\phi_0)+o_p(1). \label{essence}
\end{equation}
Let
\begin{equation}
H(\tau)=\left(\int_0^1 dW_s, \int_0^1 I(W_s\le \tau) dW_s, \int_0^1 W_sI(W_s\le \tau) dW_s\right)^\T
\label{eq:Htau}
\end{equation}
and
\begin{equation}
\Lambda(\tau)
=
\begin{pmatrix}
1 & \int_0^1 I(W_s\le \tau)ds & \int_0^1 W_sI(W_s\le \tau)ds \\
\int_0^1 I(W_s\le \tau)ds  & \int_0^1 I(W_s\le \tau)ds  & \int_0^1 W_sI(W_s\le \tau)ds \\
\int_0^1 W_sI(W_s\le \tau) ds &  \int_0^1 W_sI(W_s\le \tau) ds & \int_0^1 W_s^2I(W_s\le \tau) ds
\end{pmatrix}.\label{eq: Lambdatau}
\end{equation}
Let $\Lambda(\tau)$ be partitioned into a $2\times 2$ block matrix with the $(2,2)$-th block being $2\times 2$. Similarly partitioned is $H(\tau)=(H_{1}(\tau), H_{2}(\tau))^\T$. It follows from Eq.~(\ref{eq:exactscoreapprox}) and Eqs.~(\ref{wconv1})--(\ref{wconv3}) that the asymptotic null distribution of $T_{n}(r_n)$ can be shown to be the same as that of $\left\|(\{\Lambda^{-1}(\tau)\}_{2,2})^{1/2} \left( H_{2}(\tau)-\Lambda_{2,1}(\tau)H_{1}(\tau)\right)\right\|^2$, where $\|\cdot\|^2$ is the squared Euclidean norm of the enclosed vector. It is readily shown  that $\{\Lambda^{-1}(\tau)\}_{2,2}=\{\Lambda_{2,2}(\tau)-\Lambda_{2,1}(\tau)\Lambda_{1,2}(\tau)\}^{-1}$.  The asymptotic null distribution of $T_n$ is derived in the following theorem.
\begin{theorem}
\label{thm1}
Suppose $H_0$ holds so that $\{X_t, t=0,1,\ldots,\}$ is an IMA(1,1) process satisfying Eq.~(\ref{reparm.modelH0}), with the intercept $\phi_0=0$, $|\theta|<1$ and the  innovations  are independent and identically distributed with zero mean and finite positive variance.
 Let $r_L< r_U$ be two fixed real numbers.
 Let $$\mathcal{T}_n(\tau)= n^{-1/2}\sum_{t=2}^{n}  \frac{\eps_t}{\sigma}\sum_{j=0}^{t-2}\theta^j I\left\{ r_L<\frac{X_{t-1-j}}{n^{1/2}(1-\theta)\sigma}\leq\tau\right\},$$ for $ r_L\le \tau\le r_U$.
 Suppose $(i)$ there exists a constant $C>0$ such that, for any fixed $r_L\leq\tau_1<\tau_2\leq r_U$,
  \begin{align}
  E\left\{\left|\mathcal{T}_n(\tau_2)-\mathcal{T}_n(\tau_1)\right|^4\right\}\leq C(|\tau_2-\tau_1|^{3/2}+|\tau_2-\tau_1|/n),
  \label{tightness-c1}
  \end{align}
  and $(ii)$  uniformly for  $a\le \tau_1<\tau_2 \le b$,
  \begin{equation}
|\mathcal{T}_n(\tau_2)-T_n(\tau_1)| \le K\times L(n)(n\log\log n)^{1/2}|\tau_2-\tau_1|+o_p(1)
  \label{tightness-c2}
\end{equation}
as $n\to\infty$ where the $o_p(1)$ term holds uniformly,  $K$ is a constant that may depend on $\theta$, and  $L(\cdot)$ is some slowly varying function, i.e., for any $\lambda>0, L(\lambda x)/L(x)\to 1$ as $x\to\infty$. Then as $n\to\infty$,   $T_n=\sup \{T_{n}(r), r \in [n^{1/2}(1-\theta)\sigma r_L, n^{1/2}(1-\theta)\sigma  r_U]\}$ converges in distribution to
\begin{equation}
F(W; r_L, r_U)=\sup_{\tau\in [r_L, r_U]} \left\|\left[\{\Lambda^{-1}(\tau)\}_{2,2}\right]^{1/2} \left\{ H_{2}(\tau)-\Lambda_{2,1}(\tau)H_{1}(\tau)\right\}\right\|^2, \label{limiting-dist}
\end{equation}
whose distribution is parameter-free, although it depends on the search range of the  threshold.
\end{theorem}
 We remark that the assumption of independence and identical distribution of the innovations in the preceding theorem  can be relaxed to $\{\eps_t\}$ being a stationary, ergodic, martingale difference sequence with respect to the $\sigma$-algebra $\mathcal{F}_t$ generated by $\eps_{t-s}, s\le 0$; the proof is essentially the same.

 \begin{remark}
 Conditions (\ref{tightness-c1})--(\ref{tightness-c2}) provide a new set of  general sufficient conditions for the tightness of a sequence of stochastic processes; specifically the tightness of $\{T_{n}(n^{1/2}(1-\theta)\tau), r_L\le \tau \le r_U\}$. These sufficient conditions are motivated by the approach taken by \cite{Bil68}, Theorem 22.1, for studying the tightness of empirical processes for stationary, mixing data, and are tailor made for coping with nonstationarity under the null. To the best of our knowledge, this is the first rigorous proof of tightness for testing threshold nonlinearity against difference stationarity and constitutes a general theoretical framework that can be used in different settings.
 \end{remark}
 \noindent
 The preceding theorem assumes deterministic threshold search interval. It can be readily extended to the case that the end points are fixed quantiles of the data, which are realization specific. We omit the proof as it is based on routine analysis that builds on Theorem~\ref{thm1} and the  facts that  for any fixed $0<p<1$, $(i)$ the $p$-quantile of $\{W_s, 0\le s \le 1\}$ is $O_p(1)$, which follow from \cite{Bjo19}, Proposition 3.2, and the Markov inequality, and $(ii)$  the $p$-quantile of $\{X_t, t=0,\ldots, n\}$ is asymptotically equal to its counterpart of $\{W_s, 0\le s \le 1\}$  times $n^{1/2}(1-\theta) \sigma$.

The following result shows that Theorem~\ref{thm1} holds for normally distributed innovations.
\begin{theorem}~\label{thm2}
Conditions (\ref{tightness-c1}) and (\ref{tightness-c2}) hold if $(i)$ $|\theta|<1$ and $(ii)$ $\{\eps_t\}$ are independent and identically normally distributed with zero mean and finite positive variance.
\end{theorem}
Since the  null distribution of $T_n$ is asymptotically similar, its quantiles can be derived numerically. The tabulated quantiles of the null distribution for different threshold ranges can be found in Section~\ref{SM:tab} of the Supplementary Material.

\section{Local Power}\label{sec:LP}

In this section we derive the asymptotic distribution of the supLM statistic under a sequence of local threshold alternatives and prove its consistency in having power approaching 1 with increasing departure in some direction from the null hypothesis. The mathematical framework is as follows.  For each positive integer $n$, the system of hypothesis is:
\vskip .1 in

  \noindent $H_{0,n}$: $(X_0,\ldots,X_n)$ follow the IMA(1,1) model:   $X_t=X_{t-1}+\eps_t-\theta\eps_{t-1}.$\\
  \noindent $H_{1,n}$: $(X_0,\ldots,X_n)$ follow the TARMA(1,1) model:
  \begin{equation}
X_t=\begin{cases}
 {n^{-1/2}h_{1,0}}{}+\left(1+{n^{-1}h_{1,1}}\right)X_{t-1}+\eps_t-\theta\eps_{t-1} & \text{if $\frac{X_{t-1}}{\sigma n^{1/2}(1-\theta)} \leq\tau_0$} \\
n^{-1/2}h_{2,0}+\left(1+{n^{-1}h_{2,1}}\right)X_{t-1}+\eps_t-\theta\eps_{t-1} & \text{if $\frac{X_{t-1}}{\sigma n^{1/2}(1-\theta)}>\tau_0$,}
           \end{cases} \label{TARMAalt}
\end{equation}
\noindent
where $\mathbf{h}=(h_{1,0},h_{2,0},h_{1,1},h_{2,1})^\T$ is a fixed vector with $h_{i,1}\le 0, i=1,2$ and $\tau_0$ is a fixed threshold. Note that if $h_{1,1}<0$ ($h_{2,1}<0$), then the model is locally stable in the lower (upper) regime, for all $n$ sufficiently large. In order to derive the local power, we henceforth impose the following mild regularity conditions:
\begin{itemize}
\item[C1:]  The innovations are assumed to be independent and identically distributed, with zero mean, finite positive standard deviation, $\sigma$, and probability density function
$f(\cdot/\sigma)/\sigma$, where $(i)$ $f$ is a bounded function,   $\log(f(x))$ is twice differentiable with Lipschitz continuous first and second derivatives over the support of the probability density function, $(ii)$ the moment generating function of the innovations exists and is finite over some open interval around 0,  and $(iii)$ $\mathcal{I}_f=-\int ({\ddot{f}f-\dot{f}^2}/{f^2})(x)\times  f(x) dx$ is a finite positive number, where the first (second) derivative of $f$ is  denoted by  $\dot{f}$ ($\ddot{f}$).

\item[C2:] $-{\pi}/{2}<h_{1,1},h_{2,1}\le 0$ and $h_{1,1}+ h_{2,1} <0.$
\end{itemize}
Note that $\mathcal{I}_f$ is the Fisher information for the location model $f(\cdot-\mu)$ where $\mu$ is the location parameter.
Let $P_{0,n}$ and $P_{1,n}$ be the probability measures induced by $(X_0,\ldots,X_n)$ under $H_{0,n}$ and $H_{1,n}$, respectively.
Condition (C1) holds for many commonly used innovation distributions.
Condition (C2) ensures that the local alternative first-order TARMA model is asymptotically locally stable in at least one regime. These two conditions are imposed to ensure that  $\{P_{1,n}\}$ is contiguous  to $\{P_{0,n}\}$.
Finally, let $\rho$ be the correlation between $\eps_t$ and $(\dot{f}/{f})(\eps_t)$, i.e., $\rho \surd{\mathcal{I}_f} =\int x \dot{f}(x)dx$.
\begin{theorem}\label{thm:3}
 Suppose all the conditions stated in Theorem~\ref{thm1} hold. Assume (C1) and (C2) hold. Under $H_{1,n}$ and as $n\to\infty$, $T_n=\sup \{T_{n}(r), r \in [n^{1/2}(1-\theta)\sigma r_L, n^{1/2}(1-\theta)\sigma  r_U]\}$, where $r_L, r_U$ are two fixed numbers,  converges in distribution to $F(W; r_L, r_U)$ defined in Eq.~(\ref{limiting-dist}) but with $W$ now being a threshold diffusion process satisfying the following stochastic differential equation (SDE):
\begin{equation}
dW_s=dW_s^\dagger+
\begin{cases} \rho\surd{\mathcal{I}_f}\left\{\frac{h_{1,0}}{\sigma(1-\theta)}+h_{1,1}W_s\right\} ds,  & \text{if}\,\,\, W_s\leq\tau_0, \\
\rho\surd{\mathcal{I}_f}\left\{\frac{h_{2,0}}{\sigma(1-\theta)}+h_{2,1}W_s\right\} ds, & \text{otherwise},
\end{cases} \label{TDif}
\end{equation}
where $W_0=0$ almost surely and  $\{dW_s^\dagger, s\ge 0\}$ is a standard Brownian motion.
\end{theorem}
\noindent
Henceforth in this section, $W$ denotes the threshold diffusion satisfying Eq.~(\ref{TDif}).  Note that if $h_{i,0}=h_{i,1}=0, i=1,2$, then we get back the limiting null distribution for $T_n$. Otherwise, $W$ is a threshold diffusion process \citep{Su15}.  Thus, the building block $W$ determining the limiting distribution of the supLM statistic changes from a standard Brownian motion under $H_{0,n}$ to a threshold diffusion under $H_{1,n}$, if $\rho\not =0$. Consequently, the proposed test would have power to detect the local threshold alternatives. Since the functional $F(\cdot; r_L, r_U)$ is quite complex, we examine an example to demonstrate the consistency of the proposed test.

\subsection{An asymptotically ergodic, symmetric TARMA(1,1) alternative.}\label{example}
Suppose the parameters in Eq.~(\ref{TARMAalt}) are such that $\tau_0=0$, $\rho\surd{\mathcal{I}_f}h_{1,1}= \rho\surd{\mathcal{I}_f}h_{2,1}=-1/2$ and
$\rho\surd{\mathcal{I}_f}\frac{h_{1,0}}{\sigma (1-\theta)}= 2h>0$ and
$h_{2,0}=-h_{1,0}$.
Consequently, $W$ is driven by the following SDE:
\begin{equation}
dW_s=\{2hI(W_s\le 0)-2hI(W_s>0)-W_s/2\}ds+ dW_s^\dagger,
\label{STDif}
\end{equation}
with initial condition $W_0=0$ and  parameter $h>0$. It is an ergodic diffusion whose stationary marginal probability density function is given by
$$
\pi(x)=k^{-1}\exp[ -\{(x-h)^2I(x\le 0)+(x+h)^2I(x>0)\}/2 ], \quad -\infty<x<\infty,
$$
where
$$
k=2\sqrt{2\pi}\Phi(-h),
$$
and
$\Phi(\cdot)$ is the standard normal cumulative distribution function \citep[see][Theorem 1]{Su15}. It
is well known that
$$
\frac{1}{h}\phi(h)\le \Phi(-h) \le \frac{h}{h^2+1} \phi(h),
$$
where $\phi(\cdot)$ is the standard normal probability density function. Thus, the stationary probability density function equals $\pi(x)=(h/2)\{1+o(1)\} \exp(-x^2/2)\exp(-|x|h)$, which is asymptotically the Laplace distribution with scale parameter  $h^{-1}$ (mean 0 and  variance $2/h^2$),  as $h\to\infty$.
As $h\to\infty$, the $W$ process approaches
stationarity at an increasing rate, implying that solving Eq.~(\ref{STDif}) via discretization can be achieved with step size inversely proportional to $h$. ($\W_h$ below stands for a random variable having the Laplace distribution with scale parameter $h^{-1}$.)
Consequently, it follows from ergodicity and the preceding discussion that, letting $\D_h$ be a $3\times 3$ diagonal matrix with  $1,1,h$ as its diagonal,
$$
\D_h\Lambda_0 \D_h \to
\begin{pmatrix}
1 &\frac{1}{2} & -\frac{1}{2}\\
\frac{1}{2}  & \frac{1}{2}  & -\frac{1}{2} \\
-\frac{1}{2} & -\frac{1}{2} & 1
\end{pmatrix},
$$
a non-singular matrix; this is because, as $h\to \infty$,
$(i)$ $\int_0^1 I(W_s\le 0)ds\approx E\{I(\W_h\le 0)\}=1/2$, $(ii)$ $h\int_0^1 W_sI(W_s\le 0)ds\approx h E\{W_hI(\W_h\le 0)\}=-1/2$ and $(iii)$ $h^2 \int_0^1 W_s^2I(W_s\le 0)ds\approx h^2 E\{\W_h^2I(\W_h\le 0)\}=1$.
It can be similarly checked that
\begin{align}
\D_h H(0) &=
\begin{pmatrix}
\int_0^1 dW_s^\dagger +\int_0^1\{ -2h \times \sign(W_s)-W_s/2\}ds\\
\int_0^1 I(W_s\le 0)dW_s^\dagger + \int_0^1 (2h-W_s/2)I(W_s\le 0)ds\\
h\int_0^1 W_s I(W_s\le 0)dW_s^\dagger + h\int_0^1 (2h-W_s/2)W_sI(W_s\le 0)ds
\end{pmatrix}\\
&= (O_p(1), 2h+O_p(1) , -h+O_p(1))^\T. \nonumber
\end{align}
Assuming that $r_L\le 0\le r_U$, then
$$F(W; r_L,r_U)\ge \left\|\{(\Lambda_0^{-1})_{2,2}\}^{1/2} \left\{ H_{2}(0)-\Lambda_{2,1,0}H_{1}(0)\right\}\right\|^2\to\infty,$$
in probability as $h\to\infty$. Thus the supLM test statistic has power approaching 1 in rejecting the null hypothesis, as $h\to \infty$.

\section{Finite sample performance} \label{sec:sim}
In this section we compare the finite sample performance of the proposed test with a number of existing relevant tests. To better approximate the finite sample distribution of $T_n$, we have simulated the null distributions for the sample sizes in use. Moreover, since we have found that the finite sample distribution of $T_n$ changes appreciably only when $|\theta|$ is close to one, we have adopted the following, conservative, approach: if $|\hat\theta|>0.3$, we use the quantiles of the simulated null with $\theta = \sign(\hat\theta)\cdot0.9$. Furthermore, wild bootstrap obtained by bootstrapping with randomly signed residuals is also added to improve the empirical size of the test. We denote our asymptotic test and its wild bootstrap version by sLM and sLMb, respectively.
\par
The set of competing tests can be divided into those whose alternative is a threshold autoregressive model and those that do not specify explicitly a nonlinear alternative. The former set includes the tests proposed by \cite{Kap06} (KS), \cite{End98} (EG) and  \cite{Bec04} (BBC), with their bootstrap variants (if implemented) signified by appending the extension b to their abbreviations. The latter set includes the ADF test of  \cite{Dic79}, the class of M tests of \cite{Ng01} ($\bar{\mathrm{M}}^{\mathrm{g}}$),  the $\bar{\mathrm{MP}}_{\mathrm{T}}^{\mathrm{GLS}}$ test of \cite{Ng01} (${\mathrm{MP}}_{\mathrm{T}}$) and the GLS detrended version of the ADF test (ADF$^{\mathrm{g}}$), and the test $\mathrm{M}^{\mathrm{GLS}}$ of \cite{Per07} ($\mathrm{M}^{\mathrm{g}}$). Note that we have obtained the results for all the tests proposed in the above references. Of these tests, we report only the best performing ones. Unreported results are available upon request.
\par
The sample sizes considered are 100, 300 and 500. The rejection percentages are derived with a nominal size $\alpha = 5\%$ and based upon 10000 replications. In order to reduce the computational burden, for the bootstrap tests we select 1000 replications and $B=1000$ bootstrap resamples. The threshold search ranges from the 25\% to the 75\% of the sample distribution.
We simulate data from the following first-order TARMA model
\begin{equation}
X_t=
\begin{cases}
 \phi_{1,0}+\phi_{1,1}X_{t-1} +\eps_t- \theta \eps_{t-1}, & \text{if } X_{t-1} \le 0, \\
 \phi_{2,0}+\phi_{2,1}X_{t-1} +\eps_t- \theta \eps_{t-1}, & \text{otherwise},
\end{cases}
\label{tarma.sim}
\end{equation}
where $(\phi_{1,0}, \phi_{1,1}, \phi_{2,0}, \phi_{2,1})=\tau\times (0, 0.7,-0.02,0.99)+(1-\tau)\times(0,1,0,1)$ with $\tau$ increasing from 0 to 1.5 with increments 0.5. When $\tau=0$, the model is an IMA(1,1) model with zero intercept. When $\tau>0$, the model becomes a stationary first-order TARMA model that is increasingly distant from the IMA(1,1) model with increasing $\tau$. As for the MA parameter we set $\theta=-0.9,-0.5,0,0.5,0.9$. The empirical sizes of the tests are displayed in Table~\ref{tabT:size}. Note that we have partitioned the set of 11 tests according to their nature: the first 9 are asymptotic and the last 2 are bootstrap tests. Clearly, the ADF, the KS, the BBC and the EG tests are severely oversized as $\theta$ approaches unity. Moreover, the wild bootstrap sLMb test is the only test that shows a correct size in all the settings, whereas both the sLM and the M class of tests show some bias, albeit small. Note that, when $\theta=0$ the TARMA model reduces to a TAR model. In this case, the auxiliary model of the KS, BBC, EG tests is correctly specified and their size is correct; however, when $\theta$ becomes positive their size is severely biased and this raises issues concerning their practical utility.
\begin{table}
\spacingset{1}
\small
\centering
\caption{Rejection percentages from the TARMA model of Eq.(\ref{tarma.sim}), with nominal size at $\alpha = 5\%$. Sizes over 15\% are highlighted in bold font.}\label{tabT:size}
\begin{tabular}{rrrrrrrrrrrr}
       & \multicolumn{9}{c}{asymptotic} & \multicolumn{2}{c}{bootstrap}\\
  \cmidrule(lr){2-10} \cmidrule(lr){11-12}
$\theta$ & sLM & $\bar{\mathrm{M}}^{\mathrm{g}}$ & $\mathrm{M}^{\mathrm{g}}$ & $\mathrm{MP}_\mathrm{T}$ & ADF & ADF$^{\mathrm{g}}$ & KS & BBC & EG & sLMb & KSb  \\
  \cmidrule(lr){2-10} \cmidrule(lr){11-12}
  \multicolumn{10}{l}{$n=100$} & \multicolumn{2}{c}{}\\
\cmidrule(lr){1-1}
  -0.9 & 2.2  & 7.7 &     7.0  & 7.1 &     2.5  &     3.8  &     8.1   &    11.2 &      7.1   & 5.1 &     4.9   \\
  -0.5 & 1.6  & 6.3 &     6.1  & 5.8 &     4.8  &     5.1  &     7.0   &     6.1 &      5.7   & 5.0 &     5.6   \\
   0.0 & 1.6  & 5.1 &     5.1  & 4.6 &     5.3  &     5.6  &     8.1   &     2.7 &      5.0   & 4.5 &     5.3   \\
   0.5 & 1.7  & 5.6 &     5.9  & 5.1 &     6.7  &     7.4  &{\bfseries 64.5}   &    10.2 &{\bfseries  57.5}   & 5.2 &{\bfseries 58.6}   \\
   0.9 & 11.3 & 6.5 &{\bfseries 17.7}  & 6.4 &{\bfseries 77.9}  &{\bfseries 17.8}  &{\bfseries 100.0}  &{\bfseries 92.4} &{\bfseries 100.0}   & 5.7 &{\bfseries 99.8}  \\

\multicolumn{10}{l}{$n=300$} & \multicolumn{2}{c}{}\\
\cmidrule(lr){1-1}
 -0.9 & 5.5 & 6.7 & 6.3 & 6.1 &    3.3  &     4.2  &     6.3   &    14.0 &      6.5   & 5.3 &      3.8   \\
 -0.5 & 4.7 & 5.2 & 5.1 & 4.8 &    4.5  &     4.5  &     5.1   &     8.5 &      5.4   & 5.0 &      4.5    \\
  0.0 & 2.9 & 4.9 & 4.9 & 4.4 &    5.1  &     4.6  &     6.9   &     3.2 &      4.4   & 5.6 &      4.3    \\
  0.5 & 2.3 & 5.5 & 5.4 & 5.1 &    5.4  &     5.8 &{\bfseries  74.5}   &{\bfseries 19.0} &{\bfseries  61.1}   & 4.9 &{\bfseries  67.7}    \\
  0.9 & 4.9 & 1.9 & 2.4 & 1.9 &{\bfseries 86.0} &{\bfseries 15.8} &{\bfseries 100.0}   &{\bfseries 99.7} &{\bfseries 100.0}   & 4.9 &{\bfseries 100.0}    \\
\multicolumn{10}{l}{$n=500$} & \multicolumn{2}{c}{}\\
\cmidrule(lr){1-1}
 -0.9 & 8.1 & 6.4 & 6.1 & 6.0 &    7.4  & 4.7  &      5.7 &{\bfseries 16.0} &      6.1 & 5.5 &    4.0    \\
 -0.5 & 5.3 & 5.5 & 5.3 & 5.0 &    5.1  & 4.8  &      5.2 &     9.2 &      5.4 & 4.7 &    4.2   \\
  0.0 & 3.5 & 4.9 & 4.8 & 4.5 &    4.9  & 4.6  &      7.3 &     3.5 &      5.0 & 3.8 &    4.5   \\
  0.5 & 2.5 & 5.2 & 5.1 & 4.8 &    5.1  & 5.3  &{\bfseries  78.4} &{\bfseries 23.7} &{\bfseries  62.3} & 4.5 &{\bfseries 71.7}   \\
  0.9 & 3.3 & 1.3 & 1.4 & 1.4 &{\bfseries 83.2} & 14.5 &{\bfseries 100.0} &{\bfseries 99.9} &{\bfseries 100.0} & 5.4 &{\bfseries 100.0}  \\
  \cmidrule(lr){2-10} \cmidrule(lr){11-12}
\end{tabular}
\end{table}
\begin{table}
\spacingset{1}
\centering \small
\caption{Size corrected power of the asymptotic and bootstrap tests at nominal size $\alpha=5\%$
}\label{tabT:SCpower}
\begin{tabular}{crrrrrrrrrrrr}
$n=300$       & \multicolumn{9}{c}{asymptotic} & \multicolumn{2}{c}{bootstrap}\\
 \cmidrule(lr){1-1} \cmidrule(lr){2-10} \cmidrule(lr){11-12}
 $\tau\;;\;\theta$ & sLM & $\bar{\mathrm{M}}^{\mathrm{g}}$ & $\mathrm{M}^{\mathrm{g}}$ & $\mathrm{MP}_\mathrm{T}$ & ADF & ADF$^{\mathrm{g}}$ & KS & BBC & EG & sLMb & KSb  \\
  \cmidrule(lr){2-10} \cmidrule(lr){11-12}
  0.0;-0.9 &  5.0 &  5.0 &  5.0 & 5.0  &  5.0 &  5.0 &  5.0 &  5.0 &  5.0 &  5.0 &  5.0  \\
  0.5;-0.9 & 25.7 & 17.6 & 17.8 & 18.2 & 10.2 & 19.4 &  5.1 & 16.5 &  1.6 & 23.7 &  8.3  \\
  1.0;-0.9 & 52.5 & 26.7 & 26.9 & 27.7 & 15.8 & 30.4 & 17.4 & 31.9 &  3.8 & 54.3 & 27.0  \\
  1.5;-0.9 & 77.1 & 33.5 & 34.0 & 35.1 & 22.1 & 38.2 & 36.9 & 50.1 &  8.2 & 75.6 & 45.5  \\
  \cmidrule(lr){2-10} \cmidrule(lr){11-12}
  0.0;-0.5 &  5.0 &  5.0 &  5.0 & 5.0  &  5.0 &  5.0 &  5.0 &  5.0 &  5.0 &  5.0 &  5.1  \\
  0.5;-0.5 & 21.7 & 22.8 & 22.8 & 22.7 & 11.6 & 22.4 & 11.2 & 15.6 &  3.3 & 23.5 &  9.1  \\
  1.0;-0.5 & 48.3 & 34.5 & 34.9 & 34.9 & 18.2 & 35.0 & 32.3 & 31.1 &  8.2 & 47.8 & 29.1  \\
  1.5;-0.5 & 72.6 & 45.0 & 45.8 & 46.1 & 25.9 & 45.7 & 55.5 & 50.1 & 17.1 & 74.9 & 53.6  \\
  \cmidrule(lr){2-10} \cmidrule(lr){11-12}
  0.0;0.0  &  5.0 &  5.0 &  5.0 & 5.0  &  5.0 &  5.0 &  5.0 &  5.0 &  5.0 &  5.1 &  5.1  \\
  0.5;0.0  & 22.4 & 25.8 & 26.1 & 26.9 & 11.0 & 26.6 & 37.9 & 15.2 & 22.6 & 22.5 & 40.7  \\
  1.0;0.0  & 50.5 & 41.3 & 42.0 & 41.7 & 18.0 & 42.0 & 66.7 & 33.6 & 43.5 & 46.9 & 69.7  \\
  1.5;0.0  & 75.3 & 54.7 & 55.7 & 55.7 & 27.7 & 55.7 & 84.8 & 55.8 & 65.8 & 73.8 & 84.7  \\
  \cmidrule(lr){2-10} \cmidrule(lr){11-12}
  0.0;0.5  &  5.0 &  5.0 &  5.0 & 5.0  &  5.0 &  5.0 &  5.0 &  5.0 &  5.0 &  5.1 &  0.0  \\
  0.5;0.5  & 20.6 & 25.0 & 25.1 & 24.5 & 12.9 & 25.6 & 42.9 & 18.8 & 35.3 & 21.8 &  0.0  \\
  1.0;0.5  & 50.1 & 39.5 & 40.1 & 39.2 & 26.2 & 40.9 & 70.9 & 45.1 & 65.8 & 49.2 &  0.0  \\
  1.5;0.5  & 76.9 & 49.8 & 51.9 & 49.9 & 43.1 & 53.1 & 88.9 & 72.5 & 88.0 & 77.3 &  0.0  \\
  \cmidrule(lr){2-10} \cmidrule(lr){11-12}
  0.0;0.9  &  5.0 &  5.0 &  5.0 &  5.0 &  5.0 &  5.0 &  5.0 &  5.0 &  5.0 &  4.9 &  0.0  \\
  0.5;0.9  & 24.8 & 16.2 & 19.9 & 16.0 & 14.9 & 18.2 &  6.4 & 34.6 & 29.6 & 14.3 &  0.0  \\
  1.0;0.9  & 62.8 & 22.9 & 34.5 & 22.5 & 32.8 & 26.5 & 12.4 & 63.6 & 52.4 & 36.1 &  0.0  \\
  1.5;0.9  & 86.3 & 25.3 & 44.9 & 25.0 & 47.8 & 29.5 & 23.5 & 77.2 & 65.8 & 61.7 &  0.0  \\
  \cmidrule(lr){2-10} \cmidrule(lr){11-12}
\end{tabular}
\end{table}
The size-corrected power of the tests is presented in Table~\ref{tabT:SCpower}. Here, the sample size is 300; see Section~\ref{SM:power} of the Supplementary Material for results for $n=100, 500$. The rows for $\tau=0$ correspond to the size and other rows give size-corrected power. The size correction for bootstrap tests is achieved by calibrating the $p$-values. In some cases, the corrected size deviates from the nominal 5\% due to discretization effects on the empirical distribution of bootstrap $p$-values. Clearly, the supLM tests are almost always more powerful than the other tests, especially as $\tau$ increases. For instance, when $\tau=1.5$ the sLM test has almost double the power of M tests in several instances. As mentioned before, the case $\theta=0$ (central panel) corresponds to a TAR model and this is one of two instances where the KS tests are slightly more powerful than the supLM tests. The power of the bootstrap version of the KS test is zero in three cases, due to its 100\% oversize. See the Supplementary Material for further simulation results.
\subsection{Measurement error and heteroskedasticity}
In this section we assess the effect of measurement error and heteroskedasticity on the behaviour of the tests. We simulate from the following IMA(1,1) model
\begin{equation}\label{eq:ima11}
X_t= X_{t-1} + \theta \eps_{t-1}+ \eps_{t},
\end{equation}
where $\theta=-0.9$ (model M1), -0.5 (model M2), 0.5 (model M3), 0.9 (model M4). We add measurement noise as follows
\begin{equation}\label{eq:mod2}
  Y_t = X_t + \eta_t,
\end{equation}
where the measurement error $\eta_t\sim N(0,\sigma^2_\eta)$ is such that the signal to noise ratio SNR~$=\sigma^2_X/\sigma^2_\eta$ is equal to $\{+\infty, 50,10,5\}$. Here, $\sigma^2_X$ is the variance of $X_t$ computed by means of simulation. The case without noise (SNR~$=+\infty$) is taken as the benchmark. The empirical sizes (rejection percentages) for models M1--M4 are presented in Table~\ref{tab:mes300} for $n=300$ and the results for $n=100$ and 500 can be found in Section~\ref{SM:MCmerr} of the Supplementary Material. Clearly, the measurement noise has little effect upon the size of the supLM tests. On the contrary, the size bias of the tests KS, BBC and EG increases appreciably when $\theta$ is positive (Models M3--M4). Worst still the bias does not reduce when the sample size increases.
\begin{table}
\spacingset{1}
\centering
\caption{Empirical size (rejection percentage) at nominal $\alpha=5\%$ and $n=300$ for the IMA(1,1) models M1--M4 with increasing levels of measurement error.}\label{tab:mes300}
\begin{tabular}{rrrrrrrrrrrrr}
&& \multicolumn{9}{c}{asymptotic} & \multicolumn{2}{c}{bootstrap}\\
\cmidrule(lr){3-11}\cmidrule(lr){12-13}
& \textsc{snr} & sLM & $\bar{\mathrm{M}}^{\mathrm{g}}$ & $\mathrm{M}^{\mathrm{g}}$ & $\mathrm{MP}_\mathrm{T}$ & ADF & ADF$^{\mathrm{g}}$ & KS & BBC & EG & sLMb & KSb  \\
\cmidrule(lr){3-11}\cmidrule(lr){12-13}
\multirow{4}{5pt}{M1}
& $\infty$ & 4.4 & 7.1 & 6.7 & 6.8 &  3.4 &  4.8 &   6.2 & 12.5 &   6.7 & 5.0 &  5.3 \\
& 50       & 3.6 & 6.4 & 6.4 & 5.4 &  4.8 &  4.9 &   5.8 & 10.4 &   6.5 & 3.8 &  4.7 \\
& 10       & 2.8 & 5.0 & 5.0 & 4.2 &  6.1 &  4.7 &   4.9 &  5.7 &   5.1 & 5.0 &  4.0 \\
& 5        & 5.5 & 5.3 & 5.0 & 4.8 &  5.2 &  4.8 &   4.9 &  3.1 &   3.1 & 5.3 &  3.8 \\
\cmidrule(lr){3-11}\cmidrule(lr){12-13}
\multirow{4}{5pt}{M2}
& $\infty$ & 4.0 & 6.4 & 6.2 & 5.6 &  5.6 &  5.4 &   4.2 &  5.5 &   5.6 & 5.2 &  3.4 \\
& 50       & 4.7 & 6.3 & 5.9 & 5.9 &  5.8 &  5.4 &   4.0 &  4.8 &   5.4 & 6.1 &  3.2 \\
& 10       & 5.9 & 6.3 & 6.1 & 5.1 &  6.6 &  5.3 &   3.6 &  4.1 &   4.6 & 6.4 &  2.3 \\
& 5        & 5.4 & 5.5 & 5.3 & 5.1 &  6.3 &  5.2 &   5.4 &  2.5 &   5.6 & 5.4 &  3.8 \\
\cmidrule(lr){3-11}\cmidrule(lr){12-13}
\multirow{4}{5pt}{M3}
& $\infty$ & 2.8 & 5.8 & 5.8 & 4.4 &  5.6 &  6.3 &  67.8 & 14.1 &  59.9 & 5.2 & 62.0 \\
& 50       & 3.4 & 5.6 & 5.7 & 4.2 &  5.7 &  5.9 &  68.2 & 15.7 &  60.9 & 5.0 & 62.7 \\
& 10       & 2.4 & 6.2 & 6.0 & 5.0 &  5.8 &  7.0 &  74.2 & 19.7 &  67.4 & 4.8 & 66.7 \\
& 5        & 2.5 & 5.6 & 5.5 & 4.2 &  5.2 &  6.7 &  84.3 & 28.4 &  77.6 & 5.3 & 76.7 \\
\cmidrule(lr){3-11}\cmidrule(lr){12-13}
\multirow{4}{5pt}{M4}
& $\infty$ & 6.1 & 1.2 & 2.1 & 0.9 & 86.6 & 15.4 & 100.0 & 98.5 & 100.0 & 4.3 & 99.5 \\
& 50       & 5.8 & 1.2 & 1.8 & 1.1 & 87.8 & 15.6 & 100.0 & 98.8 & 100.0 & 3.6 & 99.6 \\
& 10       & 4.2 & 2.5 & 2.7 & 1.4 & 89.8 & 17.1 & 100.0 & 99.3 & 100.0 & 2.2 & 99.9 \\
& 5        & 6.7 & 3.5 & 4.5 & 2.9 & 94.9 & 19.5 & 100.0 & 99.8 & 100.0 & 3.6 &100.0 \\
\cmidrule(lr){3-11}\cmidrule(lr){12-13}
\end{tabular}
\end{table}

Since daily financial time series are characterized by volatility and this is known to bias the size of unit root tests \citep[see e.g][and references therein]{Bos18}, we simulate from the following models

\begin{center}
\small
\begin{tabular}{ll}
M5. ARIh &
 $
 \Delta  X_t=-0.6\Delta X_{t-1}+\eps_{t} $, where $V(x_t) = (1.5)^2 V(x_s)$, \\
 &  $t = 1,\dots,[n/2]$ , $s = [n/2]+1,\dots,n$.\\
M6. IMA-GARCH &
  $
  \begin{array}{ll}
    \Delta X_t =  -0.6\eps_{t-1} +  \sqrt{h_t}\,\eps_t,  & \text{where } h_t = 0.05 + 0.30 \eps^2_{t-1} + 0.65 h_{t-1} \\
  \end{array}
  $
\\
M7. ARI-GARCH &
  $
  \begin{array}{ll}
    \Delta X_t =  0.3 \Delta X_{t-1} +  \sqrt{h_t}\,\eps_t,  & \text{where } h_t = 0.05 + 0.30 \eps^2_{t-1} + 0.65 h_{t-1} \\
  \end{array}
  $
\end{tabular}
\end{center}
Model M5 is an integrated linear AR(1) where the variance of the series in the first half of the sample is different from that of the second half. Models M6 and M7 are integrated MA and AR with GARCH innovations. Measurement errors are added as described previously. The results of Table~\ref{tab:mesh300} confirm that the supLM tests are well behaved in the presence of heteroskedasticity and measurement error, with the sLMb wild bootstrap test being more so. The KS, BBC, and EG tests are severely affected by the combined presence of heteroskedasticity and measurement error. Furthermore, the size bias gets worse with increasing sample size (see the Supplementary Material,  Section~\ref{SM:MCmerr}). The class of M tests is also robust in this respect but they can display low power in a number of instances, especially when the DGP is nonlinear. See also \cite{Cha20} and Section~\ref{SM:MCmerr} of the Supplementary Material.
\begin{table}
\spacingset{1}
\centering
\caption{Empirical size (rejection percentage) at nominal $\alpha=5\%$ and $n=300$ for the heteroskedastic models M5--M7 with increasing levels of measurement error.}\label{tab:mesh300}
\begin{tabular}{rrrrrrrrrrrrr}
&& \multicolumn{9}{c}{asymptotic} & \multicolumn{2}{c}{bootstrap}\\
\cmidrule(lr){3-11}\cmidrule(lr){12-13}
& \textsc{snr} & sLM & $\bar{\mathrm{M}}^{\mathrm{g}}$ & $\mathrm{M}^{\mathrm{g}}$ & $\mathrm{MP}_\mathrm{T}$ & ADF & ADF$^{\mathrm{g}}$ & KS & BBC & EG & sLMb & KSb  \\
\cmidrule(lr){3-11}\cmidrule(lr){12-13}
\multirow{4}{5pt}{M5}
& $\infty$& 2.6 & 4.9 &  4.9 & 3.9 &  7.5 & 5.4 &  65.8 &  4.7 &  62.2 &  3.7 &  57.3 \\
& 50      & 2.7 & 5.9 &  5.9 & 4.7 &  6.9 & 6.4 &  82.1 &  8.0 &  81.3 &  3.5 &  72.0 \\
& 10      & 3.5 & 6.1 &  6.0 & 4.9 &  8.7 & 7.0 &  98.9 & 30.9 & 100.0 &  5.6 &  92.3 \\
& 5       & 4.0 & 5.7 &  5.4 & 4.7 & 19.2 & 9.5 & 100.0 & 67.2 & 100.0 &  7.4 &  98.9 \\
\cmidrule(lr){3-11}\cmidrule(lr){12-13}
\multirow{4}{5pt}{M6}
& $\infty$& 7.7 & 4.8 &  4.6 & 4.1 &  9.8 & 5.8 &  82.0 & 48.4 &  82.3 &  6.2 &  76.4 \\
& 50      & 7.4 & 5.0 &  5.0 & 3.8 & 10.3 & 5.9 &  91.9 & 53.5 &  92.7 &  7.0 &  86.0 \\
& 10      & 5.7 & 4.9 &  4.7 & 3.2 & 15.1 & 7.0 &  99.6 & 70.0 & 100.0 &  6.6 &  95.5 \\
& 5       & 5.2 & 3.5 &  3.7 & 2.4 & 27.6 & 7.9 & 100.0 & 90.0 & 100.0 &  6.9 &  98.4 \\
\cmidrule(lr){3-11}\cmidrule(lr){12-13}
\multirow{4}{5pt}{M7}
& $\infty$&11.1 & 5.2 &  5.0 & 4.6 &  7.6 & 4.5 &  10.6 & 12.5 &   8.4 &  4.2 &   8.6 \\
& 50      & 9.2 & 5.6 &  5.2 & 5.1 &  7.2 & 4.6 &  36.8 & 15.5 &  19.5 &  8.4 &  30.3 \\
& 10      & 7.0 & 6.5 &  6.2 & 5.8 & 10.0 & 6.3 &  96.2 & 40.5 & 100.0 &  9.1 &  89.4 \\
& 5       & 5.1 & 5.4 &  5.5 & 4.4 & 23.2 & 7.5 & 100.0 & 71.1 & 100.0 &  7.4 &  97.3 \\
\cmidrule(lr){3-11}\cmidrule(lr){12-13}
\end{tabular}
\end{table}
%
\section{A real application: new light on the PPP hypothesis}\label{sec:real}
%
In this section we apply our supLM tests to the real exchange rates of the panel of 17 European countries of the eurozone, plus the series of the aggregated eurozone. The idea is to contribute to the widely debated issue of the purchasing power parity (PPP) and show that the threshold ARMA model can be a useful tool to this aim. As mentioned in the introduction,  macroeconomic theory suggests that price gaps (measured in a common currency) for the same goods in different countries should rapidly disappear. However, existing unit-root tests have generally failed to reject the null hypothesis of a random walk, suggesting the existence of a strong persistence of the gaps. As also pointed out in \cite{Tay01} the  divergence between macroeconomic theory and econometric findings can be ascribed to multiple reasons. First, the way economic data are produced or aggregated can result in severely biased inference.
This is also noted in \cite{Pel15} where the authors show that real exchange rates based on the consumer price index do not preserve the possible stationarity properties of the ratios. Also, deviations from the equilibrium can be produced by the presence of non-traded goods in the basket and price frictions can slow down the mean-reversion mechanism. A third possible reason is that the commonly presumed linear price dynamics is a mis-specification. Indeed, the presence of trading costs implies that the mechanisms governing price adjustments are nonlinear and threshold autoregressive models provide a solution to the problem by allowing a ``band of inaction'' random walk regime, where arbitrage does not occur, and other regimes where mean reversion takes place so that the model is globally stationary \citep[see][and references therein for further discussion]{Bec04}. For a review on how TAR models are used to analyse the exchange rates dynamics see also \cite{Han11}. Among other approaches, \cite{Bec08b} and \cite{Gou06} introduced the switching models to incorporate a random threshold that delineates the regimes where arbitrage takes place. A critical investigation on the practical usefulness of combining unit-root tests and other stationarity tests in the PPP debate was put forward by  \cite{Can01a}.
\par
Measurement noise is ubiquitous and can be a further hindrance to the empirical verification of a threshold regulated mechanism governing the dynamics of real exchange rates. To the best of our knowledge, our attempt will represent the first time that measurement noise is incorporated in the PPP debate. By leveraging the threshold ARMA framework, we focus on the daily $\log$ real exchange rates from the exclusive adoption of the Euro for the founding countries from \texttt{2002-01-01} to \texttt{2021-08-16} ($n=5120$). The countries are: Estonia (EE), Spain (ES), Luxembourg (LU), Ireland (IE), Italy (IT), Slovenia (SI), Portugal (PT), Netherland (NL), Malta (MT), Belgium (BE), Cyprus (CY), Slovakia (SK), Finland (FI), France (FR), Greece (GR), Austria (AT), Germany (DE). The series denoted with XM indicates the Eurozone as a whole. We also add to the panel the series for Great Britain (GB) and USA (US). Overall, the panel comprises 20 time series whose time plots are presented in Figure~\ref{fig:S1} of the Supplementary Material. The real exchange rates are produced by the Bank of International Settlements (BIS) by taking the geometric weighted average of a basket of bilateral exchange rates (60 economies), adjusted with the corresponding relative consumer prices. Such weights are constructed from manufacturing trade flows so as to encompass both third-market competition and direct bilateral trade through a double-weighting scheme. See \cite{Kla06} and \url{https://www.bis.org/} for more details on the construction of the indexes.
\par
Previous studies on the Eurozone focused on monthly data and failed to reveal convincing evidence for the PPP hypothesis. In particular, \cite{Bos18} suggested the presence of a non-stationary volatility over the period 1973-2015 and proposed an adaptive likelihood ratio test for the null hypothesis of a unit root against a linear heteroskedastic AR model. However, we do not find signs of non-stationary volatility over the period 2002-2021. As we will show, the key factors behind our results are the specification that includes a threshold regulation with measurement error and the long daily series.
\par

\begin{figure}
\centering
\includegraphics[width=0.6\linewidth]{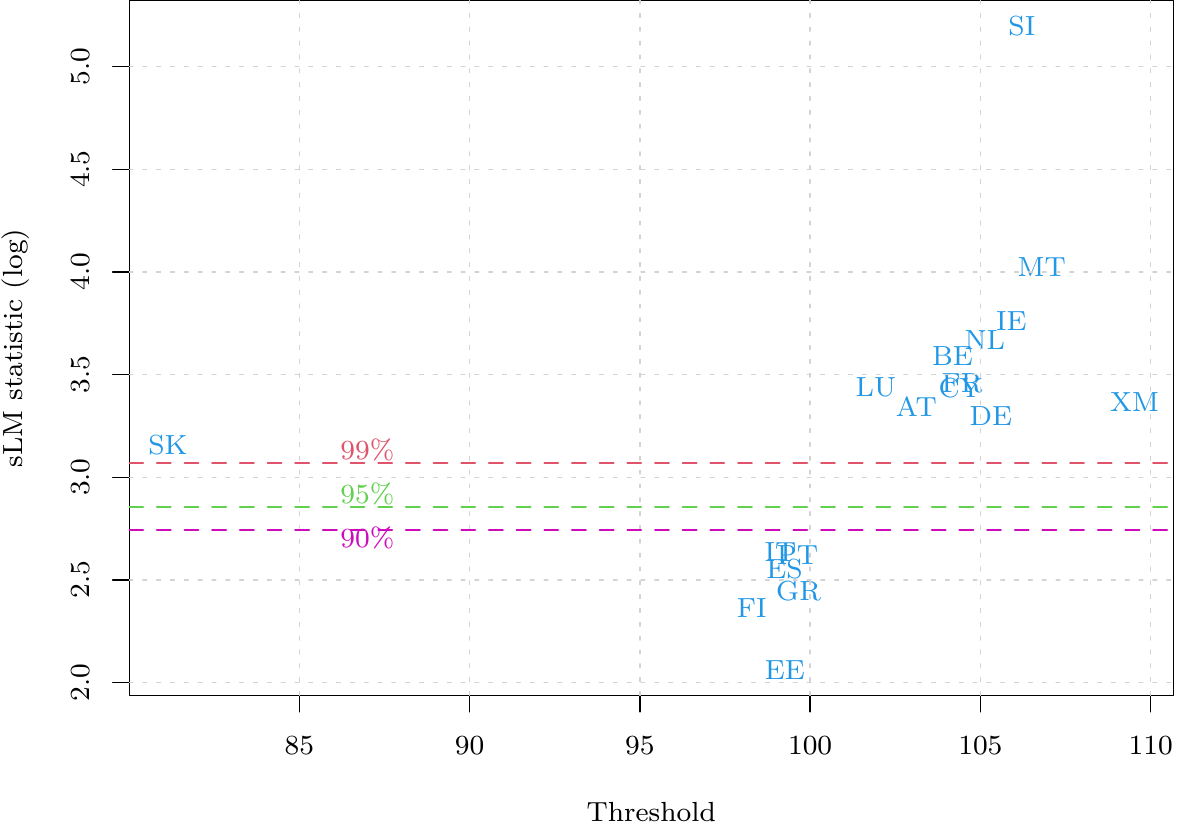}
\caption{Values of the sLM statistic (in log scale) versus threshold for the eurozone countries. The critical values of the null distribution at levels 90\%, 95\% and 99\% are added as purple, green and red dashed lines, respectively.}\label{fig:1}
\end{figure}

\begin{table}
\spacingset{1}
\centering \small
\caption{$p$-values of the supLM tests applied to the panel of 20 series. The $p$-values for the asymptotic test have been obtained from the quantiles of the simulated null distribution. The third row reports the sample percentile corresponding to the threshold.}\label{tab:ppp1}
\begin{tabular}{rrrrrrrrrrr}
 & XM & US & EE & ES & LU & IE & IT & SI & PT & NL \\
   \cmidrule(lr){2-11}
   sLM & 0.001 & 0.272 & 0.381 & 0.076 & 0.001 & 0.001 & 0.052 & 0.001 & 0.056 & 0.001 \\
  sLMb & 0.027 & 0.561 & 0.704 & 0.290 & 0.014 & 0.009 & 0.248 & 0.060 & 0.238 & 0.009 \\
   \cmidrule(lr){2-11}
\end{tabular}
\begin{tabular}{rrrrrrrrrrr}

 & MT & BE & CY & SK & FI & FR & GR & GB & AT & DE \\
   \cmidrule(lr){2-11}
   sLM & 0.001 & 0.001 & 0.001 & 0.001 & 0.167 & 0.001 & 0.122 & 0.001 & 0.001 & 0.001 \\
  sLMb & 0.002 & 0.009 & 0.046 & 0.114 & 0.432 & 0.023 & 0.373 & 0.048 & 0.025 & 0.023 \\
   \cmidrule(lr){2-11}
\end{tabular}
\end{table}
Table~\ref{tab:ppp1} reports the results of the application of the supLM tests on the 20 series. The threshold is searched from the 1st to the 99th percentiles of the sample and the $p$-values for the asymptotic test were obtained from the quantiles of the simulated null distribution. The asymptotic $p$-values indicate a clear rejection of the null hypothesis in 13 out of 20 series. The wild-bootstrap $p$-values are more conservative and account for the possible presence of heteroskedasticity. At level 5\%, they reject the null hypothesis for 11 series. Note that none of the M tests rejects the null hypothesis at the 5\% level (see Table~\ref{tab:S0} of the Supplement). In many situations when the supLM tests reject the null hypothesis the threshold is located in the extreme upper tail of the sample distribution; see also Figure~\ref{fig:S2} of the Supplementary Material for details. The first conclusion that can be drawn is that the mean reversion mechanism is present in most of the series but a large sample size is needed in order to detect it. Another important piece of information, presented in Figure~\ref{fig:1}, is brought by the plot of the values of the sLM statistic versus the threshold $r$. Clearly, two clusters of countries are identifiable. The first one includes those of northern Europe (plus Cyprus and Malta) for which the sLM statistic is well above the 99\%th critical value of the null distribution (red dashed line) and with a higher threshold (around 105). The second cluster identifies the Mediterranean countries plus Finland and  Estonia and is characterized by lower thresholds and values of the statistic just below the 90\%th percentile of the null distribution (purple dashed line). This result can reflect the different sectorial composition of trade of the two groups: the northern countries export is mainly directed at medium to high-tech industries while the Mediterranean countries (plus Finland and Estonia) are more focused on low-tech export \citep[see e.g.][for further details]{Dim08}. Finally, we have tested for regulation from below by applying the test to $-X_t$. The results are presented in Figure~\ref{fig:S8} of the Supplementary Material and show no apparent threshold effects in the lower tail, the only exception being the Great Britain.
\par
\begin{figure}
\centering
\includegraphics[width=0.7\linewidth]{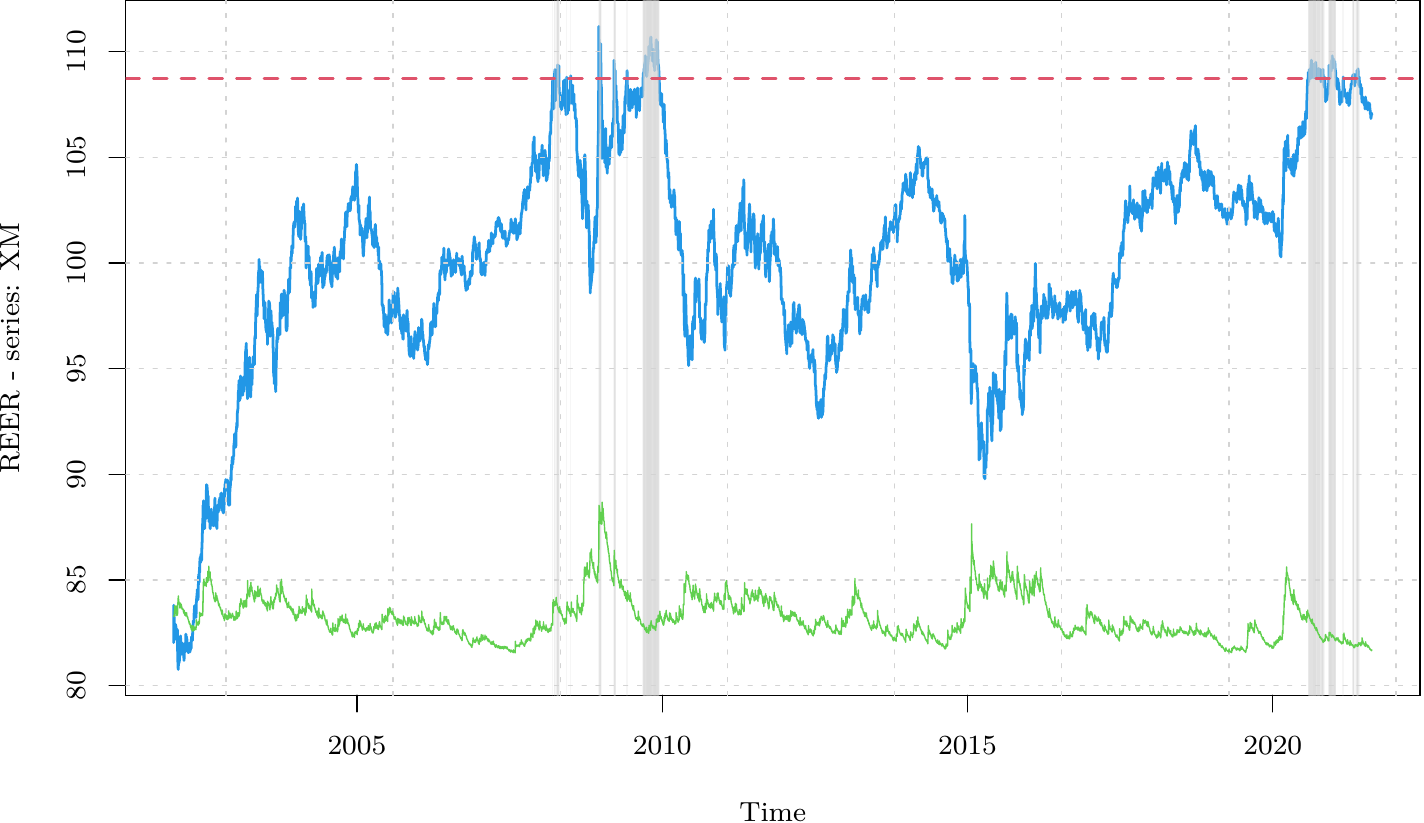}
 \caption{Time series of the real daily exchange rates for the Eurozone series (XM, in blue). The estimated threshold $\hat r=108.74$ that minimises the AIC criterion is indicated with a red dashed line and the gray shaded area indicates the days associated with the upper regime. The estimated volatility from Eq.~(\ref{garch.fit}) is also shown as a green line.}\label{fig:2}
\end{figure}
\begin{figure}
\centering
\includegraphics[width=0.4\linewidth]{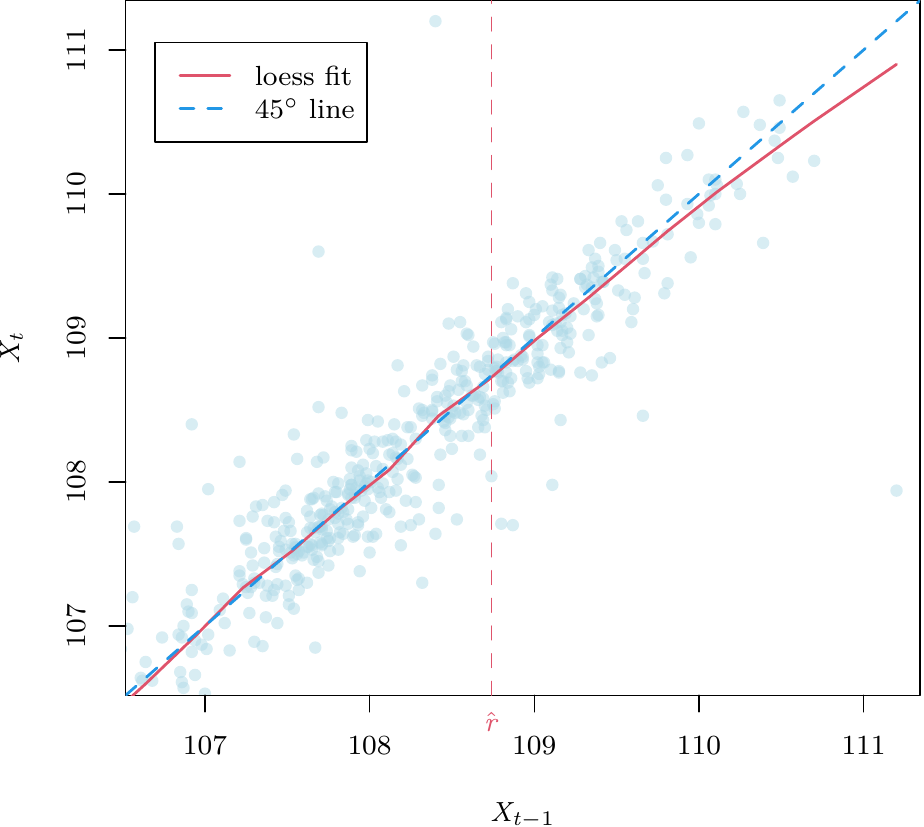}
 \caption{Lag plot of $X_t$ vs $X_{t-1}$ for the daily time series of real exchange rates for the Eurozone (XM).  The 45$^\circ$ line and the nonparametric fit (loess) are shown in the blue dashed line and the red continuous line, respectively. The estimated threshold $\hat r$ is indicated by the vertical red dashed line.}\label{fig:3}
\end{figure}

We now turn to estimating a TARMA model on the global Eurozone series (XM). In Figure~\ref{fig:2} we present the time series (in natural scale) with the estimated threshold that minimizes the AIC criterion $\hat r=108.74$ indicated by a red dashed line. The threshold grid ranges from the 1st to the 99th percentiles of the data and the estimated threshold falls around the 96th percentile. It lends support to the suggestion that the mean reversion mechanism is triggered upon crossing a threshold located in the extreme upper tail of the sample distribution. For this reason, a large sample size is needed and in our case the upper regime contains 204 observations. The gray shaded area denotes the days associated with the upper regime.
\begin{equation}
X_t=
\begin{cases}
 \underset{(0.19)}{0.67} + \underset{(0.04)}{0.86} X_{t-1}  - \underset{(0.07)}{0.41} \eps_{t-1} +\eps_t, & \text{if } X_{t-1} > 108.74\\

\underset{(0.0004)}{0.0135} + \underset{(0.0009)}{0.9971} X_{t-1} + \underset{(0.014)}{0.008} \eps_{t-1} +\eps_t, & \text{if } X_{t-1} \leq 108.74
\end{cases}\label{tarma.fit}
\end{equation}
The parameter estimates are presented in Equation~(\ref{tarma.fit}) (with standard errors in parenthesis). The fitted model, which is globally stationary and ergodic, has a near unit-root lower regime, and an upper TARMA(1,1) regime where the slope of the AR part is clearly smaller than 1.  Figure~\ref{fig:2} shows that the ``intervention'' regime is visited mostly in 2008-2009 and between August 2020 and January 2021. A further confirmation of the goodness of the fit is shown in Figure~\ref{fig:3}, where we present the lag plot of $X_t$ vs $X_{t-1}$, zoomed in the proximity of the threshold, and with the 45$^\circ$ dashed line in blue and the nonparametric fit (loess) in continuous red line. Clearly, the nonparametric fit deviates from the 45$^\circ$ line starting from values close to the estimated threshold.
\par
We note that the MA parameter $\theta$ has played a crucial role in greatly enhancing the fitting capability of the model while retaining parsimony. Now we examine the TARMA model fitted above. While there is no autocorrelation in its residuals (see Figure~\ref{fig:S3} of the Supplement), the test based upon the entropy measure $S_k$, introduced in \cite{Gia15} rejects both the null hypothesis of independence  and that of linearity of the residuals (see Figure~\ref{fig:S4} of the Supplement). Further, an inspection of the correlograms of the squared residuals (Figure~\ref{fig:S5} of the Supplement) and a test for ARCH effects suggest the possible presence of conditional heteroscedasticity. Hence, we improve the TARMA(1,1) model by fitting a GARCH(1,2) model to the residuals $\hat e_t$ and obtain the following:
\begin{align}\label{garch.fit}
\hat e_t &= \sqrt{h_t}z_t \nonumber\\
h_t &= \underset{(0.004)}{0.056} \hat e^2_{t-1} + \underset{(0.01)}{0.49} h_{t-1} +  \underset{(0.01)}{0.45} h_{t-2}.
\end{align}
Here, the conditional distribution of the innovations is assumed to follow the Generalized Error distribution with estimated shape $1.33$ (with standard error 0.03), which accounts for the non-Gaussianity, as shown in Figure~\ref{fig:S6} of the Supplement. The entropy-based test for serial independence applied upon the standardized residuals of the fit shows no residual dependence of any kind (see Figure~\ref{fig:S7} of the Supplement), and this is also confirmed by the battery of diagnostic tests implemented in the R package \texttt{rugarch} \cite{Gha20}  and applied to the fitted GARCH model. (The results are available upon request.) As a further confirmation of the results, we have computed both the supLM and the M tests on simulated series from the fitted models. In particular, the empirical sizes reported in Table~\ref{tab:Sfit1} of the Supplement have been obtained from an ARIMA-GARCH model with GED innovations whose parameters have been estimated on the Eurozone series. While the asymptotic sLM test shows some size bias, probably due to the presence of volatility and heavy tails, the wild bootstrap sLMb test has always a correct size and behaves like the M tests. Table~\ref{tab:Sfit2} of the Supplement reports the empirical power computed on simulated series from the fitted TARMA-GARCH process with GED innovations. (See Eq.~(\ref{tarma.fit}) and (\ref{garch.fit}).) Here, the asymptotic sLM test is the most powerful whereas the wild bootstrap test sLMb and M tests have similar power, the former being slightly more conservative so that a rejection can be taken as a genuine indication of a stationary DGP.
Finally, the estimated volatility from Eq.~(\ref{garch.fit}) is added as a green line to Figure~\ref{fig:2}. The inspection of the plot reveals that the series is characterized by consistent fluctuations of conditional variance but the threshold crossings are not trivially associated to periods of high volatility.
%

\section{Conclusion}\label{sec:conc}

In this paper, we argue that the ubiquity of measurement error implies that to test for regulation in dynamics, it is more appropriate and perhaps even crucially important to formulate the test within a TARMA specification. We adopt  the TARMA(1,1) model as the general hypothesis and the IMA(1,1) model as the null hypothesis. As far as we know, this is the first time that a TARMA specification is used in the present context although it was previously utilized in a very different context, namely for linearity testing under stationarity \citep{Li11}. We derive a Lagrange multiplier test which is asymptotically similar given the threshold search range. Empirical studies confirm that the proposed approach enjoys much higher power in detecting regulation in dynamics than existing tests that do not address measurement errors. The IMA(1,1) model can mimic well other integrated processes with short-memory first differences. In particular, empirical results reported in \cite{Cha20} and in the Supplementary Material indicate that our new tests generally perform well under heteroskedasticity, even when the null hypothesis entails a non-stationary process different from the IMA(1,1) model, and remain powerful for other forms of regulation. The application to the real exchange rates of a panel of time series of the Eurozone sheds new light upon the PPP debate: the evidence points to a threshold-regulated mean reversion mechanism that takes place in the extreme upper tail of the distribution and, as such, a large sample size is needed before a  definitive conclusion can be drawn. This could be the reason behind the failure of many approaches to detect the mean-reversion. Also, the deviations from the random walk regime tend to be compensated within days or weeks and this seems to be consistent with the general expectation. Moreover, the estimated thresholds identify two clusters of countries with different sectorial compositions of trade, thereby prompting further interesting macroeconomic investigations.

\medskip
\subsection*{SUPPLEMENTARY MATERIAL}
 The Supplementary material (pdf format) contains all the proofs, further results from the real data analysis, the tabulated quantiles of the null distribution and further Monte Carlo investigations.

\bibliographystyle{plainnat}
\bibliography{Tarma16}

\newpage
\setcounter{table}{0}
\setcounter{section}{0}
\setcounter{page}{1}
\renewcommand{\thesection}{\Alph{section}}

\begin{center}
\section*{\Large Supplement for:\\ Testing for threshold regulation in presence of measurement error with an application to the PPP hypothesis.}
\end{center}

\bigskip

\begin{center}
\Large
 Kung-Sik Chan, Simone Giannerini, Greta Goracci and Howell Tong
\end{center}

\bigskip\bigskip

\noindent
\textbf{Abstract}\\
\smallskip
This supplement has seven sections. In Section~\ref{sec:Merr} it is shown that a TAR process plus measurement error becomes a TARMA process. Section~\ref{sec:proofs} contains all the proofs. In Section~\ref{sec:ppp} we report further results from the analysis of the panel of 20 time series of daily real exchange rates, plus model diagnostics fromhe TARMA-GARCH fit on the global Eurozone time series. Section~\ref{SM:tab} reports the tabulated quantiles of the asymptotic null distribution of the supLM test for different values of the threshold range. In Section~\ref{SM:power} and Section~\ref{SM:MCmerr} we report the simulation results of the main text for sample sizes equal to 100 and 500. In Section~\ref{SM:MCmerr} we also present Monte Carlo results regarding the empirical power in presence of measurement error and heteroskedasticity. In Section~\ref{SM:MCadd} we focus on the null hypothesis of an IMA(1,1) model against the alternative of a TARMA(1,1) model with an IMA(1,1) regime.

\bigskip\bigskip

\section{TAR process plus measurement error}\label{sec:Merr}
Consider a first-order nonlinear autoregressive model
\begin{equation}
\label{eqn:NAR}
\tilde{X}_t =\phi_0(\tilde{X}_{t-1})+\phi_1(\tilde{X}_{t-1})\tilde{X}_{t-1}+ e_t,
\end{equation}
where $e_t$ is independent of past $\tilde{X}$'s, and $\phi_i(\cdot), i=0,1,$ are  functions which are constant functions for a \emph{linear} AR(1) model and  piecewise constant  functions  for a  TAR model.  Suppose we observe $X_t=\tilde{X}_{t}+\eta_t$ where the independent and identically distributed  noise $\eta$'s  are independent of the $e$'s and $\tilde{X}$'s. Then,
\begin{equation}
\label{eqn:NAR1}
X_t =\phi_0(\tilde{X}_{t-1})+\phi_1(\tilde{X}_{t-1})X_{t-1}+ e_t+\eta_t- \phi_1(\tilde{X}_{t-1}) \eta_{t-1}.
\end{equation}
Let $\Delta_t= e_t+\eta_t- \phi_1(\tilde{X}_{t-1}) \eta_{t-1}$. It is readily checked that $\Delta_t$ is of zero mean and 1-dependent, i.e., $\Delta_t$ and $\Delta_s$ are uncorrelated if $|t-s|>1$. If $\{\Delta_t\}$ is furthermore (second-order) stationary, which holds if $\phi_1(\cdot)$ is  constant  or if  $\{\tilde{X}_t\}$ is stationary, then $\{\Delta_t\}$ can be represented as an MA(1) process \citep[Proposition 3.2.1]{Bro01}. Thus, if $\{\tilde{X}_t\}$ is a random walk,  $\{X_t\}$ is an IMA(1,1) process. Also, if $\{\tilde{X}_t\}$ is a first-order TAR process and the $\eta$'s are of bounded, small support around 0, $\phi_i(\tilde{X}_t)=\phi_i(X_t)$  except in the proximity of the threshold, so $\{X_t\}$ is nearly a first-order TARMA process. The preceding argument can be readily extended to show that a TAR process of order $p$ corrupted with additive measurement noise  may be approximated by a TARMA model of order ($p,p)$.

\section{Proofs}\label{sec:proofs}

\subsection{Proof of Theorem~\ref{thm1}}

Recall $r_n=(1-\theta)\tau\sigma n^{1/2}$.
 Let $\Gamma$ be a $3\times3$ diagonal matrix with diagonal equal to $(\tilde{I}_{1,1}^{-1/2}, \tilde{I}_{1,1}^{-1/2}, 1)$. Define the process
\begin{equation}
H_n =\left\{H_{n}(\tau):=\Gamma^{-1}
\begin{pmatrix}
 n^{-1/2}   \frac{\partial \ell(r_n)}{\partial \phi_0}\\
 n^{-1/2} \frac{\partial \ell(r_n)}{\partial \phi_{1,0}}\\
 n^{-1}  \frac{\partial \ell(r_n)}{\partial \phi_{1,1}}
\end{pmatrix}, \tau\in [r_L,r_U]\right\}.
\end{equation}
Step 1 below shows that
\begin{equation}
H_n  \rightsquigarrow H=\left\{H(\tau):=
\begin{pmatrix}
\int_0^1 dW_s\\ \int_0^1 I(W_s\le \tau) dW_s\\ \int_0^1 W_sI(W_s\le \tau) dW_s
\end{pmatrix},
\tau\in [r_L,r_U]\right\} \label{weakconvergence}
\end{equation}
in the  space of $D_{R^3}[r_L, r_U]$ consisting of CADLAG functions, i.e.,  functions from $[r_U, r_L] \to R^3$ whose component functions are continuous from the right and have left-hand limits, equipped with the supremum  metric.
Step 2 below proves the uniform convergence of $K_n^{-1}\hat{I}_n(r_n)K_n^{-1}\to I(\tau)$, and consequently
$$
Q_n\hat{I}_{2,2,n}(r_n)Q_n-Q_n\hat{I}_{2,1,n}(r_n)\left\{P_n\hat{I}_{1,1,n}(r_n)P_n\right\}^{-1}P_n \hat{I}_{1,2,n}(r_n)Q_n \to I_{2,2}(\tau) -I_{2,1}(\tau)I_{1,1}^{-1}I_{1,2}(\tau).
$$
Step 3 below establishes(19). 
Hence,
\begin{align*}
Q_n\frac{\partial \hat{\ell} }{\partial \bpsi_2}(r_n) =
\begin{pmatrix}
-\tilde{I}_{2,1}(\tau)\tilde{I}_{1,1}^{-1}, \mathrm{I}
\end{pmatrix}
\begin{pmatrix}
  n^{-1/2} \frac{\partial \ell}{\partial \phi_0} \\
 n^{-1/2}  \frac{\partial \ell(r_n)}{\partial \phi_{1,0}}\\
{n}^{-1} \frac{\partial \ell(r_n)}{\partial \phi_{1,1}}
\end{pmatrix}
+ o_p(1)
,\end{align*}
%
 where the $o_p(1)$ term holds uniformly for $r_n\in R_n$, and $\mathrm{I}$ is the $2\times 2$ identity matrix.
 It follows from (13) that 
\begin{align*}
&T_{n}(r_n)=\\
&\left\{Q_n\frac{\partial \hat{\ell} }{\partial \bpsi_2}(r_n) \right\}^\T\left[Q_n\hat{I}_{2,2,n}(r_n)Q_n-Q_n\hat{I}_{2,1,n}(r_n)\left\{P_n\hat{I}_{1,1,n}(r_n)P_n\right\}^{-1}P_n \hat{I}_{1,2,n}(r_n)Q_n\right]^{-1}\\
& \qquad \times \left\{Q_n\frac{\partial \hat{\ell} }{\partial \bpsi_2^\T}(r_n) \right\}
\\
&\rightsquigarrow
\left\|
\left\{(I_{2,2}(\tau) -I_{2,1}(\tau)I_{1,1}^{-1}I_{1,2}(\tau)\right\}^{-1/2}\left( \begin{array}{ll}
-\tilde{I}_{2,1}(\tau)\tilde{I}_{1,1}^{-1}, \mathrm{I}
\end{array}
\right) \Gamma H(\tau)
\right\|^2\\
&=\left\|
\left\{\Lambda_{2,2}(\tau) -\Lambda_{2,1}(\tau)\Lambda_{1,2}(\tau)\right\}^{-1/2}\left\{ H_{2}(\tau)-\Lambda_{2,1}(\tau)H_{1}(\tau)\right\}
\right\|^2
\end{align*}
as a process in $D_R[r_L, r_U]$, thereby  the desired result follows from the fact that the supremum function is a continuous functional from $D_R[r_L, r_U]$ to $R$.

\subsubsection{Step 1: Validity of (\ref{weakconvergence})}

The proof consists of verifying  the tightness of $H_n$ and its finite dimensional convergence to $H$. We first consider tightness. Clearly, $\{ n^{-1/2} \frac{\partial \ell(\tau)}{\partial \phi_{0}}\}$ is tight since it does not depend on $\tau$. Because $\sup_\tau | n^{-1/2} \frac{\partial \ell(\tau)}{\partial \phi_{1,0}}-\mathcal{T}_n(\tau)|=o_p(1)$,  they share the same tightness property.
 The tightness of $\{ n^{-1/2} \frac{\partial \ell(\tau)}{\partial \phi_{1,1}}, r_L \le \tau\le r_U \}$  can be similarly inferred from that of
  $$\widetilde{\mathcal{T}}_n(\tau)= n^{-1/2}\sum_{t=2}^{n}  \frac{\eps_t}{\sigma}\sum_{j=0}^{t-2}\theta^j \frac{X_{t-1-j}}{n^{1/2}\sigma} I\left\{ r_L<\frac{X_{t-1-j}}{n^{1/2}(1-\theta)\sigma}\leq\tau\right\},$$
  for $r_L\le \tau\le r_U$. But then the analogous conditions (23)--(24)
 for $\widetilde{\mathcal{T}}_n$ follow trivially from  (23)--(24) 
 since $r_L, r_U$ are finite numbers.

Note the tightness of $\{\mathcal{T}_n(\tau), \tau\in [r_L,r_U]\}$ follows from $\mathcal{T}_n(a)=O_p(1)$  and verifying below that for all $\tau\in [r_L, r_U]$ and
for each positive $\epsilon$ and $\eta$, there exists a $\delta\in (0,1)$ such that for all sufficiently large $n$,
\begin{equation}
\pr\left\{\sup_{a\le \tau\le s\le \tau+\delta} |\mathcal{T}_n(s)-\mathcal{T}_n(\tau)|\ge \epsilon\right\}\le \eta;
\label{main-cond}
\end{equation}
c.f. \cite[Theorem 8.3]{Bil68}.
 Henceforth in this proof, let $\epsilon<1$ and $\eta$ be two given positive numbers. (The bound $\tau+\delta$ in (\ref{main-cond}) and similar expressions below  will be replaced by $r_U$ if it exceeds $r_U$.)
  It follows from (23) 
  that for all $\tau_2>\tau_1$ such that $\epsilon/n<(\tau_2-\tau_1)^{1/2}$, we have
\begin{equation}
E|\mathcal{T}_n(\tau_2)-\mathcal{T}_n(\tau_1)|^4\le \frac{2C}{\epsilon}(\tau_2-\tau_1)^{3/2}. \label{tC1}
\end{equation}
Let $\zeta$ be a positive number such that $\epsilon/n\le \zeta^{1/2}$. By (\ref{tC1}) and \cite[Theorem 12.2]{Bil68}, for any $\tau\in [r_L,r_U]$ and  positive integer $m$,
\begin{equation}
\pr\left\{\max_{1\le i \le m} |\mathcal{T}_n(\tau+i \zeta)-\mathcal{T}_n(\tau)|\le \lambda\right\} \le \frac{C}{\epsilon\lambda^4}(m\zeta)^{3/2},  \label{max1}
\end{equation}
where $C$ is another constant. By (24), 
there exists a sufficiently large constant $K\ge 1$ such that the LHS of (24) 
is bounded by $K \times L(n)\sqrt{n\log\log n}\zeta+\epsilon$, uniformly for $\tau\in [r_L,r_U]$
and $\zeta\le 1$, except outside  an event, denoted by $A$, which is  of probability not less than $1-\eta$.   On event $A$,
\begin{equation}
\sup_{\tau\le s\le \tau+m\zeta} |\mathcal{T}_n(s)-\mathcal{T}_n(\tau)|\le 3 \max_{i\le m} |\mathcal{T}_n(\tau+i\zeta)-\mathcal{T}_n(\tau)|+K\zeta L(n) \sqrt{n\log\log n}+\epsilon. \label{max2}
\end{equation}
Without loss of generality, the preceding equation is assumed to hold since accounting for $A^c$ will only inconsequentially inflate the RHS of (\ref{main-cond}) from $\eta$ to $2\eta$.
If
\begin{equation}
\epsilon/n\le \zeta^{1/2}\mbox{ and } K\zeta L(n) \sqrt{n\log\log n}<\epsilon,
\end{equation}
then it follows from (\ref{max1}) and (\ref{max2}) that
$$
\pr\left\{\sup_{ \tau\le s\le \tau+m\zeta} |\mathcal{T}_n(s)-\mathcal{T}_n(\tau)|\ge 5\epsilon\right\}\le
\frac{C}{\epsilon^5}(m\zeta)^{3/2}
$$
which becomes
$$
\pr\left\{\sup_{\tau\le s\le \tau+\delta} |\mathcal{T}_n(s)-\mathcal{T}_n(\tau)|\ge 5\epsilon\right\}\le
\eta\delta
$$
upon choosing $\delta=m\zeta$ such that
$C\delta^{1/2}/\epsilon^5<\eta$, which is
feasible if there exists a positive integer $m$ such that
$$
\frac{\delta K L(n)\sqrt{n\log\log n}}{\epsilon}<m\le \frac{\delta}{\epsilon^2}n^2,
$$
but the existence of such an $m$ is guaranteed for all sufficiently large $n$.
Thus, \citet[corollary to Theorem 8.3]{Bil68} entails that
(\ref{main-cond}) holds with $\epsilon$ there
replaced by $15\epsilon$ for all sufficiently large $n$. Since $\epsilon>0$ is
arbitrary,  this completes the proof of the tightness of $H_n$.

Next, we prove the finite-dimensional convergence of $H_n$ to $H$. For simplicity, we only give the proof of $n^{-1}\frac{\partial \ell(r_n)}{\partial \phi_{1,1}}
  \rightsquigarrow \int_0^1 W_sI(W_s\le \tau) dW_s$, via the following uniform approximation argument
\cite[Example 11, p.70]{Pol12}.
Let $G, G_1, G_2,\ldots$ be a sequence of random elements in a metric space $(\mathcal{X}, d)$,  with the support of  $G$ being a separable set of completely regular elements. Suppose for each $\epsilon>0, \delta>0$, there exist approximating random elements $AG, AG_1, AG_2, \ldots$ such that
\begin{description}
\item[$(i)$]  $\pr^*\{d(G, AG)>\epsilon\}<\delta$;
\item[$(ii)$]  $\lim\sup \pr^*\{d(G_n, AG_n\}>\epsilon\}<\delta$;
\item[$(iii)$] $AG_n \rightsquigarrow AG$,
\end{description}
where $\pr^*(\cdot)$ denotes the outer probability measure of the enclosed expression.
Then $G_n\rightsquigarrow G$, as $n \to\infty$.
The complete regularity condition holds here,
with the Euclidean  sample space.
We verify conditions $(i)$--$(iii)$ below.

Recall that
$$
 n^{-1}  \frac{\partial \ell}{\partial \phi_{1,1} }(r_n)=
n^{-1/2}\sum_{t=1}^n   \frac{\eps_t}{\sigma} \frac{1}{1-\theta B}\left[\frac{X_{t-1}}{n^{1/2}\sigma} I\left\{\frac{X_{t-1}}{n^{1/2}(1-\theta)\sigma}\le \tau\right\}\right].
$$
Let $$A_{n,k}=\sum_{t=k+1}^n  n^{-1/2} \frac{\eps_t}{\sigma} \sum_{j=0}^{k}\theta^j\frac{X_{t-1-j}}{n^{1/2}\sigma} I\left\{\frac{X_{t-1-j}}{n^{1/2}(1-\theta)\sigma}\le \tau\right\}.$$
We claim that  for any fixed positive integer $k$ and as $n\to \infty$,
\begin{equation}
A_{n,k} \rightsquigarrow (1-\theta^{k+1}) \times \int_0^1 W_sI(W_s\le \tau)dW_s, \label{weak-convergence}
\end{equation}
 which will be verified  later. We shall check conditions $(i)$--$(iii)$ with
 $G_n= n^{-1}  \frac{\partial \ell(r_n)}{\partial \phi_{1,1} }$,  $AG_n=A_{n,k}$,
 $G=\int_0^1 W_sI(W_s\le \tau)dW_s$ and $AG=(1-\theta^{k+1})G$; indeed condition $(iii)$ obtains due to (\ref{weak-convergence}). Clearly, $(i)$ holds by   Slutsky's theorem. It remains to show $(ii)$, which can be done by first bounding the difference
\begin{align*}
D_{n,k} &=  n^{-1}  \frac{\partial \ell(r_n)}{\partial \phi_{1,1} }-A_{n,k}\\
&= \sum_{t=1}^{k} n^{-1/2} \frac{\eps_t}{\sigma} \sum_{j=0}^{t-1}\theta^j\left[\frac{X_{t-1-j}}{n^{1/2}\sigma} I\left\{\frac{X_{t-1-j}}{n^{1/2}(1-\theta)\sigma}\le \tau\right\}\right] + \\
&\quad \sum_{t=k+1}^n  n^{-1/2} \frac{\eps_t}{\sigma} \sum_{j=k+1}^{t-1}\theta^j\left[\frac{X_{t-1-j}}{n^{1/2}\sigma} I\left\{\frac{X_{t-1-j}}{n^{1/2}(1-\theta)\sigma}\le \tau\right\}\right].
\end{align*}
The summands of $D_{n,k}$ form a martingale difference sequence with respect to the $\sigma$-algebra $\mathcal{F}_t$ generated by the innovations $\eps_{t-j}, j\ge 0$; hence $D_{n,k}$ is of zero mean and Jensen's inequality implies that its variance is bounded by
\begin{equation}\label{eqn:Dn-bound}
\frac{K}{(1-|\theta|)^2}\left \{  \frac{k(k-1)}{2n^2} +|\theta|^{k+1}\frac{1}{2}\right \},
\end{equation}
where $K>0$ is a constant such that
$E(X_t^2)=\sigma^2\{1+\theta^2+(t-1)(1-\theta)^2\}\le tK\sigma^2$. Since the true $\theta$ is less than 1 in magnitude, (\ref{eqn:Dn-bound}) indicates that for any positive $\epsilon$,  by choosing $k$ sufficiently large and then letting $n\to\infty$, $\pr(|D_{n,k}|\le \epsilon)\to 1$. Thus, $(ii)$ holds by Markov's inequality.

It remains to verify (\ref{weak-convergence}).  The claim (\ref{weak-convergence}) would follow readily from Theorem 7.10 in \citet{Kur96} were the step function $I(x \le \tau)$ a continuous function. Unfortunately, this is not the case but it is discontinuous only at $\tau$. The idea of proof is to approximate the step function by a net of smooth functions, say $G_\delta(x)$, such that $|G_\delta(x)-I(x\le \tau)|\le V_\delta (x)$ with the bound $V_\delta (x)$ being uniformly bounded, continuous functions and with  support  inside $[\tau-\delta, \tau+\delta]$. Define $$A_{n,k,\delta}=\sum_{t=k+1}^n  n^{-1/2} \frac{\eps_t}{\sigma} \sum_{j=0}^{k}\theta^j\frac{X_{t-1-j}}{n^{1/2}\sigma} G_\delta\left\{\frac{X_{t-1-j}}{n^{1/2}(1-\theta)\sigma}\right\}.$$
Then, for fixed $k$ and $\delta$, $A_{n,k,\delta}\rightsquigarrow (1-\theta^{k+1})\int_0^1 W_sG_\delta(W_s)dW_s,$  as $n\to\infty$. Consider
\begin{equation*}
A_{n,k}-A_{n,k,\delta} =
\sum_{t=k+1}^n  n^{-1/2} \frac{\eps_t}{\sigma} \sum_{j=0}^{k}\theta^j\frac{X_{t-1-j}}{n^{1/2}\sigma} \left[ I\left\{ \frac{X_{t-1-j}}{n^{1/2}(1-\theta)\sigma}\le r \right\} -G_\delta\left\{\frac{X_{t-1-j}}{n^{1/2}(1-\theta)\sigma}\right\}\right],
\end{equation*}
whose summands form a martingale difference sequence, so it is of zero mean and its variance can be bounded as follows:

\begin{align*}
&E(A_{n,k}-A_{n,k,\delta})^2\le
\frac{1}{1-|\theta|}\sum_{t=k+1}^n   n^{-1}
E\left[
\sum_{j=0}^{k} |\theta|^j\left(\frac{X_{t-1-j}}{n^{1/2}\sigma}\right)^2V_\delta^2\left\{\frac{X_{t-1-j}}{n^{1/2}(1-\theta)\sigma}\right\}
\right] \nonumber \\
&\le\frac{(1-\theta)^2\max(|r-\delta|^2, |r+\delta|^2)}{1-|\theta|}
E\left[ \sum_{t=k+1}^n   n^{-1}
\sum_{j=0}^{k} |\theta|^j V_\delta^2\left\{\frac{X_{t-1-j}}{n^{1/2}(1-\theta)\sigma}\right\}
\right].
\end{align*}
On the other hand, it follows from Theorem 7.10 in \citet{Kur96} that
\begin{equation}
\sum_{t=k+1}^n   n^{-1}
\sum_{j=0}^{k} |\theta|^j V_\delta^2\left\{\frac{X_{t-1-j}}{n^{1/2}(1-\theta)\sigma}\right\} \rightsquigarrow \frac{1-|\theta|^{k+1}}{1-|\theta|} \int_0^1 V_\delta^2(W_s)ds.
\label{expectation-bdd}
\end{equation}
Since $V^2_\delta(\cdot)$ is continuous, uniformly bounded, say, by $K>0$, and its support lies inside $[r-\delta, r+\delta]$, the expectation of the LHS of (\ref{expectation-bdd}) converges to
\begin{equation*}
 E \left\{\int_0^1 V_\delta^2(W_s)ds\right\}
  \le  K\left[\nu + \int_\nu^1 \int_{\tau-\delta}^{\tau+\delta} \frac{1}{\sqrt{2\pi}s}\exp\{-y^2/(2s)\}dyds \right],
\end{equation*}
where $0< \tau < 1$ can be chosen to be an arbitrary small, fixed number, and then the double integral having a bounded integrand can be made arbitrarily small by rendering $\delta>0$ small.
Hence, for any fixed $\epsilon>0$ and $\gamma>0$ it holds that for all sufficiently small $\delta$,
$\pr(|A_{n,k}-A_{n,k,\delta}|>\epsilon)<\gamma$  for all sufficiently large $n$.
Also, the difference $\int_0^1 W_s G_\delta(W_s)dW_s-\int_0^1 W_sI(W_s\le r)dW_s= \int_0^1 W_s \{G_\delta(W_s)-I(W_s\le r)\}dW_s$ is of zero mean and variance equal to
\begin{align*}
&\int_0^1 E\left[W^2_s \{G_\delta(W_s)-I(W_s\le \tau)\}^2\right]ds\\
&\le K \max(|\tau-\delta|^2, |\tau+\delta|^2)
\int_0^1 \pr(W_s\in [\tau-\delta, \tau+\delta])ds
\end{align*}
which can be similarly shown to be made arbitrarily small for all sufficiently small $\delta$. Thus, the claim (\ref{weak-convergence}) can be verified,  by routine arguments; c.f. \cite[Example 11, p.70]{Pol12}.

\subsubsection{Step 2: Uniform Convergence of $K_n^{-1}\hat{I}_n(r_n)K_n^{-1}\to I(\tau)$, for $r_n\in R_n$}

Recall that $\hat{I}_n(r_n)$ is the observed Fisher information matrix evaluated at $\bpsi_1=\hat{\bpsi}_1$, $\bpsi_2=0$ and threshold parameter set as $r_n$, while $I_n(r_n)$ evaluated with $\bpsi$ evaluated at the true value under $H_0$ and threshold at $r_n$.
We give the proof that  $K_n^{-1}I_n(r_n)K_n^{-1}\to I(\tau)$, uniformly for $ r_n=(1-\theta)\tau\sigma n^{1/2}\in R_n$, and then indicate how to modify the proof to lift the result for the uniform convergence of $K_n^{-1}\hat{I}_n(r_n)K_n^{-1}\to I(\tau)$. This will be shown componentwise, with greater details for the (5,5)-th component and a sketchier proof for the (4,2)th component, as a prototype of the proof for other components. Consider then
\begin{align*}
\frac{1}{n^2}\frac{\partial^2\ell }{\partial \phi_{1,1}^2 }(r_n) &=\frac{1}{n^2}\sum_{t=1}^n  \frac{1}{\sigma^2}\frac{\partial \eps_t}{\partial \phi_{1,1}}\frac{\partial \eps_t}{\partial \phi_{1,1}} + \frac{1}{n^2}\sum_{t=1}^n \frac{\eps_t}{\sigma^2} \frac{\partial^2 \eps_t}{\partial \phi_{1,1}^2}\\
&=  n^{-1}  \sum_{t=1}^n \left\{\sum_{s=0}^{t-1} \theta^{t-1-s} \frac{X_s}{n^{1/2}\sigma} I(X_s\le r_n)\right\}^2,
\end{align*}
because the second term  in the first equality is identically 0. For any real number $x$, let $[x]$ denote the greatest integer not larger than $x$.  By adapting the proof technique in Step 1, it can be readily shown that for fixed $\tau$ and as $n\to\infty$,  $$\left\{\sum_{s=0}^{[tn]} \theta^{t-1-s} \frac{X_s}{n^{1/2}\sigma} I(X_s\le r_n), 0\le t\le 1\right \}\rightsquigarrow \{W_t I(W_t \le \tau), 0\le t\le 1\}.$$
Applying Theorem 7.10 of \cite{Kur96},
we have $\frac{1}{n^2}\frac{\partial^2\ell(\bpsi_0;r_n) }{\partial \phi_{1,1}^2 }\rightsquigarrow \int_0^1 W_s^2I(W_s\le \tau)ds$. To strengthen the preceding convergence to uniform convergence for $r_n\in R_n$, it suffices to show the uniform boundedness in probability of the increment
$\Delta_n(\tau_1, \tau_2)=\left|\frac{1}{n^2}\frac{\partial^2\ell(\bpsi_0;r_{2,n}) }{\partial \phi_{1,1}^2 }-\frac{1}{n^2}\frac{\partial^2\ell(\bpsi_0;r_{1,n}) }{\partial \phi_{1,1}^2 }\right|$, where $r_{i,n}=(1-\theta)\tau_i\sigma$, for all $r_L\le c\le \tau_1 \le  \tau_2\le d\le r_U$ sufficiently close. Indeed,
\begin{align*}
&\sup_{c\le \tau_1\le\tau_2\le d}\Delta_n(\tau_1, \tau_2)  \\
&\leq \frac{2}{n}\sum_{t=1}^n \left[\sum_{s=0}^{t-1} |\theta|^{t-1-s} \frac{|X_s|}{n^{1/2}\sigma} I\left\{c< \frac{X_s}{n^{1/2}(1-\theta)\sigma} \le  d\right\}\right]\left(\sum_{s=0}^{t-1} |\theta|^{t-1-s} \frac{|X_s|}{n^{1/2}\sigma} \right),
\end{align*}
with the preceding bound converges weakly to $\int_0^1 |W_s|^2I(c<W_s\le d)$, as $n\to\infty$.

Write $\frac{1}{n^2}\frac{\partial^2\hat{\ell} }{\partial \phi_{1,1}^2 }(r_n)$ for $\frac{1}{n^2}\frac{\partial^2\ell }{\partial \phi_{1,1}^2 }(\bpsi_1=\hat{\bpsi}_1, \bpsi_2=0;r_n)$. To lift the preceding result to the uniform convergence of $\frac{1}{n^2}\frac{\partial^2\hat{\ell} }{\partial \phi_{1,1}^2 }(r_n)$ to $\int_0^1 W_s^2I(W_s\le \tau)ds$, first note that it follows from the consistency of $\hat{\theta}$ that there exists an $0<\eta<1$ such that the event  $B=\{\max(|\hat{\theta}|, |\theta|)<\eta\}$ holds with probability approaching 1 as $n \to\infty$. Without loss of generality, it is assumed  that $\max(|\hat{\theta}|, |\theta|)<\eta<1$. Therefore, for any $r_n$,
\begin{align*}
&\left|\frac{1}{n^2}\frac{\partial^2\hat{\ell} }{\partial \phi_{1,1}^2 }(r_n)-\frac{1}{n^2}\frac{\partial^2\ell }{\partial \phi_{1,1}^2 }(r_n)\right|\\
\le&  \frac{2}{n}|\hat{\theta}-\theta|\sum_{t=2}^n \left\{\sum_{s=0}^{t-2} (t-1-s)\eta^{t-2-s} \frac{|X_s|}{n^{1/2}\sigma} I(X_s\le r_n)\right\}  \left\{\sum_{s=1}^{t-1} \eta^{t-1-s} \frac{|X_s|}{n^{1/2}\sigma} I(X_s\le r_n)\right\}\\
\le&  \frac{K}{n}|\hat{\theta}-\theta|\sum_{t=2}^n   \left(\sum_{s=1}^{t-1} \tilde{\eta}^{t-1-s} \frac{|X_s|}{n^{1/2}\sigma} \right)^2=o_p(1),
\end{align*}
for some $\eta<\tilde{\eta},1$ and constant $K$, hence the desired convergence result for $\frac{1}{n^2}\frac{\partial^2\hat{\ell} }{\partial \phi_{1,1}^2 }(r_n)$.

Similarly, it can be shown that
$$
 n^{-1}  \frac{\partial^2\ell(\bpsi_0;r_n) }{\partial \phi_{1,0}^2 }\rightsquigarrow \frac{1}{(1-\theta)^2\sigma^2}\int_0^1 I(W_s\le \tau)ds.
$$
Next, consider
\begin{align*}
 n^{-1}  \frac{\partial^2\ell(\bpsi_0;r_n) }{\partial \phi_{1,0}\partial \theta } &= n^{-1}  \sum_{t=1}^n
\frac{1}{\sigma^2}\frac{\partial \eps_t}{\partial \phi_{1,0}}\frac{\partial \eps_t}{\partial \theta} + \frac{1}{n^2}\sum_{t=1}^n
\frac{\eps_t}{\sigma^2} \frac{\partial^2 \eps_t}{\partial \phi_{1,0}\partial \theta}\\
&= -\frac{1}{n\sigma^2}\sum_{t=1}^n \left(\sum_{j=0}^{t-2}\theta^j\eps_{t-1-j}\right)\left\{\sum_{j=0}^{t-1}\theta^jI\left(X_{t-1-j}\le r\right)\right\},
\end{align*}
which is shown to be $o_p(1)$ as follows. Using similar argument as in the proof of Step 1, we can approximate the right side of the last equality by
\begin{equation}
-\frac{1}{n\sigma^2}\sum_{t=[n\delta]}^n \left(\sum_{j=0}^{k}\theta^j\eps_{t-1-j}\right)\left[\sum_{j=0}^{k}\theta^jG_\tau\left\{\frac{X_{t-1-j}}{n^{1/2} (1-\theta)\sigma}\right\}\right]
\label{approxbd}
\end{equation}
where $0<\delta<1$ and $G_\tau$ is uniformly bounded and  uniformly Lipschitz continuous function, i.e., with an identical Lipschitz constant, for $r_L<\tau<r_U$; by suitably choosing $\delta, G_\tau$ and $k$ sufficiently large, the approximation error can be made arbitrarily small in magnitude, uniformly for $r_n\in R_n$,  in probability. Eqn. (\ref{approxbd}) can be further approximated by
\begin{equation}
-\frac{1}{n\sigma^2}\sum_{t=[n\delta]}^n \left(\sum_{j=0}^{k}\theta^j\eps_{t-1-j}\right)\left[\sum_{j=0}^{k}\theta^jG_\tau\left\{\frac{X_{t-2-k}}{n^{1/2} (1-\theta)\sigma}\right\}\right],
\label{approxbd1}
\end{equation}
which is of zero mean and variance of order $O(1/n)$.
The uniform Lipschitz continuity of $G_\tau(\cdot)$ implies that the approximation error is uniformly $O_p(n^{-1/2})$.
Altogether, $ n^{-1}  \frac{\partial^2\ell(\bpsi_0;r_n) }{\partial \phi_{1,0}\partial \theta }=o_p(1)$, which, as before,  can be similarly shown to hold uniformly for $r_n\in R_n$. The desired convergence for other components can similarly be established.

\subsubsection{Step 3: Proof of (\ref{eq:exactscoreapprox})}
 It follows from the discussion below (19) 
 that   it suffices to verify (20) 
 , which  we show  componentwise,
\begin{align}
 n^{-1/2} \frac{\partial \hat{\ell} (r_n)}{\partial \phi_{1,0}}&= n^{-1/2} \frac{\partial \ell (r_n)}{\partial \phi_{1,0}}-\left\{\frac{1}{(1-\theta)^2 \sigma^2}\int_0^1 I(W_s\le \tau) ds\right\}n^{1/2}\left(\hat\phi_{0}-\phi_0\right)+o_p(1) \label{eq:equality1}\\
 n^{-1}  \frac{\partial \hat{\ell}(r_n) }{\partial \phi_{1,1}}&= n^{-1}  \frac{\partial \ell(r_n) }{\partial \phi_{1,1}}-\left\{\frac{1}{(1-\theta) \sigma}\int_0^1 W_sI(W_s\le \tau) ds\right\}n^{1/2}\left(\hat\phi_{0}-\phi_0\right)+o_p(1),  \label{eq:equality12}
\end{align}
 where
 all $o_p(1)$ terms within this proof hold uniformly for $r_n\in R_n$.
We prove only (\ref{eq:equality1}), since (\ref{eq:equality12}) can be similarly proved. Within this proof, $\hat{\eps}_t$ denotes $\eps_t$ evaluated at $\bpsi_1=\hat{\bpsi}_1$, $\bpsi_2=0$, which does not depend on the threshold $r$. Also, $\frac{\partial\hat{\eps}_t}{\partial\phi_{1,0}}$ ($\frac{\partial\hat{\eps}_t}{\partial\phi_{0}}$) denotes $\frac{\partial\eps_t}{\partial\phi_{1,0}}$ ($\frac{\partial\eps_t}{\partial\phi_{0}}$) evaluated at $\bpsi_1=\hat{\bpsi}_1$, $\bpsi_2=0$, and  $r=r_n$; sometimes we write $\frac{\partial\hat{\eps}_t (r_n)}{\partial\phi_{1,0}}$, etc. to highlight that the threshold parameter is set at $r_n$. On the other hand, their counterparts without the hat sign are evaluated at the true $\bpsi$ value under $H_0$, and the threshold parameter equal to $r_n$. In particular, $\eps_t$ represent the true innovation at epoch $t$ in the rest of this  proof.
It follows from (10), (8), (9) 
and routine algebra that
\begin{align}
\hat\eps_t-\eps_t&=(\phi_0-\hat\phi_0)\sum_{j=0}^{t-1}\hat\theta^j+(\hat\theta-\theta)\sum_{j=0}^{t-1}\theta^j\eps_{t-1-j}+\hat\theta^t\eps_0; \label{eq:ehat}\\
\frac{\partial\hat\eps_t}{\partial\phi_{1,0}}-\frac{\partial\eps_t}{\partial\phi_{1,0}}&=(\hat\theta-\theta)\sum_{j=0}^{t-1}\theta^j\frac{\partial\eps_{t-1-j}}{\partial\phi_{1,0}}-\hat\theta^t\frac{\partial\eps_0}{\partial\phi_{1,0}}.
\label{eq:pehat}
\end{align}
Since $\frac{1}{\hat\sigma^2}-\frac{1}{\sigma^2}=\frac{\sigma^2-\hat{\sigma}^2}{\sigma^2\hat\sigma^2}=O_p(n^{-1/2})$, the following equality holds, up to an asymptotically negligible additive term,
\begin{align*}
 n^{-1/2} \frac{\partial\hat\ell (r_n)}{\partial\phi_{1,0}}&=- n^{-1/2} \sum_{t=1}^{n}\frac{\hat\eps_{t}}{\sigma^2}\frac{\partial\hat\eps_t}{\partial\phi_{1,0}}\\
=& n^{-1/2} \frac{\partial\ell (r_n)}{\partial\phi_{1,0}}
+\left( n^{-1/2} \sum_{t=1}^{n}\frac{\eps_{t}}{\sigma^2}\frac{\partial\eps_t}{\partial\phi_{1,0}}- n^{-1/2} \sum_{t=1}^{n}\frac{\hat\eps_{t}}{\sigma^2}\frac{\partial\hat\eps_t}{\partial\phi_{1,0}}\right).
\end{align*}
The term enclosed by round brackets equals
\begin{align*}
&\frac{1}{n^{1/2}\sigma^2}\sum_{t=1}^{n}\left\{\eps_t\frac{\partial\eps_t}{\partial\phi_{1,0}}-\hat\eps_t\frac{\partial\hat\eps_t}{\partial\phi_{1,0}}+\left(\hat\phi_0-\phi_0\right)\frac{\partial\eps_t}{\partial\phi_{1,0}}\frac{\partial\eps_t}{\partial\phi_{0}}\right\}
\\
&\qquad
-n^{1/2}\left(\hat\phi_0-\phi_0\right) n^{-1}  \sum_{t=1}^{n}\frac{1}{\sigma^2}\frac{\partial\eps_t}{\partial\phi_{1,0}}\frac{\partial\eps_t}{\partial\phi_{0}}\\
&:=A_n(r_n)-B_n(r_n).
\end{align*}
Using techniques in Step 1, we have
\[
\left\{ n^{-1}  \sum_{t=1}^{n}\frac{1}{\sigma^2}\frac{\partial\eps_t(r_n)}{\partial\phi_{1,0}}\frac{\partial\eps_t(r_n)}{\partial\phi_{0}},r_n  \in R_n\right\} \rightsquigarrow\left\{\frac{1}{(1-\theta)^2\sigma^2}\int_{0}^{1}I\left(W_s\leq\tau\right)ds,\tau\in [r_L, r_U]\right\}.
\]
 Thus, (\ref{eq:equality1}) holds if we verify that $A_n(r_n)=o_p(1)$.
Eqns.  (\ref{eq:ehat}) and (\ref{eq:pehat}) entail that
\begin{align*}
&A_n(r_n)=\\
&\frac{1}{\sqrt n\sigma^2} \sum_{t=1}^{n} \left[\frac{\partial\eps_t}{\partial\phi_{1,0}}\left(\hat\phi_0-\phi_0\right)\sum_{j=0}^{t-1}\hat\theta^j - \frac{\partial\eps_t}{\partial\phi_{1,0}}\left(\hat\theta-\theta\right)\sum_{j=0}^{t-1}\theta^j\eps_{t-1-j} - \frac{\partial\eps_t}{\partial\phi_{1,0}}\hat\theta^t\eps_0\right.\\
&\qquad\left.-\hat\eps_t\left\{(\hat\theta-\theta)\sum_{j=0}^{t-1}\theta^j\frac{\partial\eps_{t-1-j}}{\partial\phi_{1,0}}-\hat\theta^t \frac{\partial\eps_0}{\partial\phi_{1,0}}\right\}+\left(\hat\phi_0-\phi_0\right) \frac{\partial\eps_t}{\partial\phi_{1,0}}\frac{\partial\eps_t}{\partial\phi_{0}}\right].
\end{align*}
Because $\hat{\theta}$ is consistent and the true value $|\theta|<1$, there exists $0<\gamma<1$ such that the event $\mathcal{E}_n=\{ |\theta| <\gamma, |\hat{\theta}|<\gamma\}$ holds with  probability approaching 1  as $n\to\infty$. Thus, with no loss of generality,  $\mathcal{E}_n$  is assumed to hold. Consequently, there exists  a positive constant $K$ such that for all $t\ge 1$,
$$
|\sum_{j=0}^t \hat{\theta}^j -\sum_{j=0}^t \theta^j| \le |\hat{\theta} -\theta| \sum_{j=1}^t j \gamma^{j-1} \le K |\hat{\theta} -\theta|,
$$
hence
\begin{align*}
&\left|\frac{1}{n^{1/2}\sigma^2}\sum_{t=1}^n\left\{\frac{\partial\eps_t}{\partial\phi_{1,0}} \left(\hat\phi_0-\phi_0\right)\sum_{j=0}^{t-1}\hat\theta^j+\left(\hat\phi_0-\phi_0\right) \frac{\partial\eps_t}{\partial\phi_{1,0}}\frac{\partial\eps_t}{\partial\phi_{0}}\right\}\right|\\
&\le K \sigma^{-2} |\hat\phi_0-\phi_0| n^{1/2}|\hat{\theta} -\theta|
 n^{-1}  \sum_{t=1}^n\left|   \frac{\partial\eps_t}{\partial\phi_{1,0}}\right|=o_p(1),
\end{align*}
because on $\mathcal{E}_n$, $\frac{\partial\eps_t}{\partial\phi_{1,0}}$ is uniformly bounded by $(1-\gamma)^{-1}$ in magnitude.
It can be similarly proved  that
\begin{align*}
\frac{1}{n^{1/2}\sigma^2}\sum_{t=1}^{n}&\left[-\frac{\partial\eps_t}{\partial\phi_{1,0}} \left(\hat\theta-\theta\right)\sum_{j=0}^{t-1}\theta^j\eps_{t-1-j}- \frac{\partial\eps_t}{\partial\phi_{1,0}} \hat\theta^t\eps_0\right.\\
&\left.-\hat\eps_t\left\{(\hat\theta-\theta)\sum_{j=0}^{t-1}\theta^j\frac{\partial\eps_{t-1-j}}{\partial\phi_{1,0}}- \hat\theta^t\frac{\partial\eps_0}{\partial\phi_{1,0}}\right\}\right]=o_{p}(1),
\end{align*}
hence $A_n(r_n)=o_p(1)$, which completes the proof of Step 3.
%
%
\subsection{Proof of Theorem~\ref{thm2}}
We verify  (23) 
first by noting  that its LHS  equals
  \begin{align*}
&\frac{1}{\sigma^4n^2}\sum_{t_1,t_2,t_3,t_4}E\left\{\eps_{t_1}\eps_{t_2}\eps_{t_3}\eps_{t_4}\times I(t_1)
I(t_2)I(t_3)I(t_4)\right\}
\end{align*}
where
\begin{align*}
I(s)&=\sum_{j=0}^{s-2}\theta^j I\left\{\tau_1\leq \frac{X_{s-1-j}}{n^{1/2}(1-\theta)\sigma}\leq\tau_2\right\}.
\end{align*}
Below, $K, C, C_1, C_2$ denote  generic  constants that may depend on $\theta$ and may vary from occurrence to occurrence.
Assume that $t=\max_{i\in\{1,2,3,4\}}t_i$.
Denote $E_t(\cdot)$ as the conditional expectation w.r.t. $\mathcal{F}_t$, an increasing sequence of $\sigma$-algebras such  that   $X_0$ and the $\eps_t$ are measurable w.r.t. $\mathcal{F}_t$. We shall impose the condition that for all $t$,  $E_{t-1}(\eps_t)=E_{t-1}(\eps_t^3)=0$ and the magnitude of $E_{t-1}(\eps_t^\ell), \ell=2,4$ is uniformly bounded for all $t$, which are clearly satisfied under the independent and identically normally distributed innovation assumption.
The law of iterated expectations implies that
\begin{align*}
&\frac{1}{\sigma^4n^2}\sum_{t_1,t_2,t_3,t_4}E\left\{\eps_{t_1}\eps_{t_2}\eps_{t_3}\eps_{t_4}\times I(t_1)
I(t_2)I(t_3)I(t_4)\right\}\\
=&\frac{1}{\sigma^4n^2} \sum_{t=2}^{n-1}E\left\{\eps_t^4I^4(t)\right\} +\frac{K}{\sigma^4n^2}E\left\{\sum_{t=2}^{n-1}\eps_t^2I^2(t)\left\{\sum_{u<t}\eps_uI(u)\right\}^2\right\}.
\end{align*}
Below, we show that there exist two constants $C_1,C_2>0$ such that
\begin{align}
  \frac{1}{n^2}\sum_{t=2}^{n-1}E\left\{\eps_t^4I^4(t)\right\}&\leq \frac{C_1}{n}(\tau_2-\tau_1)\label{eqn:condition1}\\
  \frac{K}{n^2}E\left[\sum_{t=2}^{n-1}\eps_t^2I^2(t)\left\{\sum_{u<t}\eps_uI(u)\right\}^2\right]&\leq K(|\tau_2-\tau_1|^{3/2}+|\tau_2-\tau_1|/n)\label{eqn:condition2}
\end{align}
We verify (\ref{eqn:condition2}), by first applying Doob's inequality and then Rosenthal's inequality for martingale difference sequence \citep{Bur73, Hit90} to derive the second and fifth inequality, respectively, in the following display:
\begin{align*}
&\frac{K}{\sigma^4n^2}E\left[\sum_{t=2}^{n-1}\eps_t^2I^2(t)\left\{\sum_{2\le u<t}\eps_uI(u)\right\}^2\right] \le \frac{K}{\sigma^4n}E\left[\sup_{2\le t \le n-1}\eps_t^2I^2(t)\left\{\sum_{2\le u<t}\eps_uI(u)\right\}^2\right]\\
&\le \frac{K}{\sigma^4n}E\left[\eps_{n-1}^2I^2(n-1)\left\{\sum_{2\le u<n-1}\eps_uI(u)\right\}^2\right]
 \le \frac{K}{\sigma^4n}E^{1/2}\left\{\eps_{n-1}^4I^4(n-1)\right\}E^{1/2}\left\{\sum_{2\le u<n-1}\eps_uI(u)\right\}^4\\
&\le \frac{K}{\sigma^4n}E^{1/2}\left\{I^4(n-1)\right\}E^{1/2}\left\{\sum_{2\le u<n-1}\eps_uI(u)\right\}^4\\
&\le\frac{K}{\sigma^4n}E^{1/2}\left\{I^4(n-1)\right\}\left(E\left[\sum_{2\le u<n-1}E_{u-1}\left\{\eps_u^2I^2(u)\right\}\right]^2+ E\left\{\sup_{2\le u \le n-1} |\eps_uI(u)|^4\right\}\right)^{1/2}\\
&\le\frac{K}{\sigma^4n}E^{1/2}\left\{I^4(n-1)\right\}\left[E\left\{\sum_{2\le u<n-1}I(u)\right\}^2+ E\left\{\sum_{2\le u \le n-1} \eps_u^4I^4(u)\right\}\right]^{1/2}\\
&\le\frac{K}{\sigma^4}E^{1/2}\left\{I^4(n-1)\right\}\left[\frac{1}{n^2} E\left\{\sum_{2\le u<n-1}I(u)\right\}^2+ \frac{1}{n^2} E\left\{\sum_{2\le u \le n-1} I^4(u)\right\}\right]^{1/2} \\
&\le \frac{K}{\sigma^4} \left(|\tau_2-\tau_1|^{3/2}+|\tau_2-\tau_1|/n \right),
\end{align*}
with the last line ensuing from  the claims that $(a)$ $E\left\{I^4(n-1)\right\}$ and $E\left\{\sum_{u=2}^{n-1} I^4(u)/n\right\}$ are bounded by some  multiple of $|\tau_2-\tau_1|$ and $(b)$ $E\left\{\sum_{u=2}^{n-1}I(u)\right\}^2/n^2$ is bounded by some multiple of $|\tau_2-\tau_1|^2$, uniformly for all $a\le \tau_1< \tau_2\le b$, i.e., the constant multipliers can be chosen to depend on $-1<\theta<1$ only. For proving (a), note that under the normal innovation assumption, $ \frac{X_{s+1}}{n^{1/2}(1-\theta)\sigma} \sim N\left\{0, \frac{s(1-\theta)^2+1+\theta^2}{n (1-\theta)^2)}\right\}$, hence its probability density function is bounded by ${\sqrt{n(1-\theta)^2}}/{\sqrt{2\pi\left\{(1-\theta)^2s+1+\theta^2\right\}}}$. Then
\begin{align*}
& E\left\{I^4(n-1)\right\} \\
&\leq K E\left[\sum_{s=0}^{n-2}|\theta|^{n-2-s}I\left\{\tau_1\leq \frac{X_{s+1}}{n^{1/2}(1-\theta)\sigma}\leq\tau_2\right\}\right]\\
&\le  K \sum_{s=0}^{n-2}|\theta|^{n-2-s}\left[\int_{\tau_1}^{\tau_2}\left\{\frac{\sqrt{n(1-\theta)^2}}{\sqrt{2\pi\left\{(1-\theta)^2s+1+\theta^2\right\}}}\right\}dy\right]\\
& \le  K|\tau_2-\tau_1|\sum_{s=0}^{n-2}\frac{|\theta|^{n-2-s}\sqrt{n(1-\theta)^2}}{\sqrt{2\pi\left\{(1-\theta)^2s+1+\theta^2\right\}}}.
\end{align*}
The preceding sum is bounded for any fixed $|\theta|<1$. To see this, let $0<\alpha<1$ be fixed and we split the summation into  $S_{1,n}+S_{2n}$, the first of  which sums over $s\ge \alpha n$,  and the second  over $s<\alpha n$. Then $S_{1n}$ is bounded by $\sum_{s\ge \alpha n}^{n-2} {|\theta|^{n-2-s}\sqrt{(1-\theta)^2}}/{\sqrt{2\pi(1-\theta)^2\alpha}}$ which is  bounded for $|\theta|<1$. Also $S_{2n}$ is bounded by $|\theta|^{n-\alpha n -2} n^{1/2}\sum_{0\le s< \alpha n} |\theta|^{\alpha n -s}$ which is  clearly bounded for $|\theta|<1$. Thus, $E\left\{I^4(n-1)\right\}$ is uniformly bounded by some  multiple of $|\tau_2-\tau_1|$ and so is $E\left\{\sum_{u=2}^{n-1} I^4(u)/n\right\}$; hence claim (a) is proved.
\par
Next, we consider claim (b). For positive integers $s_1 \not = s_2$, $(X_{s_1}, X_{s_2})^\T$ normalized by $n^{1/2}(1-\theta) \sigma$ is centered bivariate normally distributed with covariance matrix given by
$$\frac{1}{n }
\begin{pmatrix}
    s_1+2\upsilon      &  \min(s_1,s_2)+\upsilon\\
\min(s_1,s_2)+\upsilon & s_2+2\upsilon
\end{pmatrix},
$$
where $\upsilon=\theta/(1-\theta)^2$ ranges strictly between $-1/4$ and $+\infty$.  Consequently their joint normal probability density function is bounded by
$n\{ s_1 s_2-\min(s_1,s_2)^2+2\upsilon\max(s_1,s_2)+3\upsilon^2\}^{-1/2}$, which is   further   bounded by $n\{ s_1 s_2-\min(s_1,s_2)^2-\max(s_1,s_2)/2+3/16\}^{-1/2}$, its value at $\upsilon=-1/4$. Consider
\begin{align*}
&\frac{1}{n^2}E\left\{\sum_{u=2}^{n-1} I(u)\right\}^2 \\
\le&\frac{1}{n^2(1-|\theta|)^2}\sum_{s_1=0}^{n-2}\sum_{s_2=0}^{n-2}\pr\left\{\tau_1\leq \frac{X_{s_1+1}}{n^{1/2}(1-\theta)\sigma}\leq\tau_2,\tau_1\leq \frac{X_{s_2+1}}{n^{1/2}(1-\theta)\sigma}\leq\tau_2\right\}
\end{align*}
The double summation can be split into the sum over the cases with $s_1=s_2$ and that over  $s_1\not =s_2$. By claim (a), the former sum is uniformly bounded by $|\tau_2-\tau_1|/n$. So, it suffices to consider the sum over $s_2\not=s_1$.
\begin{align*}
&\frac{1}{n^2}\sum_{0\le s_1 \not = s_2 \le n-2}
\pr\left\{\tau_1\leq \frac{X_{s_1}}{n^{1/2}(1-\theta)\sigma}\leq\tau_2,\tau_1\leq \frac{X_{s_2}}{n^{1/2}(1-\theta)\sigma}\leq\tau_2\right\}\\
\le & \frac{2|\tau_2-\tau_1|^2}{n}\sum_{s_2=1}^{n-1} \sum_{s1=s_2+1}^n  \left\{ s_1 s_2-\min(s_1,s_2)^2-\max(s_1,s_2)/2+3/16\right\}^{-1/2}\\
\le & \frac{2|\tau_2-\tau_1|^2}{n}\sum_{s_2=1}^{n-1} \left\{ (s_2-1)/2+3/16\right\}^{-1/2}\\
&\quad + \frac{2|\tau_2-\tau_1|^2}{n}\sum_{s_2=1}^{n-1} \sum_{s1=s_2+2}^n  \left\{ s_1 s_2-s_2^2-s_1/2+3/16\right\}^{-1/2}
\\
\le & \frac{2|\tau_2-\tau_1|^2}{n}\sum_{s_2=1}^{n-1} \left\{ (s_2-1)/2+3/16\right\}^{-1/2}+ \frac{2|\tau_2-\tau_1|^2}{n}\sum_{s_2=1}^{n-1}
{2n^{1/2}}(s_2-1/2)^{-1/2}\\
\le & K|\tau_2-\tau_1|^2
\end{align*}
where we have made repeatedly uses of the inequality that for constant $b$,  positive number $a$ and positive integer $s$, it holds that
$$
\frac{a}{2\sqrt{as+b}}\le \sqrt{as+b}-\sqrt{a(s-1)+b},
$$
whenever the square roots are well defined as real numbers.
\par
It remains to verify (24). 
It is well known that for independent and identically distributed standard normal innovations, $\max(\eps_1, \ldots, \eps_n)/ \sqrt{2\log n}\to 1$ a.s. and hence by symmetry, $\max(|\eps_1|, \ldots, |\eps_n|)/ \sqrt{2\log n}\to 1$ a.s.  Indeed, the former asymptotic rate of the maximum holds with $\sqrt{2\log n}$ replaced by $L(n)$, some slowly varying function of $n$, for very general auto-correlated innovations under the asymptotic independence of maxima assumption, see \citet{Nav03}. The aforementioned result for   stationary normal innovations was shown by \citet{Ber62} to hold under the condition that the lag $n$ autocorrelation is $o(1/n)$. Thus, under the assumption of independent and identically distributed innovations and letting $L(n)=\sqrt{n\log\log n}$, there exists a  constant $K$ such that it holds in probability that uniformly in $\tau_1 < \tau_2$ in $[a,b]$,
\begin{eqnarray*}
|\mathcal{T}_n(\tau_2)-T_n(\tau_1)|&\le& KL(n)\sum_{t=2}^{n} n^{-1/2} \sum_{j=0}^{t-2}|\theta|^j I\left\{ \tau_1<\frac{X_{t-1-j}}{n^{1/2}(1-\theta)\sigma}\leq\tau_2\right\}\nonumber \\
&\le & \frac{KL(n)}{(1-|\theta|)n^{1/2}}
\sum_{t=1}^{n} I\left\{ \tau_1<\frac{X_{t}}{n^{1/2}(1-\theta)\sigma}\leq\tau_2\right\}.
\end{eqnarray*}
Denote $N_n(c,d)=\sum_{t=1}^{n} I\left\{ c<\frac{X_t}{(1-\theta)\sigma}\leq d\right\}$. \citet[Theorem 4]{Ako93} showed that assuming $|\theta|<1$ and  if the independent and identically distributed innovations $\eps_t$ are absolutely continuous, of zero mean and has finite $p$th moment where $p>2$,  and its characteristic function $\psi(t)$ satisfies the condition that for some $\kappa>0$, $\lim_{t\to\infty} t^\kappa \psi(t)=0$, then for any $\epsilon>0$ and $\delta_n\ge n^{-1/6-2/(3p)}$, it holds in probability that
\begin{equation}
 |N_n(a, a+\delta_n)-\int_0^n I\{W(s) \in [a, a+\delta_n]\} ds | < n^{1/3+1/(3p)+\epsilon}, \label{local-time-bound}
\end{equation}
where $W$ is the standard Brownian motion. The law of iterated logarithm for the Brownian local time \citep{Kes65} entails that it holds in probability that there exists a constant $K$ such that  uniformly for all finite numbers $c\le d$, $\int_0^n I\{W(s) \in [c, d]\} ds \le K |d-c| \sqrt{n\log\log n}$, for all $n>0$. By taking $p=3$ and $\epsilon=1/36$ in (\ref{local-time-bound}),  it follows that
$N_n(0, n^{1/2}\tau_2)\le K\sqrt{n \log\log n} n^{1/2}\tau_2+n^{17/36}$ for some constant $K>2$. It is then readily seen that for any $a\le \tau_1<\tau_2 \le b$, $N_n(\sqrt{n\tau_1}, \sqrt{n\tau_2})\le  K\sqrt{n \log\log n} n^{1/2}|\tau_2-\tau_1|+2n^{17/36}$. Hence,
$$
|\mathcal{T}_n(\tau_2)-T_n(\tau_1)|
\le K \times L(n) \sqrt{n \log\log n} |\tau_2-\tau_1|+O(n^{-\kappa})
$$
for some constant $\kappa>0$, thereby establishing (24).

\subsection{Proof of Theorem~\ref{thm:3}}
Let
\begin{equation*}
h_n(X_{t-1})=\left(\frac{h_{1,0}}{n^{1/2}}+\frac{h_{1,1}}{n}X_{t-1}\right)\times I_{t-1}+\left(\frac{h_{2,0}}{n^{1/2}}+\frac{h_{2,1}}{n}X_{t-1}\right)\times (1-I_{t-1}),
\end{equation*}
where $I_t=I\left\{\frac{X_{t}}{\sigma n^{1/2}(1-\theta)}\le \tau_0\right\}$, and  $\triangle X_{t}=X_t-X_{t-1}$.  Under $H_{0,n}$, $\eps_t=\sum_{j=0}^{t-1}\theta^j\triangle X_{t-j}+\theta^t\eps_0$,
 but under $H_{1,n}$, $\eps_t=\sum_{j=0}^{t-1}\theta^j\triangle X_{t-j}-\sum_{j=0}^{t-1}\theta^jh_n(X_{t-1-j})+\theta^t\eps_0$.
 Let $d_t=\sum_{j=0}^{t-1}\theta^j\triangle X_{t-j}$.
 To simplify the proof, we shall assume  that  $\eps_0=0$, in which case $d_t=\eps_t$ under $H_{0,n}$, and indicate later how to  modify the proof for relaxing this assumption. We shall also assume that $X_0$ is fixed.
Denote by $P_{0,n}$ the probability measure induced by $X_0,\ldots, X_n$ under $H_{0,n}$ and $P_{1,n}$ that under $H_{1,n}$.
  Thus, the log-likelihood of $X_1, \ldots, X_n$  under $H_{1,n}$ (and conditional on $X_0$) is given by
$$ \ell_{1,n}=\sum_{t=1}^{n}\log\frac{1}{\sigma}f\left\{\frac{d_t-\sum_{j=0}^{t-1}\theta^jh_n(X_{t-1-j})}{\sigma}\right\},$$
which admits the second-order Taylor expansion:
\begin{align}
\ell_{1,n}&=\sum_{t=1}^{n}\log\left\{\frac{1}{\sigma}f\left(\frac{d_t}{\sigma}\right)\right\}- \sum_{t=1}^{n}\frac{1}{\sigma}\frac{\dot{f}}{f}\left(\frac{d_t}{\sigma}\right)\sum_{j=0}^{t-1}\theta^jh_n(X_{t-1-j})\nonumber \\
&\quad+\frac{1}{2}\sum_{t=1}^{n}\frac{1}{\sigma^2}I_{f}\left(\frac{d_t}{\sigma}\right) \left\{\sum_{j=0}^{t-1}\theta^jh_n(X_{t-1-j})\right\}^2+\text{remainder.}\nonumber
\end{align}
where the first term on the right side of the preceding equality is $\ell_{0,n}$ the corresponding log-likelihood under $H_{0,n}$.
The Lagrange formula for the remainder implies that there exists $0\leq\eta_n\leq1$ such that the absolute value of the remainder is bounded by
\begin{align*}
&\frac{1}{2\sigma^2}\sum_{t=1}^{n}\left|I_f\left\{\frac{d_t-\eta_n\sum_{j=0}^{t-1}\theta^jh_n(X_{t-1-j})}{\sigma}\right\}- I_f\left(\frac{d_t}{\sigma}\right)\right|\left\{\sum_{j=0}^{t-1}\theta^jh_n(X_{t-1-j})\right\}^2\\
&\leq\frac{K}{2\sigma^2}\sum_{t=1}^{n}\left|\sum_{j=0}^{t-1}\theta^jh_n(X_{t-1-j})\right| \left\{\sum_{j=0}^{t-1}\theta^jh_n(X_{t-1-j})\right\}^2=o_p(n^{-1/2}),
\end{align*}
where $K$ is the  Lipschitz constant  of the second derivative of $\log f(\cdot)$.
Consequently, the log-likelihood ratio denoted by $\Lambda_n=\log\frac{dP_{1,n}}{dP_{0,n}}=\ell_{1,n}-\ell_{0,n}$ is given by
\begin{align}
\Lambda_n
&=-\sum_{t=1}^{n}\frac{1}{\sigma}\frac{\dot{f}}{f}\left(\frac{d_t}{\sigma}\right)\sum_{j=0}^{t-1}\theta^jh_n(X_{t-1-j})\nonumber \\
&+ \frac{1}{2}\sum_{t=1}^{n} \frac{1}{\sigma^2}I_{f}\left(\frac{d_t}{\sigma}\right) \left\{\sum_{j=0}^{t-1}\theta^jh_n(X_{t-1-j})\right\}^2+o_p(1).\label{Tay2}
\end{align}
As alluded to earlier, under $H_{0,n}$ and the assumption that $\eps_0=0$, $d_t=\eps_t$.   To study the limiting distribution of $\Lambda_n$ under $H_{0,n}$, we first consider the process
$$\left\{\left(- n^{-1/2} \sum_{t=1}^{[sn]}\frac{\dot{f}}{f}\left(\frac{\eps_t}{\sigma}\right),  n^{-1/2} \sum_{t=1}^{[sn]}\frac{\eps_t}{\sigma}\right)^\T,\quad0\leq s\leq 1\right\}.$$
It converges weakly to the correlated bivariate Brownian process
$\{(\tilde{W}_s,W_s)^\T,0\leq s\leq 1\}$
which  has stationary independent increments, starts at the origin at  $s=0$, with marginal bivariate normal distribution:
\[
\begin{pmatrix}
\tilde{W}_s \\
 W_s
 \end{pmatrix}\sim N\left\{
 \begin{pmatrix}
     0 \\
     0
\end{pmatrix},
 s\begin{pmatrix}
\mathfrak{I}_f & \rho\sqrt{\mathfrak{I}_f} \\
\rho\sqrt{\mathfrak{I}_f} & 1
\end{pmatrix}\right\}, \forall \;0\le s \le 1
\]
where $\mathfrak{I}_f=E\left[\left\{\frac{\dot{f}}{f}\left(\frac{\eps_t}{\sigma}\right)\right\}^2\right]$, and $\rho=E\left\{-\frac{\eps_t}{\sigma}\frac{\dot{f}}{f}\left(\frac{\eps_t}{\sigma}\right)\right\}$.  It is readily verified that the process $U_{s,n}=\{\frac{1}{\sigma n^{1/2}}\sum_{t=1}^{[sn]}\frac{\dot{f}}{f}\left(\frac{\eps_t}{\sigma}\right)\}$ is a sequence of martingale with respect to the filtration generated by $\{\eps_t,\; t\leq [sn],\;0\leq s\leq 1 \}$. Therefore, it holds that \citep[Theorem 7.10]{Kur96}
\begin{equation*}
-\sum_{t=1}^{n}\frac{1}{\sigma}\frac{\dot{f}}{f}\left(\frac{\eps_t}{\sigma}\right)\sum_{j=0}^{t-1}\theta^jh_n(X_{t-1-j})
\rightsquigarrow \int_{0}^{1}\tilde{h}(W_t)d\tilde{W}_t.
\end{equation*}
where
$\tilde{h}(x)=\left[\left\{\frac{h_{1,0}}{\sigma(1-\theta)}+ h_{1,1}x\right\}I\left(x\leq\tau_0\right)\right]+ \left[\left\{\frac{h_{2,0}}{\sigma(1-\theta)}+h_{2,1}x\right\}I\left(x>\tau_0\right)\right]$.
Next, we will show that the second term on the right side of (\ref{Tay2}) equals minus half the  quadratic variation of the preceding Ito integral. Decompose
\begin{align}
&\frac{1}{2}\sum_{t=1}^{n}\frac{1}{\sigma^2}I_{f}\left(\frac{\eps_t}{\sigma}\right) \left\{\sum_{j=0}^{t-1}\theta^jh_n(X_{t-1-j})\right\}^2\label{eq:QV}\\
&=\frac{1}{2}\sum_{t=1}^{n}\left[\frac{1}{\sigma^2}I_{f} \left(\frac{\eps_t}{\sigma}\right)-E\left\{\frac{1}{\sigma^2}I_{f} \left(\frac{x}{\sigma}\right)\right\}\right]\left\{\sum_{j=0}^{t-1}\theta^jh_n(X_{t-1-j})\right\}^2\nonumber\\
&\qquad +\frac{1}{2}\sum_{t=1}^{n}E\left\{\frac{1}{\sigma^2} I_{f}\left(\frac{\eps_t}{\sigma}\right)\right\}\left\{\sum_{j=0}^{t-1}\theta^jh_n(X_{t-1-j})\right\}^2.
\nonumber
\end{align}
Since
$$\frac{1}{2}\sum_{t=1}^{n}\left[\frac{1}{\sigma^2} I_{f}\left(\frac{\eps_t}{\sigma}\right)-E\left\{\frac{1}{\sigma^2}I_{f}\left(\frac{\eps_t}{\sigma}\right)\right\}\right] \left\{\sum_{j=0}^{t-1}\theta^jh_n(X_{t-1-j})\right\}^2=o_p(1),$$
and $E\left[\frac{1}{\sigma^2}I_{f}\left(\frac{\eps_t}{\sigma}\right)\right]=\frac{1}{\sigma^2}\mathfrak{I}_f$, we can use techniques employed in Step 1 of the proof of Theorem~1 to express the second term on the right side of (\ref{eq:QV}) as follows:
\begin{align*}
&-\frac{1}{2\sigma^2}\mathfrak{I}_f\sum_{t=1}^{n}\left\{\sum_{j=0}^{t-1}\theta^jh_n(X_{t-1-j})\right\}^2\\
=&-\frac{1}{2\sigma^2}\mathfrak{I}_f n^{-1}  \sum_{t=1}^{n}\left[\left\{\frac{h_{1,0}(1-\theta^t)}{\sigma(1-\theta)}+\sum_{j=0}^{t-1}\theta^jh_{1,1}\frac{X_{t-1-j}}{\sigma n^{1/2}}\right\}I\left\{\frac{X_{t-1-j}}{n^{1/2}\sigma(1-\theta)}\leq\tau_0\right\}\right.\\
&\qquad +\left.\left\{\frac{h_{2,0}(1-\theta^t)}{\sigma(1-\theta)}+\sum_{j=0}^{t-1}\theta^jh_{2,1}\frac{X_{t-1-j}}{\sigma n^{1/2}}\right\}I\left\{\frac{X_{t-1-j}}{\sigma(1-\theta)}>\tau_0\right\}\right]^2\\
\rightsquigarrow&-\frac{1}{2}\mathfrak{I}_f\int_{0}^{1}\left[\left\{\frac{h_{1,0}}{\sigma(1-\theta)}+h_{1,1}W_tI\left(W_t\leq\tau_0\right)\right\} + \left\{\frac{h_{2,0}}{\sigma(1-\theta)}+h_{2,1}W_tI\left(W_t>\tau_0\right)\right\}\right]^2dt\\
=&-\frac{1}{2}\left[Y\right]_1,
\end{align*}
where $Y_s=\int_{0}^{s}\tilde{h}(W_t)d\tilde{W}_t$ and $[Y]_s$ is its quadratic variation process.

Altogether, we have shown that under $H_{0,n}$, the log-likelihood ratio $\Lambda_n$ converges in distribution to
$
\Lambda= Y_1-\frac{1}{2}[Y_1].$
Set $\Upsilon(t)=\exp\left( Y_t-\frac{1}{2}[Y]_t\right)$. We claim that $\Upsilon(1)$ has unit mean, which will be verified below. Consequently $P_{1,n}$ is contiguous to $P_{0,n}$, thanks to Le Cam's first lemma \citep[Lemma 6, p.88]{Vaa98}. Moreover, we can derive the limiting distribution of the supLM test statistic under $H_{1,n}$, as follows. Let  $[W, Y]_t$ be the quadratic covariation process of $W$ and $Y$. Since  $\{\tilde{W}_s-\rho \mathfrak{I}_f W_s, 0\le s\le 1\}$ can be readily checked to be independent of $W$,
\begin{equation*}
[W, Y]_s=
\left[\int_0^\cdot dW, \int_0^\cdot \tilde{h}(W) d \{\tilde{W}-\rho \mathfrak{I}_f W\}\right]_s+\rho \mathfrak{I}_f\left[\int_0^\cdot dW,\int_0^\cdot \tilde{h}(W)dW\right]_s = \rho \mathfrak{I}_f \int_0^s \tilde{h}(W_t)dt.
\end{equation*}
It follows from Girsanov's theorem \citep{Gir60}  that  $W^\dagger_t=W_t-[W, Y]_t, 0\le t \le 1$ is a standard Brownian motion under the limiting distribution induced by $P_{1,n}$, as $n\to\infty$, hence the threshold diffusion characterization stated in the theorem.

To verify the claim that $E\left[\Upsilon(1)\right]=1$, note that $\Upsilon(t)$ is the stochastic exponential of the process $\{Y_t\}$ and we claim that it is a martingale. Assuming this claim for the moment. Since $\Upsilon(0)=1$, $E\left[\Upsilon(t)\right]=1$, for each $t\ge 0$; in particular,  $E\left[\Upsilon(1)\right]=1$. Hence the proof is completed upon verifying the martingale claim. From \cite{Kaz77}, it suffices to verify  that
\begin{equation}
E\left[\exp\left\{\frac{1}{2}\int_{0}^{1}\tilde{h}^2(W_s)ds\right\}\right]<\infty.
\label{integralcriterion}
\end{equation}
Let $K_1=\max(h_{1,1}^2,h_{2,1}^2)/{2}$ and $K_2=\max(|h_{1,0}h_{1,1}|,|h_{2,0}h_{2,1}|)/\{\sigma(1-\theta)\}$. Under (C2), $K_1>0$. Then,
$$
E\left\{\exp\left(\frac{1}{2}\int_{0}^{1}\delta_s^2ds\right)\right\}
\le \exp\left[\left\{\frac{h_{1,0}}{\sigma(1-\theta)}\right\}^2 + \left\{\frac{h_{2,0}}{\sigma(1-\theta)}\right\}^2\right]\int_{0}^{\infty}\exp\left(K_1y+K_2\sqrt{y}\right)dF(y),
$$
where $F(\cdot)$ is cdf of  $\int_{0}^{1}W_s^2ds$. Eqn. (\ref{integralcriterion}) holds  if and only if $\int_{1}^{\infty}\exp\left(K_1y+K_2\sqrt{y}\right)dF(y)$ is finite. Since $\sqrt{y}=o(y)$, it suffices to prove that $\int_{1}^{\infty}\exp\left(K_1y\right)dF(y)<\infty$.  Consider
\begin{align*}
&\int_{1}^{\infty}e^{K_1y}dF(y)
=\int_{1}^{\infty}\left(e^{K_1y}-1\right)dF(y)+\int_{1}^{\infty}dF(y)\\
&=\int_{1}^{\infty}\int_{0}^{y}\frac{1}{K_1}e^{K_1x}dxdF(y)+1 =\frac{1}{K_1}\int_{1}^{\infty}\int_{\max(x,1)}^{\infty}dF(y)e^{K_1x}dx+1.
\end{align*}
Therefore, we need only  verify that the integral
\begin{equation}
\int_{0}^{\infty}\bar F(x)e^{K_1x}dx<\infty, \label{integral2}
\end{equation}
where $\bar F(x)=1-F(x)$.
From \citep[Lemma 2]{Li92} with $\theta=0$, it follows that, as $x\to+\infty$,
\begin{equation*}
\bar F(x)\sim Kx^{-1/2}e^{-\frac{x}{2\lambda}},\quad\text{with } \lambda=\frac{4}{\pi^2} \text{ and some constant } K.
\end{equation*}
Thus the integrand in (\ref{integral2}) is  $\sim K x^{-1/2}\exp\left\{\left(-\frac{\pi^2}{8}+K_1\right)x\right\}$, hence (\ref{integral2}) holds  if $\max(h_{1,1}^2,h_{2,1}^2)<\frac{\pi^2}{4}$.
This completes the proof.
\par
We now sketch how to modify the proof for the general case of unknown $\eps_0$. Then, the (conditional) log-likelihood of $X_0, \ldots, X_n$ under $H_{1,n}$ given $X_0$ equals
\begin{align*}
\ell_{1,n} &= \log\int\exp\left(\sum_{t=1}^{n}\log\left[\frac{1}{\sigma}f\left\{\frac{d_t-\sum_{j=0}^{t-1}\theta^jh_n(X_{t-1-j}) +\theta^t\tilde{\eps}_0}{\sigma}\right\}\right]\right)\frac{1}{\sigma}f\left(
 \frac{\tilde{\eps}_0}{\sigma} \right)
 d\tilde{\eps}_0
\end{align*}
whereas that under $H_{0,n}$ equals
\begin{align*}
\ell_{1,n}&=\log\int\exp\left[\sum_{t=1}^{n}\log\left\{\frac{1}{\sigma}f\left(\frac{d_t+\theta^t \tilde{\eps}_0}{\sigma}\right)\right\}\right]\frac{1}{\sigma}f\left(
 \frac{\tilde{\eps}_0}{\sigma} \right)
 d\tilde{\eps}_0.
\end{align*}
Let $\eps_0$ be the true innovation at epoch $t$. Then,
under $H_{0,n}$,
\begin{align}
&\sum_{t=k}^{n}\log\left[\frac{1}{\sigma}f\left\{\frac{d_t-\sum_{j=0}^{t-1}\theta^jh_n(X_{t-1-j})+\theta^t\tilde{\eps}_0}{\sigma}\right\}\right]\\
&= \sum_{t=k}^{n}\log\left[\frac{1}{\sigma}f\left\{\frac{\eps_t-\sum_{j=0}^{t-1}\theta^jh_n(X_{t-1-j})}{\sigma}\right\}\right] +R_{k,n},
\nonumber
\end{align}
where $|R_{1,k,n}|\le K\frac{|\theta|^k}{1-|\theta|}(|\tilde{\eps}_0|+|\eps_0|)$ with $K$ being some  constant, thanks to (C1). Thus,
\begin{align*}
&\ell_{1,n} =\sum_{t=k}^{n}\log\left[\frac{1}{\sigma}f\left\{\frac{\eps_t-\sum_{j=0}^{t-1}\theta^jh_n(X_{t-1-j})}{\sigma}\right\}\right] \nonumber \\
& + \log\int\exp\left(R_{1,k,n} + \sum_{t=1}^{k-1}\log\left[\frac{1}{\sigma}f\left\{\frac{d_t-\sum_{j=0}^{t-1}\theta^jh_n(X_{t-1-j}) +\theta^t\tilde{\eps}_0}{\sigma}\right\}\right]\right)\frac{1}{\sigma}f\left(
 \frac{\tilde{\eps}_0}{\sigma} \right)
 d\tilde{\eps}_0 \nonumber.
\end{align*}
Similarly,
\begin{equation*}
\ell_{0,n} =\sum_{t=k}^{n}\log\left\{\frac{1}{\sigma}f\left(\frac{\eps_t}{\sigma}\right)\right\}
+ \log\int\exp\left[R_{0,k,n} + \sum_{t=1}^{k-1}\log\left\{\frac{1}{\sigma} f\left(\frac{d_t+\theta^t\tilde{\eps}_0}{\sigma}\right)\right\}\right]\frac{1}{\sigma}f\left(
 \frac{\tilde{\eps}_0}{\sigma} \right)
 d\tilde{\eps}_0,
\end{equation*}
where $R_{0,k,n}$ shares the same bound as $R_{1,k,n}$.
The difference between the second terms on the right side of  the preceding expressions for $\ell_{i,n}, i=0,1$ can be made smaller in magnitude than any given positive number, in probability, by first taking $k$  large and then $n$ sufficiently large, thanks to (C1). Hence, the log-likelihood ratio $\ell_{1,n}-\ell_{0,n}$ is asymptotically the same as the case when $\eps_{0}=0$. This completes the proof.

\section{Real data analysis: further results}\label{sec:ppp}
\begin{figure}[H]
  \centering
\includegraphics[width=0.99 \linewidth]{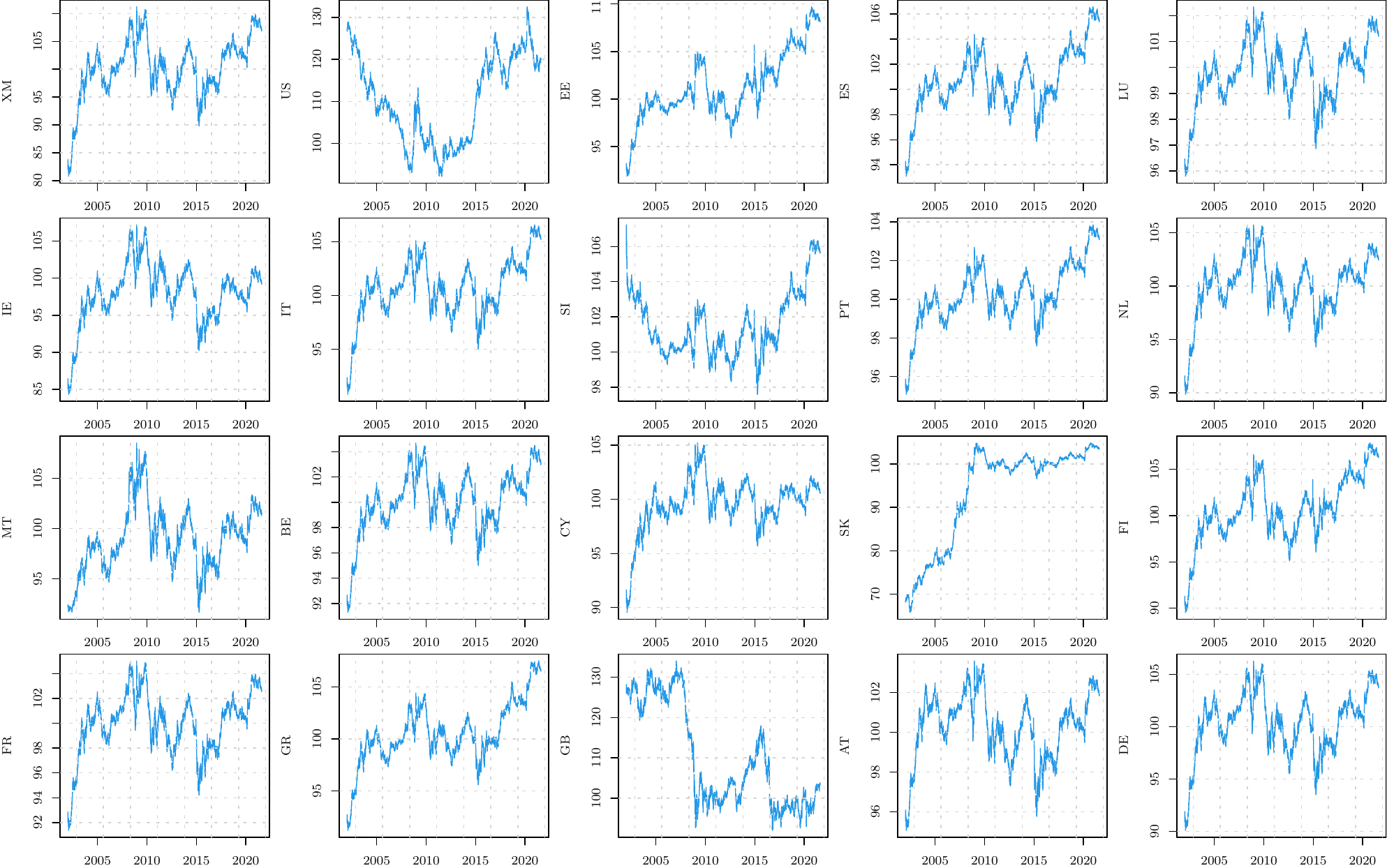}
 \caption{Time series of daily real exchange rates for a panel of European countries (plus GB and USA), from January, 2002 to August, 2021.}\label{fig:S1}
\end{figure}

 In Figure~\ref{fig:S1} we report the panel of 20 time series of daily real exchange rates for the Euro zone plus Great Britain and USA.

\begin{figure}[H]
  \centering
\includegraphics[width=0.99 \linewidth]{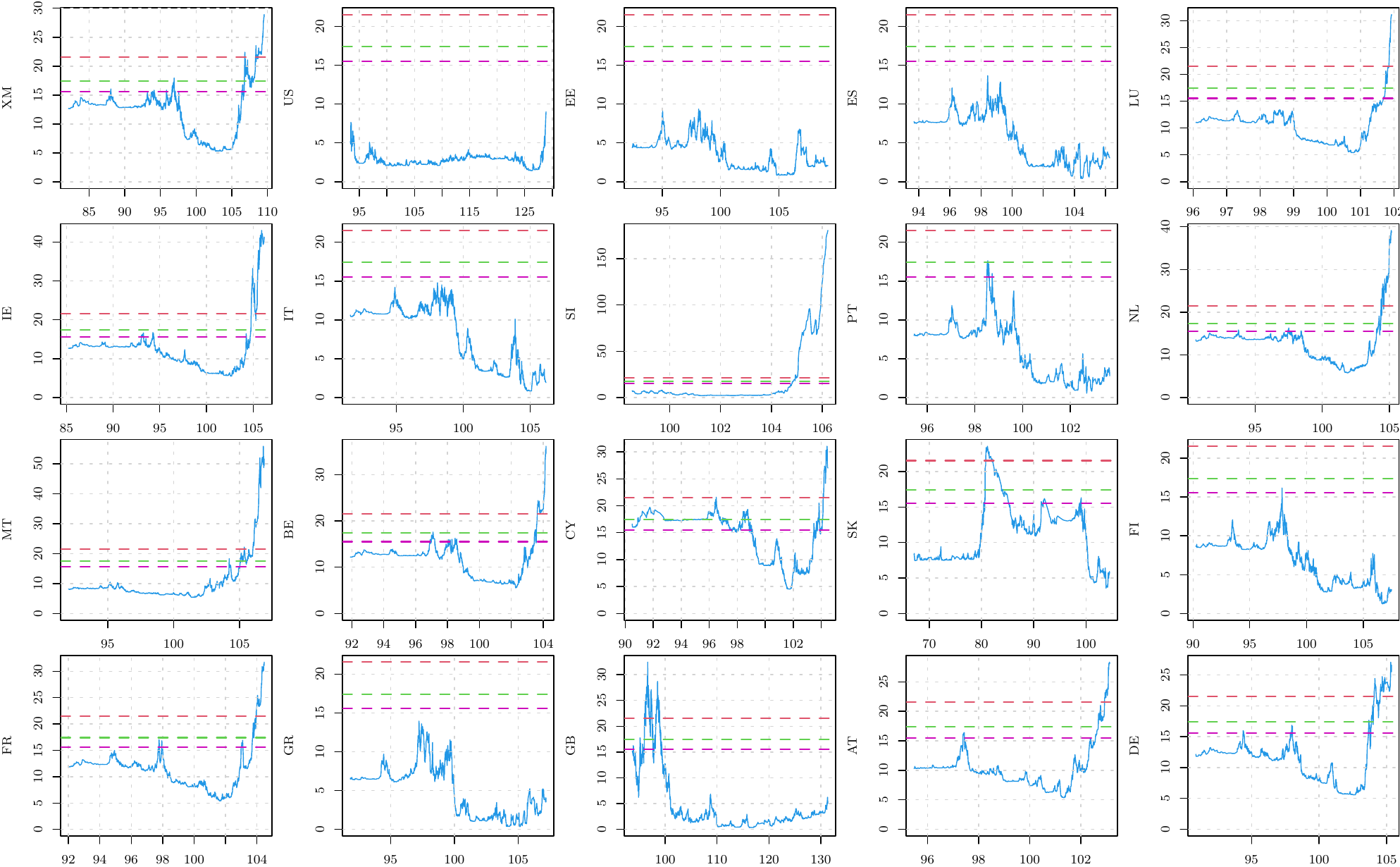}
 \caption{Values of the LM statistic $T(r)$ over the threshold grid for the 20 series. The critical values of the null distribution of the sLM  statistic at levels 90\%, 95\% and 99\% are indicated as purple, green and red dashed lines, respectively.}\label{fig:S2}
\end{figure}

 Figure~\ref{fig:S2} shows the plot of the values of the LM statistic $T_r$ over the threshold grid for the 20 series. The critical values of the null distribution of the sLM statistic at levels 90\%, 95\% and 99\% are indicated  as purple, green and red dashed lines, respectively.  Figure~\ref{fig:S3} is as Figure~\ref{fig:S2} but refers to testing for regulation from below. Table~\ref{tab:S0} presents the results of the application of the M and ADF tests over the 20 series. We have added an asterisk when the test rejects at $5\%$ level. Clearly, none of the M tests is able to reject the random walk hypothesis. Only the classic ADF test rejects for some series and its results are consistent with those of the supLM test statistics.

\begin{figure}[H]
  \centering
\includegraphics[width=0.99 \linewidth]{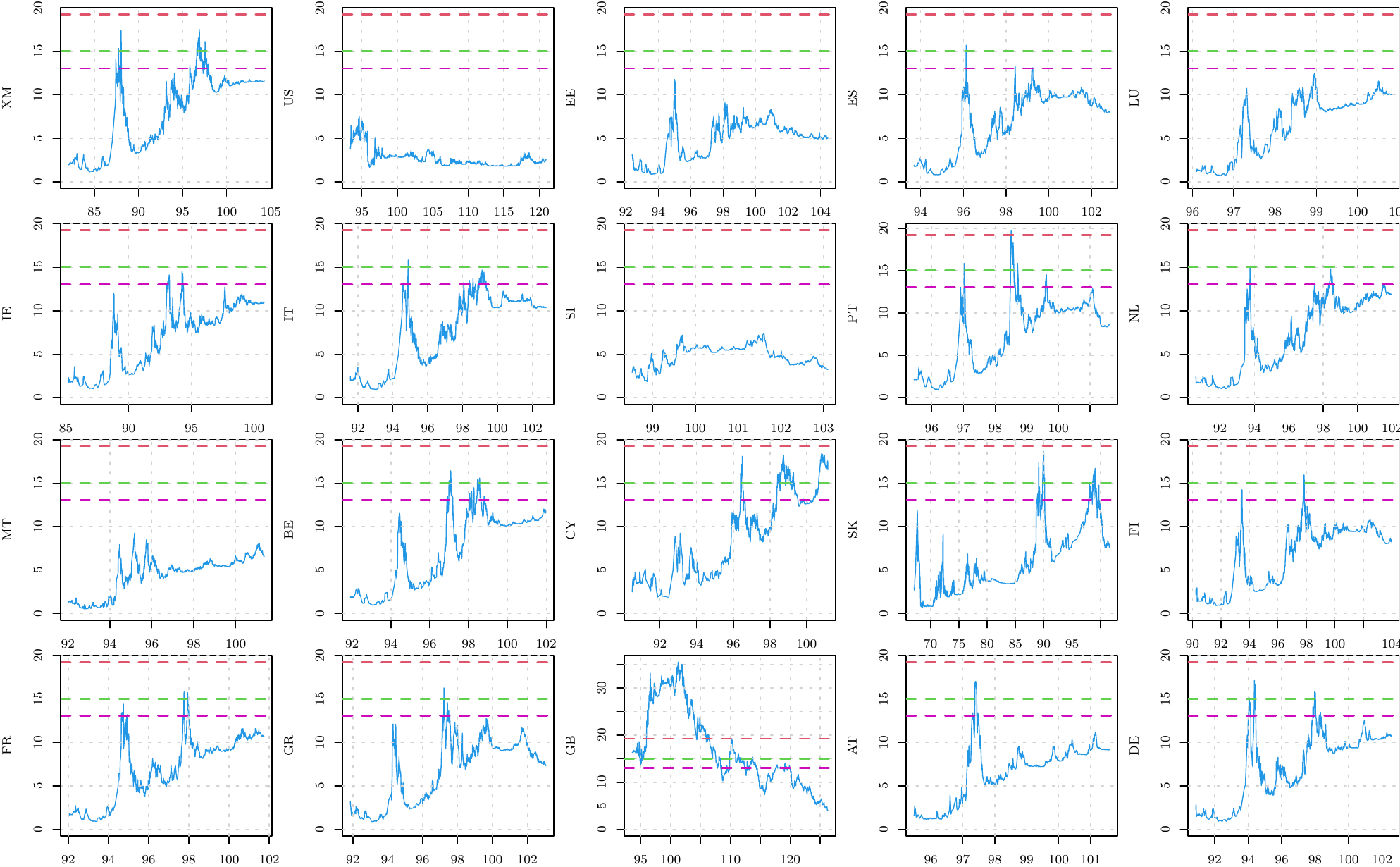}
 \caption{Testing for threshold regulation from below. Values of the LM statistic $T(r)$ over the threshold grid for the 20 series. The critical values of the null distribution of the sLM  statistic at levels 90\%, 95\% and 99\% are indicated as purple, green and red dashed lines, respectively.}\label{fig:S8}
\end{figure}

\begin{table}[H]
\spacingset{1}
\centering
\begin{tabular}{rrrrlr}
  \cmidrule(lr){2-6}
& $\bar{\mathrm{M}}^{\mathrm{g}}$ & $\mathrm{M}^{\mathrm{g}}$ & $\mathrm{MP}_\mathrm{T}$ & ADF & ADF$^{\mathrm{g}}$  \\
  \cmidrule(lr){2-6}
    XM &  0.19 &  0.19 &  47.01  & -3.59$^*$ &  0.17 \\
    US & -1.02 & -1.02 &  23.01  & -1.72  & -0.69 \\
    EE &  0.69 &  0.69 &  54.52  & -2.10  &  0.62 \\
    ES &  0.39 &  0.39 &  40.98  & -2.77  &  0.31 \\
    LU & -0.25 & -0.25 &  27.47  & -3.34$^*$ & -0.17 \\
    IE & -0.15 & -0.15 &  39.68  & -3.58$^*$ & -0.13 \\
    IT &  0.31 &  0.31 &  44.93  & -3.26$^*$ &  0.26 \\
    SI & -1.16 & -1.16 &  21.15  & -2.42  & -0.79 \\
    PT &  0.51 &  0.51 &  53.54  & -2.83  &  0.47 \\
    NL & -0.04 & -0.04 &  38.15  & -3.66$^*$ & -0.03 \\
    MT & -1.51 & -1.51 &  13.06  & -2.84  & -0.70 \\
    BE &  0.08 &  0.08 &  39.62  & -3.50$^*$ &  0.07 \\
    CY & -0.12 & -0.03 &  40.53  & -4.07$^*$ & -0.10 \\
    SK &  0.63 &  0.63 & 206.61  & -2.47  &  1.15 \\
    FI &  0.35 &  0.35 &  46.61  & -2.92$^*$ &  0.30 \\
    FR & -0.05 & -0.05 &  35.20  & -3.47$^*$ & -0.04 \\
    GR &  0.68 &  0.68 &  58.25  & -2.54  &  0.64 \\
    GB & -0.50 & -0.50 &  26.15  & -1.59  & -0.34 \\
    AT & -0.45 & -0.45 &  24.08  & -3.24$^*$ & -0.29 \\
    DE & -0.03 & -0.03 &  35.23  & -3.46$^*$ & -0.02 \\
  \cmidrule(lr){2-6}
\end{tabular}
 \caption{Values of the M and ADF test statistics for the 20 time series. The asterisk indicates rejection of the null hypothesis at $5\%$ level.}\label{tab:S0}
\end{table}

\subsection{TARMA-GARCH modelling of the Eurozone time series}\label{sec:XM}
\begin{figure}[H]
\centering
\includegraphics[width=0.4\linewidth]{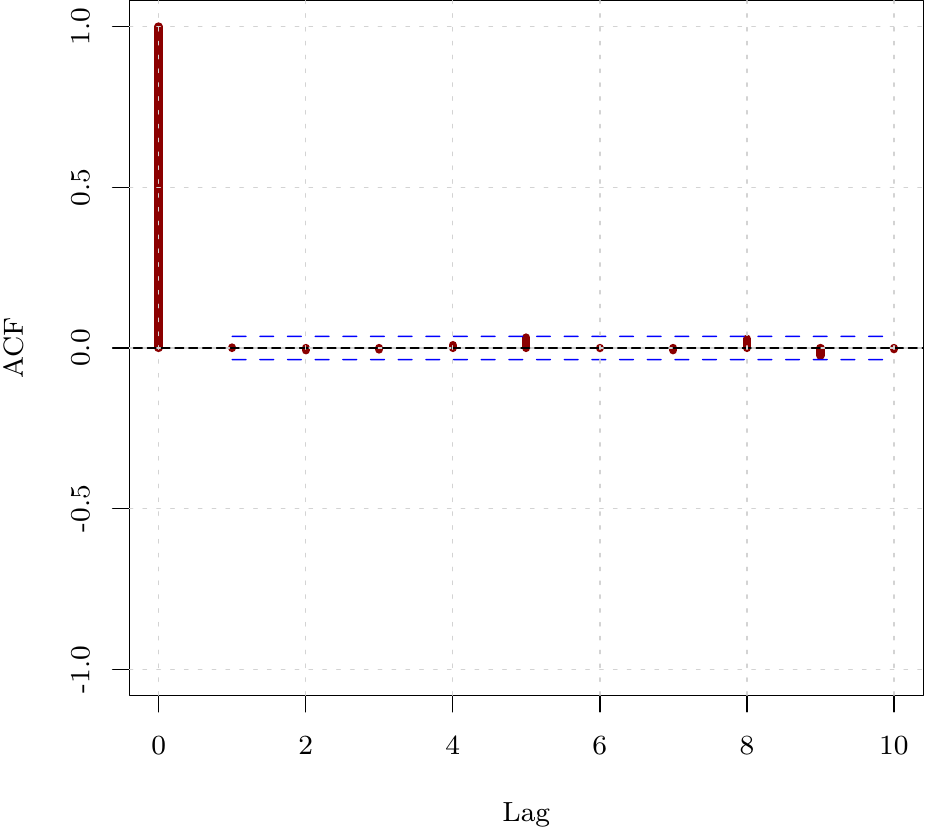}
\includegraphics[width=0.4\linewidth]{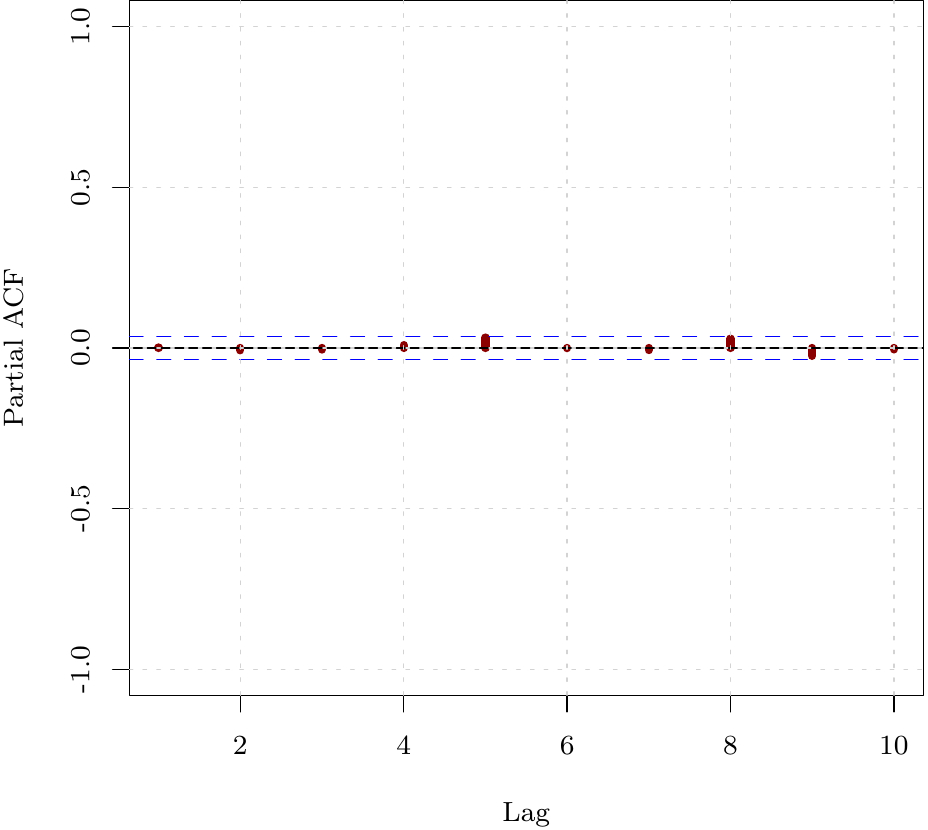}
 \caption{Sample autocorrelations and sample partial autocorrelations of the residuals from the TARMA fit for the time series of daily real exchange rates for the Eurozone.}\label{fig:S3}
\end{figure}

\begin{figure}[H]
\centering
\includegraphics[width=0.4\linewidth]{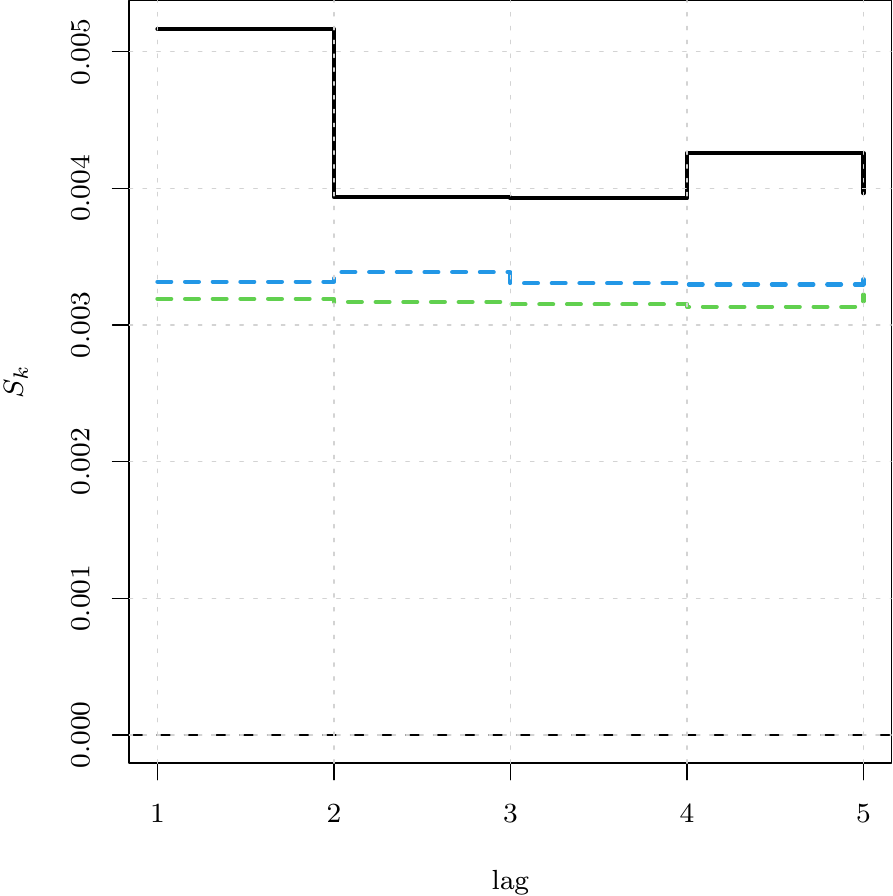}
\includegraphics[width=0.4\linewidth]{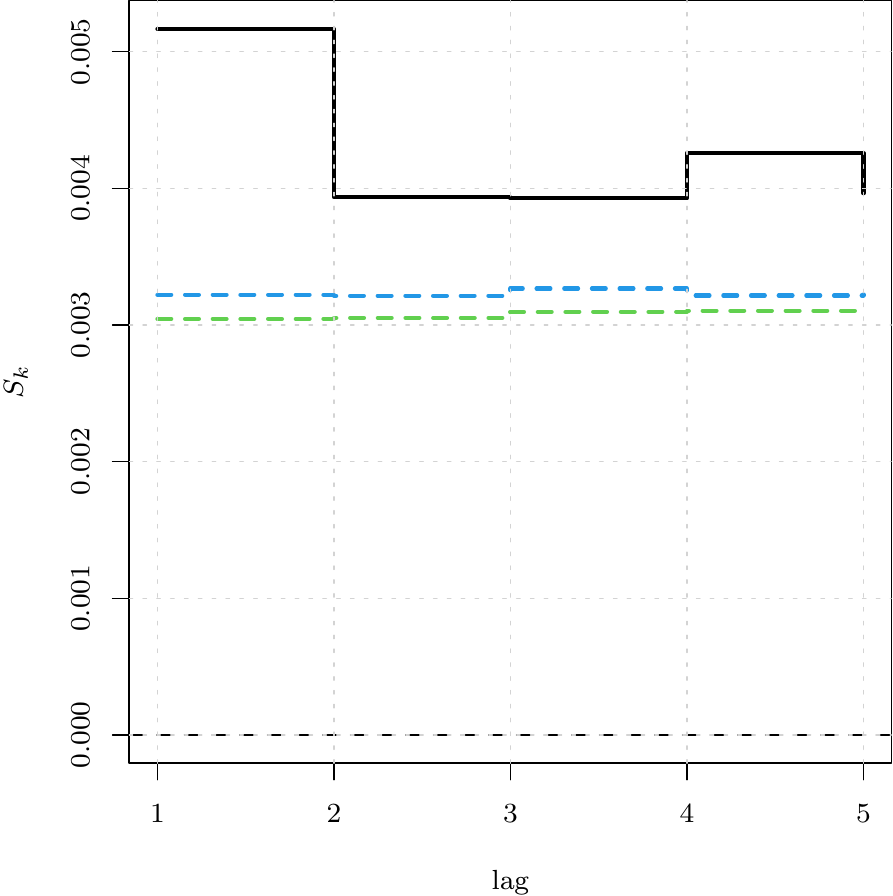}
\caption{Entropy based metric up to lag 5, computed on the residuals of the TARMA model of Eq.~(\ref{tarma.fit}). The dashed lines correspond to bootstrap rejection bands at levels 95\%(green) and 99\%(blue) under the null hypothesis of independence (left panel) and linearity (right panel).}\label{fig:S4}
\end{figure}

\begin{figure}[H]
\centering
\includegraphics[width=0.4\linewidth]{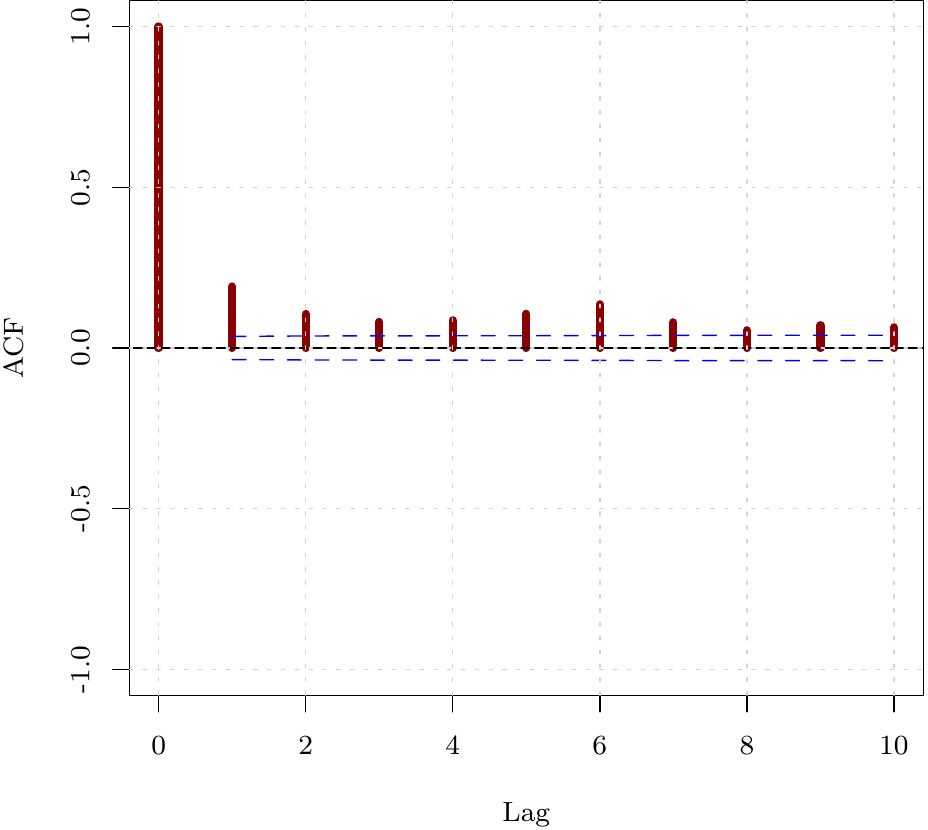}
\includegraphics[width=0.4\linewidth]{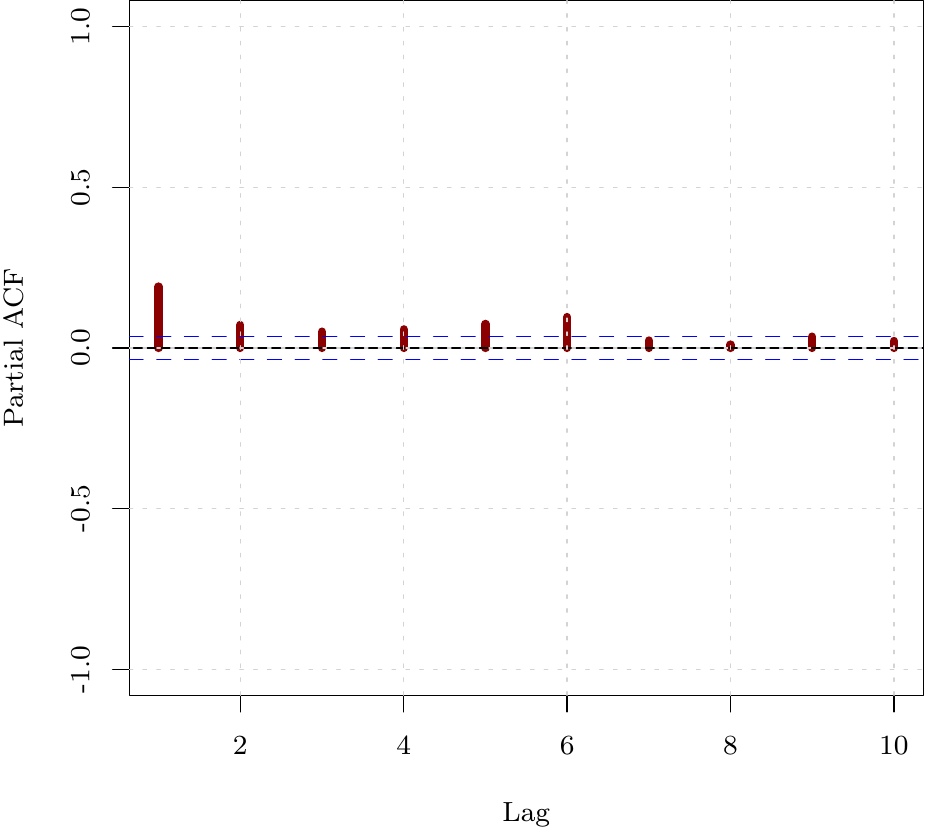}
 \caption{Sample autocorrelations and sample partial autocorrelations of the squared residuals from the TARMA fit for the time series of daily real exchange rates for the Eurozone.}\label{fig:S5}
\end{figure}

\begin{figure}[H]
  \centering
\includegraphics[width=0.4\linewidth]{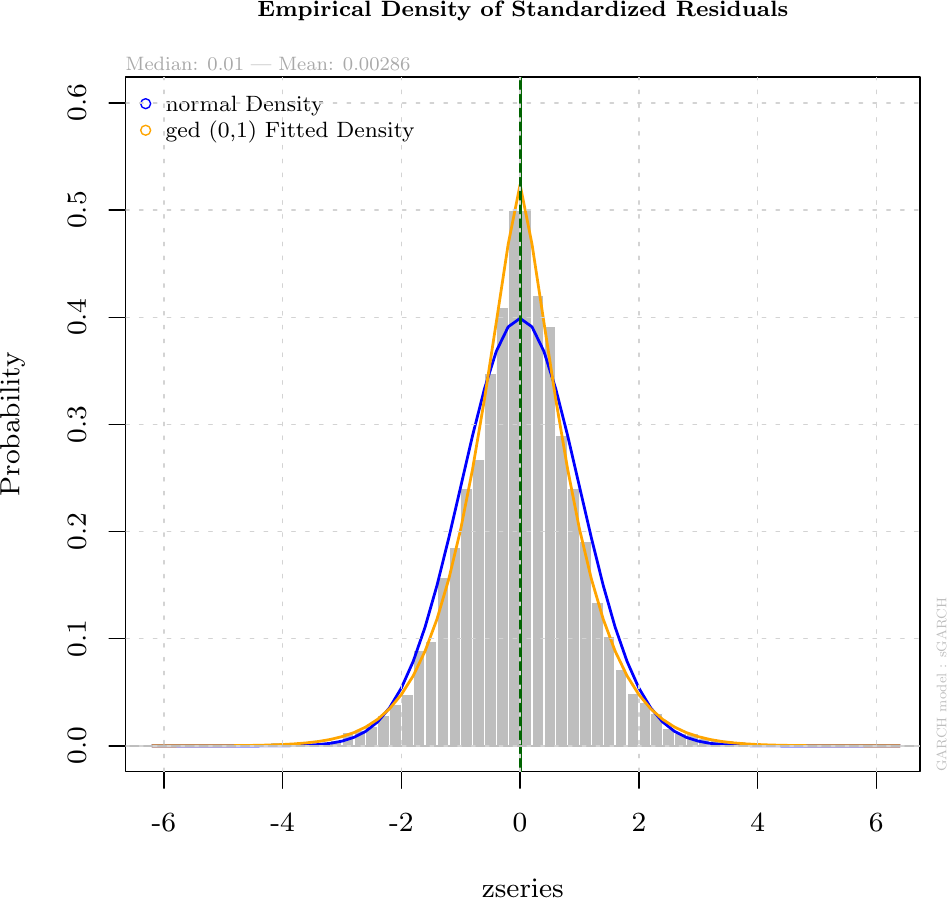}
\caption{Histogram of the standardized residuals of the TARMA-GARCH model with the GED and a Gaussian fit superimposed in yellow and blue, respectively.}\label{fig:S6}
\end{figure}

\begin{figure}[H]
  \centering
\includegraphics[width=0.4\linewidth]{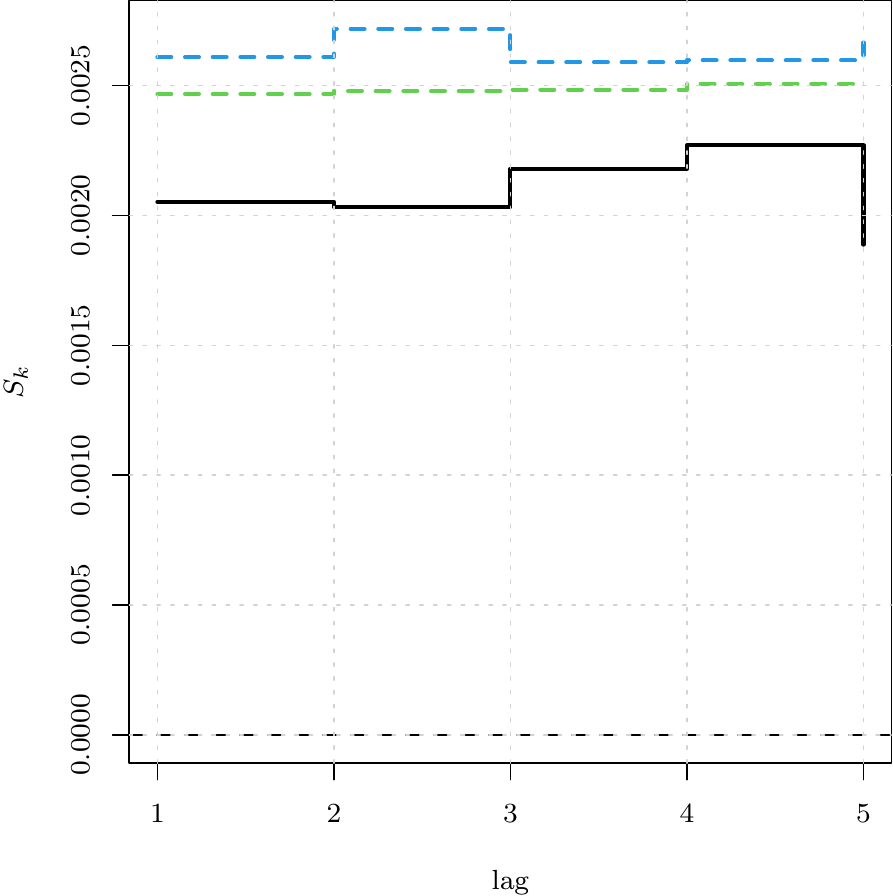}
\caption{Entropy based metric up to lag 5, computed on the residuals of the TARMA-GARCH model of Eq.~(\ref{tarma.fit}) and (\ref{garch.fit}). The dashed lines correspond to bootstrap rejection bands at levels 95\%(green) and 99\%(blue) under the null hypothesis of independence.}\label{fig:S7}
\end{figure}

In this subsection we present additional results on the TARMA-GARCH modelling of the time series of real exchange rates of the Eurozone area as a whole (XM). Figures~\ref{fig:S3}-\ref{fig:S5} display residuals diagnostics of the TARMA model and show the presence of volatility, whereas Figures~\ref{fig:S6}-\ref{fig:S7} show the adequacy of the TARMA-GARCH fit with GED innovations.


In Table~\ref{tab:Sfit1} we present the empirical size of both supLM and M tests computed on simulated series from an ARIMA-GARCH process with GED innovations, whose parameters are estimated on the Eurozone time series. The length of the series is $n=1000, 5000, 10000$ and the threshold is searched between the $1^{\text{st}}$ and the $99^{\text{th}}$ sample quantiles. While the asymptotic sLM test shows some size bias, probably due to the presence of both volatility and heavy tails, the wild bootstrap sLMb test has always the correct size and behaves like the M tests.
\begin{table}[H]
\spacingset{1}
  \centering
\caption{Empirical size (rejection percentages) for varying sample size $n$, at nominal $\alpha=5\%$, for simulated series of an ARIMA-GARCH process with GED innovations whose parameters have been estimated upon the Eurozone time series.}\label{tab:Sfit1}
\begin{tabular}{rrrrrrrr}
  $n$ & sLM & sLMb & MZa & MZa3 & MPT & ADFols & ADF \\
  \cmidrule(lr){2-8}
   1000 & 16.3 & 5.3 & 6.2 & 6.1 & 5.3 & 6.7 & 5.3 \\
   5000 & 25.6 & 3.9 & 5.2 & 5.2 & 4.6 & 7.2 & 4.6 \\
  10000 & 27.7 & 5.5 & 6.8 & 6.8 & 6.1 & 6.8 & 6.0 \\
  \cmidrule(lr){2-8}
\end{tabular}
\end{table}

Table~\ref{tab:Sfit2} reports the empirical power computed on simulated series from the fitted TARMA-GARCH process with GED innovations (see Eq.~(\ref{tarma.fit}) and (\ref{garch.fit})). Here, the asymptotic sLM test is the most powerful whereas the wild bootstrap test sLMb and M tests have similar power, the former being slightly more conservative.
 \begin{table}[H]
\spacingset{1}
  \centering
\caption{Empirical power (rejection percentages) for varying sample size $n$, at nominal $\alpha=5\%$, for simulated series of a TARMA-GARCH process with GED innovations whose parameters are as in Eq.~(\ref{tarma.fit}) and (\ref{garch.fit}) and have been estimated upon the Eurozone time series.}\label{tab:Sfit2}
\begin{tabular}{rrrrrrrr}
  $n$ & sLM & sLMb & MZa & MZa3 & MPT & ADFols & ADF \\
  \cmidrule(lr){2-8}
   1000 & 29.0 & 17.0 & 16.8 & 16.4 & 15.3 & 12.6 & 15.4 \\
   5000 & 87.2 & 58.1 & 68.0 & 67.8 & 65.9 & 72.2 & 66.6 \\
  10000 & 99.9 & 89.4 & 91.3 & 91.2 & 90.3 & 98.9 & 90.5 \\
  \cmidrule(lr){2-8}
\end{tabular}
\end{table}

\section{Tabulated quantiles of the null distribution}\label{SM:tab}

In Table~\ref{tab:perc} we tabulate the quantiles of the asymptotic null distribution of the sLM statistic for different threshold ranges, determined by $[\pi,1-\pi]$, where $\pi$ indicates the $(\pi\times 100)$th percentile of the observed data. Since the test is similar with respect to the MA parameter $\theta$ we set it to 0 and produced the table by simulating 50000 trajectories of length 5000 from a random walk.
%
\begin{table}[H]
\spacingset{1}
\centering
\caption{Quantiles of the asymptotic null distribution of the sLM statistic for different levels of the threshold range determined by $[\pi,1-\pi]$, where $\pi$ denotes the sample $(\pi\times 100)$-th percentile.}\label{tab:perc}
\begin{tabular}{rrrrr}

  $\pi$   & 90\% & 95\% & 99\% & 99.9\% \\
  \cmidrule(lr){2-5}
  0.01 & 15.22 & 17.12 & 21.33 & 26.73 \\
  0.05 & 14.21 & 16.13 & 20.23 & 25.22 \\
  0.10 & 13.54 & 15.50 & 19.61 & 25.41 \\
  0.15 & 12.98 & 14.87 & 19.02 & 24.48 \\
  0.20 & 12.52 & 14.54 & 18.70 & 24.22 \\
  0.25 & 12.10 & 14.02 & 18.15 & 23.91 \\
  0.30 & 11.63 & 13.54 & 17.67 & 22.76 \\
  0.35 & 11.16 & 12.99 & 17.08 & 22.28 \\
  0.40 & 10.37 & 12.29 & 16.37 & 21.85 \\
  \cmidrule(lr){2-5}
\end{tabular}
\end{table}

\section{Supplementary Monte Carlo results for $n=100,500$}\label{SM:power}

Table~\ref{tabT:100} and Table~\ref{tabT:500} report the empirical power of the tests under the alternative hypothesis described in Section~6 of the main text. They are the analogs of Table~2 but for sample size equal to 100 and 500, respectively.
\begin{table}[H]
\spacingset{1}
\centering \small
\caption{n = 100.  Rejection percentages. Simulation from the TARMA(1,1) model of Eq.~(28)
.}\label{tabT:100}
\begin{tabular}{crrrrrrrrrrr}
$n=100$       & \multicolumn{9}{c}{asymptotic} & \multicolumn{2}{c}{bootstrap}\\
 \cmidrule(lr){1-1} \cmidrule(lr){2-10} \cmidrule(lr){11-12}
 $\tau\;;\;\theta$ & sLM & $\bar{\mathrm{M}}^{\mathrm{g}}$ & $\mathrm{M}^{\mathrm{g}}$ & $\mathrm{MP}_\mathrm{T}$ & ADF & ADF$^{\mathrm{g}}$ & KS & BBC & EG & sLMb & KSb  \\
  \cmidrule(lr){2-10} \cmidrule(lr){11-12}
  0.0;-0.9 &  5.0 & 5.0  & 5.0  & 5.0  & 5.0 & 5.0  & 5.0  & 5.0  & 5.0  &   5.1 &  5.0  \\
  0.5;-0.9 &  9.5 & 8.7  & 9.1  & 9.1  & 6.0 & 9.9  & 2.2  & 7.1  & 1.8  &   9.4 &  4.2  \\
  1.0;-0.9 & 16.0 & 9.6  & 10.4 & 10.3 & 7.1 & 12.1 & 2.8  & 9.0  & 1.5  &  17.3 &  3.9  \\
  1.5;-0.9 & 28.1 & 10.7 & 12.1 & 11.6 & 8.3 & 15.0 & 5.8  & 12.7 & 2.1  &  29.6 &  7.9  \\
  0.0;-0.5 &  5.0 & 5.0  & 5.0  & 5.0  & 5.0 & 5.0  & 5.0  & 5.0  & 5.0  &   5.1 &  4.9  \\
  0.5;-0.5 &  9.2 & 10.7 & 11.1 & 11.3 & 5.9 & 11.3 & 3.3  & 8.0  & 2.4  &   7.1 &  2.7  \\
  1.0;-0.5 & 16.2 & 13.1 & 13.9 & 13.8 & 7.2 & 14.1 & 5.8  & 11.1 & 2.6  &  15.1 &  4.8  \\
  1.5;-0.5 & 25.3 & 16.2 & 17.4 & 16.6 & 8.3 & 17.5 & 11.2 & 15.2 & 3.9  &  24.5 &  9.5  \\
  0.0;0.0  &  5.0 & 5.0  & 5.0  & 5.0  & 5.0 & 5.0  & 5.0  & 5.0  & 5.0  &   4.9 &  5.1  \\
  0.5;0.0  &  7.4 & 12.1 & 12.6 & 11.9 & 6.2 & 12.2 & 13.4 & 7.6  & 9.6  &   8.0 & 12.6  \\
  1.0;0.0  & 13.5 & 17.8 & 18.9 & 17.7 & 7.7 & 18.1 & 26.2 & 10.7 & 15.8 &  15.8 & 26.2  \\
  1.5;0.0  & 20.6 & 22.8 & 24.4 & 22.9 & 9.7 & 23.3 & 38.6 & 15.8 & 22.1 &  23.7 & 43.5  \\
  0.0;0.5  &  5.0 & 5.0  & 5.0  & 5.0  & 5.0 & 5.0  & 5.0  & 5.0  & 5.0  &   4.8 &  0.0  \\
  0.5;0.5  &  7.5 & 11.2 & 11.8 & 11.3 & 6.9 & 11.5 & 18.8 & 8.8  & 14.3 &   7.3 &  0.0  \\
  1.0;0.5  & 12.4 & 15.7 & 17.2 & 15.9 & 9.1 & 16.7 & 29.6 & 14.1 & 24.4 &  10.3 &  0.0  \\
  1.5;0.5  & 20.4 & 18.1 & 21.0 & 18.7 & 12.3& 19.8 & 42.3 & 21.0 & 36.0 &  19.6 &  0.0  \\
  0.0;0.9  &  5.0 & 5.0  & 5.0  & 5.0  & 5.0 & 5.0  & 5.0  & 5.0  & 5.0  &   4.7 &  0.0  \\
  0.5;0.9  &  7.4 & 8.1  & 9.0  & 8.0  & 5.8 & 8.1  & 2.8  & 10.0 & 8.6  &   4.4 &  0.0  \\
  1.0;0.9  & 12.7 & 9.9  & 12.7 & 9.8  & 7.0 & 9.6  & 3.4  & 13.6 & 12.0 &  10.0 &  0.0  \\
  1.5;0.9  & 22.4 & 12.2 & 15.2 & 12.2 & 8.0 & 12.1 & 4.6  & 15.7 & 13.9 &  20.8 &  0.0  \\
 \cmidrule(lr){1-1}   \cmidrule(lr){2-10} \cmidrule(lr){11-12}
\end{tabular}
\end{table}

\begin{table}[H]
\spacingset{1}
\centering \small
\caption{n = 500.  Rejection percentages. Simulation from the TARMA(1,1) model of Eq.~(28)
.}\label{tabT:500}
\begin{tabular}{rrrrrrrrrrrr}
$n=500$       & \multicolumn{9}{c}{asymptotic} & \multicolumn{2}{c}{bootstrap}\\
 \cmidrule(lr){1-1} \cmidrule(lr){2-10} \cmidrule(lr){11-12}
 $\tau\;;\;\theta$ & sLM & $\bar{\mathrm{M}}^{\mathrm{g}}$ & $\mathrm{M}^{\mathrm{g}}$ & $\mathrm{MP}_\mathrm{T}$ & ADF & ADF$^{\mathrm{g}}$ & KS & BBC & EG & sLMb & KSb  \\
  \cmidrule(lr){2-10} \cmidrule(lr){11-12}
  0.0;-0.9 & 5.0  & 5.0  & 5.0  & 5.0  & 5.0  & 5.0  & 5.0  & 5.0  & 5.0  &   5.1 &  5.1  \\
  0.5;-0.9 & 40.5 & 26.5 & 26.0 & 26.9 & 16.5 & 28.1 & 13.2 & 26.1 & 3.0  &  39.9 & 17.3  \\
  1.0;-0.9 & 77.3 & 41.4 & 40.9 & 42.1 & 29.7 & 44.2 & 43.0 & 55.7 & 10.5 &  75.5 & 47.1  \\
  1.5;-0.9 & 94.5 & 53.8 & 53.2 & 54.4 & 45.3 & 57.2 & 71.4 & 78.4 & 24.0 &  93.5 & 76.1  \\
  0.0;-0.5 &  5.0 &  5.0 &  5.0 &  5.0 &  5.0 &  5.0 &  5.0 &  5.0 &  5.0 &   4.9 &  5.0  \\
  0.5;-0.5 & 39.6 & 32.7 & 32.6 & 33.2 & 16.8 & 33.7 & 21.6 & 27.0 &  6.0 &  41.5 & 23.1  \\
  1.0;-0.5 & 76.4 & 51.0 & 51.0 & 51.5 & 30.7 & 52.1 & 57.8 & 57.7 & 18.7 &  77.1 & 57.2  \\
  1.5;-0.5 & 94.4 & 65.4 & 65.9 & 65.8 & 46.9 & 66.5 & 83.9 & 81.8 & 40.3 &  95.3 & 84.4  \\
  0.0;0.0  &  5.0 &  5.0 &  5.0 &  5.0 &  5.0 &  5.0 &  5.0 &  5.0 &  5.0 &   4.8 &  5.1  \\
  0.5;0.0  & 39.0 & 38.6 & 38.8 & 38.4 & 17.7 & 38.6 & 56.6 & 25.5 & 32.9 &  40.0 & 57.1  \\
  1.0;0.0  & 77.9 & 61.7 & 62.2 & 61.5 & 33.1 & 61.8 & 85.7 & 59.6 & 66.5 &  81.0 & 86.8  \\
  1.5;0.0  & 95.7 & 75.7 & 76.4 & 75.4 & 53.0 & 75.7 & 96.3 & 85.7 & 90.6 &  96.4 & 96.1  \\
  0.0;0.5  & 5.0  & 5.0  & 5.0  & 5.0  & 5.0  & 5.0  & 5.0  & 5.0  & 5.0  &   5.1 &  0.0  \\
  0.5;0.5  & 40.6 & 39.7 & 39.4 & 40.3 & 21.8 & 40.5 & 63.7 & 36.0 & 53.1 &  44.2 &  0.0  \\
  1.0;0.5  & 83.0 & 59.8 & 59.6 & 60.1 & 46.4 & 61.2 & 91.1 & 77.3 & 88.6 &  84.8 &  0.0  \\
  1.5;0.5  & 97.7 & 70.6 & 71.3 & 71.2 & 72.2 & 73.1 & 98.8 & 96.3 & 99.1 &  97.9 &  0.0  \\
  0.0;0.9  & 5.0  & 5.0  & 5.0  & 5.0  & 5.0  & 5.0  & 5.0  & 5.0  & 5.0  &   5.1 &  0.0  \\
  0.5;0.9  & 46.0 & 24.2 & 27.2 & 23.7 & 36.6 & 29.4 & 20.2 & 71.6 & 64.8 &  33.8 &  0.0  \\
  1.0;0.9  & 84.2 & 29.8 & 39.8 & 29.4 & 77.7 & 35.9 & 49.3 & 96.8 & 93.8 &  56.6 &  0.0  \\
  1.5;0.9  & 95.2 & 30.5 & 49.0 & 30.1 & 92.8 & 36.8 & 75.6 & 99.5 & 98.3 &  75.1 &  0.0  \\
 \cmidrule(lr){1-1} \cmidrule(lr){2-10} \cmidrule(lr){11-12}
\end{tabular}
\end{table}

\newpage
\section{Supplementary Monte Carlo results: measurement error and heteroskedasticity}\label{SM:MCmerr}
Tables~\ref{tab:mes100} and \ref{tab:mes500} show the empiricalsi sizes of the tests for the IMA(1,1) model of Eq.~(\ref{eq:ima11}) plus measurement error, for sample sizes of $100$ and 500, respectively. We investigated the power by simulating from the following TAR(1) model:
\begin{equation}\label{eq:tar1}
X_t=\left\{
      \begin{array}{ll}
\phi_{1,0}+ \phi_{1,1} X_{t-1}+ \eps_{t}, & \text{if } X_{t-1}\leq 0 \\
\phi_{1,0}+ \phi_{1,1} X_{t-1}+ \eps_{t}, & \text{if } X_{t-1}>0
      \end{array}\right.
\end{equation}
\noindent
where the parameters are:
\begin{center}
\spacingset{1}
$
\begin{array}{rrrrr}

  & \phi_{1,0}&\phi_{1,1}&\phi_{2,0}&\phi_{2,1} \\
\cmidrule(lr){2-5}
\text{M8} & 0.0 & 0.6 & 0.0 & 0.35 \\
\text{M9} & 0.0 & 0.6 & 0.0 &-0.35 \\
\text{M10}& 0.0 &-0.6 & 0.0 &-0.35 \\
\text{M11}& 0.5 &-2.0 &-0.5 & 1.00 \\
\cmidrule(lr){2-5}
\end{array}
$
\end{center}

All three parameterizations are ergodic and M8--M10 are also geometrically ergodic. We have added measurement noise as follows
\begin{equation}\label{eqSM:mod2}
  Y_t = X_t + \eta_t,
\end{equation}
where the measurement error $\eta_t\sim N(0,\sigma^2_\eta)$ is such that the signal to noise ratio SNR~$=\sigma^2_X/\sigma^2_\eta$ is equal to $\{+\infty, 50,10,5\}$. Here, $\sigma^2_X$ is the variance of $X_t$ computed by means of simulation. The case without noise (SNR~$=+\infty$) is taken as the benchmark.
\begin{table}
\spacingset{1}
\centering
\caption{Empirical size (rejection percentage) at nominal $\alpha=5\%$ and $n=100$ for the IMA(1,1) models M1--M4 with increasing levels of measurement error.}\label{tab:mes100}
\begin{tabular}{rrrrrrrrrrrrr}
&& \multicolumn{9}{c}{asymptotic} & \multicolumn{2}{c}{bootstrap}\\
\cmidrule(lr){3-11}\cmidrule(lr){12-13}
& \textsc{snr} & sLM & $\bar{\mathrm{M}}^{\mathrm{g}}$ & $\mathrm{M}^{\mathrm{g}}$ & $\mathrm{MP}_\mathrm{T}$ & ADF & ADF$^{\mathrm{g}}$ & KS & BBC & EG & sLMb & KSb  \\
\cmidrule(lr){3-11}\cmidrule(lr){12-13}
\multirow{4}{5pt}{M1}
& $\infty$ & 4.9 & 7.9 &  6.6 & 6.0 &  2.8 &  2.9 &   9.0 &  7.3 &   8.3 & 5.7 &  5.9 \\
& 50       & 4.9 & 5.5 &  5.3 & 5.2 &  4.2 &  4.1 &   7.7 &  5.9 &   7.9 & 5.2 &  5.6 \\
& 10       & 6.1 & 5.5 &  5.6 & 5.3 &  5.1 &  4.9 &   6.1 &  4.6 &   7.7 & 6.6 &  4.6 \\
& 5        & 6.4 & 2.8 &  2.8 & 2.8 &  7.6 &  2.7 &   5.2 &  1.7 &   7.4 & 5.8 &  4.1 \\
\cmidrule(lr){3-11}\cmidrule(lr){12-13}
\multirow{4}{5pt}{M2}
& $\infty$ & 5.1 & 5.9 &  6.1 & 5.3 &  4.3 &  4.7 &   6.5 &  3.7 &   6.1 & 4.9 &  4.3 \\
& 50       & 5.9 & 6.2 &  6.2 & 5.3 &  4.0 &  5.3 &   6.2 &  3.7 &   5.5 & 6.1 &  4.1 \\
& 10       & 6.9 & 6.2 &  5.9 & 5.3 &  2.6 &  4.9 &   5.6 &  2.8 &   4.6 & 7.3 &  4.0 \\
& 5        & 5.2 & 3.8 &  3.7 & 2.8 &  4.8 &  4.5 &   5.0 &  1.5 &   3.9 & 4.5 &  3.5 \\
\cmidrule(lr){3-11}\cmidrule(lr){12-13}
\multirow{4}{5pt}{M3}
& $\infty$ & 1.1 & 6.1 &  6.4 & 5.1 &  5.9 &  7.7 &  56.3 &  7.1 &  55.5 & 6.3 & 48.5 \\
& 50       & 0.5 & 6.6 &  6.9 & 5.3 &  5.8 &  7.5 &  59.5 &  7.8 &  57.5 & 5.1 & 50.2 \\
& 10       & 0.7 & 5.8 &  6.3 & 4.8 &  6.5 &  8.0 &  66.5 &  8.4 &  64.6 & 5.0 & 55.9 \\
& 5        & 1.1 & 5.8 &  6.5 & 5.1 & 10.1 &  8.0 &  76.4 & 11.2 &  77.4 & 5.8 & 67.1 \\
\cmidrule(lr){3-11}\cmidrule(lr){12-13}
\multirow{4}{5pt}{M4}
& $\infty$ & 4.8 & 5.7 & 17.6 & 5.4 & 77.5 & 15.8 &  99.9 & 83.8 & 100.0 & 4.0 & 97.9 \\
& 50       & 4.4 & 5.8 & 16.8 & 5.2 & 78.2 & 16.5 &  99.9 & 85.0 & 100.0 & 4.3 & 98.5 \\
& 10       & 6.2 & 6.5 & 19.2 & 5.8 & 80.7 & 15.4 & 100.0 & 86.7 & 100.0 & 5.1 & 98.2 \\
& 5        & 5.3 & 8.8 & 25.8 & 7.9 & 83.6 & 19.3 & 100.0 & 85.4 & 100.0 & 3.5 & 98.5 \\
\cmidrule(lr){3-11}\cmidrule(lr){12-13}
\end{tabular}
\end{table}
\begin{table}
\spacingset{1}
\centering
\caption{Empirical size (rejection percentage at nominal $\alpha=5\%$ and $n=500$ for the IMA(1,1) models M1--M4 with increasing levels of measurement error.}\label{tab:mes500}
\begin{tabular}{rrrrrrrrrrrrr}
&& \multicolumn{9}{c}{asymptotic} & \multicolumn{2}{c}{bootstrap}\\
\cmidrule(lr){3-11}\cmidrule(lr){12-13}
& \textsc{snr} & sLM & $\bar{\mathrm{M}}^{\mathrm{g}}$ & $\mathrm{M}^{\mathrm{g}}$ & $\mathrm{MP}_\mathrm{T}$ & ADF & ADF$^{\mathrm{g}}$ & KS & BBC & EG & sLMb & KSb  \\
\cmidrule(lr){3-11}\cmidrule(lr){12-13}
\multirow{4}{5pt}{M1}
& $\infty$ & 4.2 & 6.6  & 6.5 & 6.1 &  7.3 &  5.0 &   4.9 &  13.2 &   5.5  & 4.5  &  4.0 \\
& 50       & 2.8 & 5.1  & 5.1 & 5.4 &  5.1 &  4.2 &   4.8 &  10.9 &   5.0  & 4.2  &  3.5 \\
& 10       & 4.1 & 5.5  & 5.4 & 5.2 &  5.8 &  4.8 &   3.2 &   3.0 &   3.5  & 3.6  &  2.3 \\
& 5        & 2.8 & 5.3  & 4.9 & 5.2 &  5.0 &  4.5 &   7.7 &   2.4 &   7.0  & 2.9  &  6.0 \\
\cmidrule(lr){3-11}\cmidrule(lr){12-13}
\multirow{4}{5pt}{M2}
& $\infty$ & 3.7 & 4.8  & 4.7 & 4.2 &  4.9 &  4.2 &   5.8 &   7.5 &   5.9  & 5.5  &  4.7 \\
& 50       & 3.8 & 4.2  & 4.1 & 4.1 &  4.7 &  3.8 &   5.1 &   5.5 &   5.2  & 5.9  &  3.7 \\
& 10       & 5.3 & 4.1  & 4.0 & 3.3 &  4.7 &  3.6 &   4.8 &   2.1 &   4.9  & 4.9  &  3.9 \\
& 5        & 4.3 & 4.9  & 4.9 & 4.4 &  5.5 &  4.4 &  17.4 &   2.5 &  11.9  & 4.8  & 14.2 \\
\cmidrule(lr){3-11}\cmidrule(lr){12-13}
\multirow{4}{5pt}{M3}
& $\infty$ & 4.6 & 5.3  & 5.2 & 4.8 &  5.4 &  5.2 &  73.1 &  17.8 &  62.1  & 6.7  & 64.2 \\
& 50       & 5.1 & 5.5  & 5.3 & 4.5 &  5.6 &  5.5 &  76.9 &  21.0 &  66.5  & 6.8  & 68.6 \\
& 10       & 5.2 & 5.5  & 5.4 & 5.0 &  5.7 &  5.7 &  87.0 &  30.0 &  80.1  & 7.6  & 80.4 \\
& 5        & 5.3 & 5.1  & 5.1 & 4.6 &  8.2 &  6.4 &  94.1 &  51.3 &  92.1  & 7.3  & 88.6 \\
\cmidrule(lr){3-11}\cmidrule(lr){12-13}
\multirow{4}{5pt}{M4}
& $\infty$ & 4.4 & 1.1  & 1.4 & 0.8 & 82.8 & 14.8 & 100.0 &  99.7 & 100.0  & 4.6  & 99.8 \\
& 50       & 5.9 & 1.8  & 1.9 & 1.2 & 85.3 & 14.7 & 100.0 & 100.0 & 100.0  & 5.5  & 99.8 \\
& 10       & 6.5 & 1.1  & 1.3 & 1.0 & 93.6 & 17.4 & 100.0 &  99.9 & 100.0  & 5.5  & 99.8 \\
& 5        & 7.6 & 3.1  & 3.3 & 2.3 & 97.9 & 23.7 & 100.0 & 100.0 & 100.0  & 5.1  & 99.9 \\
\cmidrule(lr){3-11}\cmidrule(lr){12-13}
\end{tabular}
\end{table}
\begin{table}
\spacingset{1}
\centering
\caption{Empirical size (rejection percentage) at nominal $\alpha=5\%$ and $n=100$ for the heteroskedastic models M5--M7 with increasing levels of measurement error.}\label{tab:mesh100}
\begin{tabular}{rrrrrrrrrrrrr}
&& \multicolumn{9}{c}{asymptotic} & \multicolumn{2}{c}{bootstrap}\\
\cmidrule(lr){3-11}\cmidrule(lr){12-13}
& \textsc{snr} & sLM & $\bar{\mathrm{M}}^{\mathrm{g}}$ & $\mathrm{M}^{\mathrm{g}}$ & $\mathrm{MP}_\mathrm{T}$ & ADF & ADF$^{\mathrm{g}}$ & KS & BBC & EG & sLMb & KSb  \\
\cmidrule(lr){3-11}\cmidrule(lr){12-13}
\multirow{4}{5pt}{M5}
& $\infty$ & 0.3 & 4.4  & 4.5 & 3.1 &  7.0 &  7.0 & 64.7 &  4.3 &  64.1  & 4.5  & 57.0  \\
& 50       & 0.9 & 4.4  & 4.7 & 3.4 &  6.9 &  8.1 & 69.2 &  4.9 &  69.7  & 4.3  & 61.1  \\
& 10       & 1.1 & 6.2  & 6.4 & 4.2 &  8.0 &  9.7 & 85.4 &  6.6 &  94.1  & 4.6  & 75.9  \\
& 5        & 0.6 & 5.5  & 6.2 & 4.0 & 11.7 & 10.1 & 97.3 & 12.6 & 100.0  & 4.4  & 87.6  \\
\cmidrule(lr){3-11}\cmidrule(lr){12-13}
\multirow{4}{5pt}{M6}
& $\infty$ & 2.0 & 4.9  & 5.5 & 3.8 & 11.7 &  7.3 & 73.5 & 29.1 &  77.9  & 4.7  & 66.5  \\
& 50       & 1.8 & 4.1  & 5.0 & 3.6 & 12.5 &  6.8 & 77.9 & 26.9 &  81.9  & 4.6  & 69.8  \\
& 10       & 0.8 & 4.2  & 4.8 & 3.4 & 16.2 &  8.7 & 89.5 & 32.5 &  96.3  & 4.9  & 83.1  \\
& 5        & 0.8 & 4.9  & 5.9 & 3.9 & 23.1 & 10.3 & 97.2 & 38.0 & 100.0  & 4.7  & 92.1  \\
\cmidrule(lr){3-11}\cmidrule(lr){12-13}
\multirow{4}{5pt}{M7}
& $\infty$ & 9.4 & 4.2  & 4.3 & 4.5 &  8.0 &  4.2 & 10.7 &  8.7 &   8.9  & 5.3  &  9.1  \\
& 50       & 7.1 & 5.7  & 6.2 & 4.5 &  7.4 &  6.3 & 13.9 &  6.3 &   9.8  & 7.0  & 10.6  \\
& 10       & 3.0 & 6.5  & 7.0 & 4.5 &  7.5 &  7.0 & 53.9 & 11.7 &  49.6  & 8.1  & 46.8  \\
& 5        & 1.2 & 6.9  & 7.7 & 5.2 & 11.6 &  8.9 & 89.2 & 15.4 & 100.0  & 6.9  & 83.8  \\
\cmidrule(lr){3-11}\cmidrule(lr){12-13}
\end{tabular}
\end{table}
\begin{table}
\spacingset{1}
\centering
\caption{Empirical size (rejection percentage) at nominal $\alpha=5\%$ and $n=500$ for the heteroskedastic models M5--M7 with increasing levels of measurement error.}\label{tab:mesh500}
\begin{tabular}{rrrrrrrrrrrrr}
&& \multicolumn{9}{c}{asymptotic} & \multicolumn{2}{c}{bootstrap}\\
\cmidrule(lr){3-11}\cmidrule(lr){12-13}
& \textsc{snr} & sLM & $\bar{\mathrm{M}}^{\mathrm{g}}$ & $\mathrm{M}^{\mathrm{g}}$ & $\mathrm{MP}_\mathrm{T}$ & ADF & ADF$^{\mathrm{g}}$ & KS & BBC & EG & sLMb & KSb  \\
\cmidrule(lr){3-11}\cmidrule(lr){12-13}
\multirow{4}{5pt}{M5}
& $\infty$ &  1.9 & 6.9  & 6.9 & 5.5 &  6.5 &  7.2 &  70.0 &  4.0 &  63.8 &  1.8  & 61.4 \\
& 50       &  3.0 & 7.9  & 7.7 & 6.4 &  6.3 &  8.1 &  89.8 & 10.9 &  93.8 &  3.5  & 81.4 \\
& 10       &  4.4 & 7.5  & 7.5 & 5.8 &  8.4 &  9.2 & 100.0 & 55.0 & 100.0 &  6.5  & 95.7 \\
& 5        &  4.7 & 6.4  & 5.9 & 4.5 & 20.9 & 10.6 & 100.0 & 94.2 & 100.0 &  6.6  & 99.8 \\
\cmidrule(lr){3-11}\cmidrule(lr){12-13}
\multirow{4}{5pt}{M6}
& $\infty$ &  7.1 & 5.5  & 5.5 & 5.4 & 10.7 &  6.0 &  85.3 & 52.0 &  83.0 &  3.9  & 80.3 \\
& 50       &  7.8 & 6.4  & 6.2 & 5.2 & 12.2 &  6.4 &  94.8 & 62.2 &  96.8 &  6.4  & 89.8 \\
& 10       &  7.2 & 5.1  & 4.8 & 4.2 & 19.6 &  6.9 &  99.9 & 85.2 & 100.0 &  7.4  & 96.7 \\
& 5        &  6.6 & 4.2  & 4.0 & 3.4 & 38.9 &  7.9 & 100.0 & 98.7 & 100.0 &  8.5  & 99.3 \\
\cmidrule(lr){3-11}\cmidrule(lr){12-13}
\multirow{4}{5pt}{M7}
& $\infty$ & 14.6 & 5.1  & 5.0 & 5.3 &  7.5 &  4.0 &  11.0 & 13.9 &   9.7 &  5.8  &  8.5 \\
& 50       & 13.4 & 5.8  & 5.5 & 5.9 &  7.4 &  4.6 &  61.2 & 27.9 &  40.4 & 13.1  & 54.4 \\
& 10       & 11.0 & 5.6  & 5.3 & 5.9 & 12.1 &  5.9 &  99.9 & 68.9 & 100.0 & 13.0  & 94.4 \\
& 5        &  7.6 & 5.3  & 5.0 & 4.7 & 28.7 &  6.9 & 100.0 & 95.5 & 100.0 &  9.5  & 99.4 \\
\cmidrule(lr){3-11}\cmidrule(lr){12-13}
\end{tabular}
\end{table}
\begin{table}
\spacingset{1}
\centering
\caption{Empirical power (rejection percentage) at nominal $\alpha=5\%$ and $n=100$ for the TAR(1) models M8--M11 with increasing levels of measurement error.}\label{tab:mep100}
\begin{tabular}{rrrrrrrr}
&& \multicolumn{5}{c}{asymptotic} & \multicolumn{1}{c}{boot.}\\
\cmidrule(lr){3-7}\cmidrule(lr){8-8}
& \textsc{snr} & sLM & $\bar{\mathrm{M}}^{\mathrm{g}}$ & $\mathrm{M}^{\mathrm{g}}$ & $\mathrm{MP}_\mathrm{T}$ & ADF$^{\mathrm{g}}$ & sLMb \\
\cmidrule(lr){3-7}\cmidrule(lr){8-8}
\multirow{4}{5pt}{M8}
& $\infty$ & 23.5 & 40.0 & 70.5 & 36.6 & 47.7 & 55.2   \\
& 50       & 21.8 & 39.2 & 71.3 & 37.3 & 47.2 & 55.4   \\
& 10       & 12.2 & 36.8 & 66.8 & 34.9 & 46.2 & 39.9   \\
& 5        &  6.7 & 19.1 & 49.4 & 17.5 & 27.1 & 26.4   \\
\cmidrule(lr){3-7}\cmidrule(lr){8-8}
\multirow{4}{5pt}{M9}
& $\infty$ & 12.4 & 29.8 & 67.2 & 28.1 & 40.1 & 22.8   \\
& 50       & 10.1 & 29.1 & 67.0 & 28.1 & 41.3 & 21.5   \\
& 10       &  8.6 & 32.3 & 68.5 & 31.0 & 44.5 & 16.8   \\
& 5        &  7.5 & 33.2 & 69.3 & 31.0 & 46.0 & 11.9   \\
\cmidrule(lr){3-7}\cmidrule(lr){8-8}
\multirow{4}{5pt}{M10}
& $\infty$ & 43.3 &  7.9 & 47.8 &  7.5 & 19.2 & 32.7   \\
& 50       & 41.1 &  7.9 & 49.5 &  7.8 & 19.2 & 29.3   \\
& 10       & 30.4 &  7.1 & 45.8 &  6.6 & 16.8 & 22.4   \\
& 5        & 19.8 & 11.9 & 46.3 & 11.4 & 28.4 & 13.9   \\
\cmidrule(lr){3-7}\cmidrule(lr){8-8}
\multirow{4}{5pt}{M11}
& $\infty$ & 97.6 & 43.6 & 66.4 & 40.1 & 50.3 & 97.8   \\
& 50       & 97.2 & 43.4 & 66.1 & 39.7 & 49.7 & 97.2   \\
& 10       & 94.9 & 41.7 & 64.9 & 37.4 & 48.9 & 94.6   \\
& 5        & 87.7 & 40.6 & 65.9 & 37.5 & 47.9 & 90.7   \\
\cmidrule(lr){3-7}\cmidrule(lr){8-8}
\end{tabular}
\end{table}
\begin{table}
\spacingset{1}
\centering
\caption{Empirical power (rejection percentage) at nominal $\alpha=5\%$ and $n=300$ for the TAR(1) models M8--M11 with increasing levels of measurement error.}\label{tab:mep300}
\begin{tabular}{rrrrrrrr}
&& \multicolumn{5}{c}{asymptotic} & \multicolumn{1}{c}{boot.}\\
\cmidrule(lr){3-7}\cmidrule(lr){8-8}
& \textsc{snr} & sLM & $\bar{\mathrm{M}}^{\mathrm{g}}$ & $\mathrm{M}^{\mathrm{g}}$ & $\mathrm{MP}_\mathrm{T}$ & ADF$^{\mathrm{g}}$ & sLMb \\
\cmidrule(lr){3-7}\cmidrule(lr){8-8}
\multirow{4}{5pt}{M8}
& $\infty$ & 96.6 & 43.8 & 77.8 & 40.7 & 58.2 &  97.2   \\
& 50       & 96.2 & 42.7 & 77.0 & 40.5 & 59.0 &  96.7   \\
& 10       & 91.2 & 38.9 & 75.3 & 36.0 & 55.9 &  91.2   \\
& 5        & 80.0 & 42.4 & 75.0 & 40.3 & 62.1 &  80.8   \\
\cmidrule(lr){3-7}\cmidrule(lr){8-8}
\multirow{4}{5pt}{M9}
& $\infty$ & 52.5 & 32.4 & 70.1 & 30.1 & 53.1 &  41.6   \\
& 50       & 50.6 & 32.1 & 69.4 & 30.0 & 53.6 &  40.3   \\
& 10       & 42.0 & 33.0 & 73.0 & 31.9 & 54.8 &  30.9   \\
& 5        & 36.0 & 30.0 & 70.9 & 28.2 & 54.7 &  18.5   \\
\cmidrule(lr){3-7}\cmidrule(lr){8-8}
\multirow{4}{5pt}{M10}
& $\infty$ & 71.1 &  7.0 & 47.9 &  6.6 & 25.7 &  32.8   \\
& 50       & 65.1 &  6.7 & 45.0 &  5.9 & 26.5 &  25.6   \\
& 10       & 63.2 &  8.1 & 61.8 &  8.5 & 28.1 &  24.4   \\
& 5        & 42.3 &  8.5 & 46.0 &  7.9 & 27.5 &   8.3   \\
\cmidrule(lr){3-7}\cmidrule(lr){8-8}
\multirow{4}{5pt}{M11}
& $\infty$ &100.0 & 55.7 & 78.5 & 53.2 & 67.0 &  99.9   \\
& 50       &100.0 & 56.0 & 77.8 & 53.6 & 67.5 & 100.0   \\
& 10       &100.0 & 55.0 & 78.4 & 50.4 & 67.1 &  99.7   \\
& 5        & 99.7 & 52.6 & 78.4 & 49.7 & 68.4 &  99.2   \\
\cmidrule(lr){3-7}\cmidrule(lr){8-8}
\end{tabular}
\end{table}
\begin{table}
\spacingset{1}
\centering
\caption{Empirical power (rejection percentage) at nominal $\alpha=5\%$ and $n=500$ for the TAR(1) models M8--M11 with increasing levels of measurement error.}\label{tab:mep500}
\begin{tabular}{rrrrrrrr}
&& \multicolumn{5}{c}{asymptotic} & \multicolumn{1}{c}{boot.}\\
\cmidrule(lr){3-7}\cmidrule(lr){8-8}
& \textsc{snr} & sLM & $\bar{\mathrm{M}}^{\mathrm{g}}$ & $\mathrm{M}^{\mathrm{g}}$ & $\mathrm{MP}_\mathrm{T}$ & ADF$^{\mathrm{g}}$ & sLMb \\
\cmidrule(lr){3-7}\cmidrule(lr){8-8}
\multirow{4}{5pt}{M8}
&$\infty$ & 99.7 & 48.0 & 80.6 & 44.9 & 66.1 & 99.8  \\
& 50      & 99.5 & 48.0 & 78.9 & 45.0 & 66.7 & 99.3  \\
& 10      & 92.5 & 46.8 & 77.6 & 44.7 & 68.2 & 91.3  \\
& 5       & 73.5 & 51.0 & 77.4 & 47.4 & 71.6 & 68.3  \\
\cmidrule(lr){3-7}\cmidrule(lr){8-8}
\multirow{4}{5pt}{M9}
&$\infty$& 63.8 & 36.9 & 74.6 & 34.9 & 58.4 & 50.5  \\
& 50     & 58.2 & 36.5 & 74.1 & 34.9 & 59.4 & 43.1  \\
& 10     & 45.2 & 20.8 & 67.2 & 18.4 & 36.2 & 24.1  \\
& 5      & 45.1 & 25.1 & 74.0 & 25.2 & 55.4 & 20.6  \\
\cmidrule(lr){3-7}\cmidrule(lr){8-8}
\multirow{4}{5pt}{M10}
&$\infty$& 79.2 &  7.9 & 49.3 &  7.8 & 33.8 & 32.8  \\
& 50     & 72.1 &  9.2 & 43.2 &  8.5 & 35.3 & 23.4  \\
& 10     & 62.1 & 12.6 & 32.4 & 12.2 & 42.5 & 10.3  \\
& 5      & 67.3 &  8.9 & 43.6 &  8.3 & 39.1 &  9.3  \\
\cmidrule(lr){3-7}\cmidrule(lr){8-8}
\multirow{4}{5pt}{M11}
&$\infty$& 99.9 & 65.5 & 83.8 & 61.5 & 75.3 & 99.9  \\
& 50     &100.0 & 64.7 & 83.7 & 60.3 & 74.8 & 99.9  \\
& 10     & 99.9 & 60.4 & 81.6 & 59.8 & 74.0 & 99.8  \\
& 5      & 99.4 & 61.9 & 79.8 & 59.9 & 76.7 & 98.1  \\
\cmidrule(lr){3-7}\cmidrule(lr){8-8}
\end{tabular}
\end{table}

The empirical power of the tests is reported in Tables~\ref{tab:mep100}--\ref{tab:mep500}. Note that the power is not size-corrected as it is not clear what will be the reference model under $H_0$ in the presence of measurement error. For this reason, we reported the tests that did not suffer from  biased size as the results for the other tests are not informative due to their severe oversize. The effect of measurement error on the power of the supLM tests varies with the values of the parameters. For models M8--M10 there is some power loss whereas there is no loss for model M11. The M tests appear to be less sensitive to the presence of measurement error but their power is generally much lower than that of the supLM tests.

\section{Supplementary Monte Carlo results: TARMA(1,1) with a IMA(1,1) regime}\label{SM:MCadd}
We simulated data from the following  TARMA(1,1) model
\begin{equation}
X_t= \begin{cases}
 \phi_{1,0}+\phi_{1,1}X_{t-1} +\eps_t- \theta \eps_{t-1}, & \text{if } X_{t-d} \le r, \\
                      X_{t-1} +\eps_t- \theta \eps_{t-1}, & \text{otherwise},
\end{cases} \label{tarma.ima}
\end{equation}
where $(\phi_{1,0}, \phi_{1,1}, \phi_{2,0}, \phi_{2,1})=\tau\times (0, 0.7,-0.02,0.99)+(1-\tau)\times(0,1,0,0) + (0,0,0,1)$ with $\tau$ increasing from 0 to 1.5 with increments 0.5. When $\tau=0$, the model is an IMA(1,1) model with zero intercept, while the model becomes a stationary TARMA(1,1) model with $\tau>0$, of  increasing disparity from the IMA(1,1) model with increasing $\tau$. Note that the upper regime corresponds to a IMA(1,1) process. The resulting parameter set is reported in Table~\ref{SMtab:param}. As for the MA parameter we chose $\theta=-0.9,-0.5,0,0.5,0.9$.
\begin{table}[H]
\spacingset{1}
  \centering
$  \begin{array}{rrrrr}
    \tau   & \phi_{10}& \phi_{11} & \phi_{20}& \phi_{21}\\
\cmidrule(lr){2-5}
    0.0 & 0.00     & 1.00  & 0 & 1   \\
    0.5 &-0.01     & 0.85  & 0 & 1   \\
    1.0 &-0.02     & 0.70  & 0 & 1   \\
    1.5 &-0.03     & 0.55  & 0 & 1   \\
\cmidrule(lr){2-5}
  \end{array}
$  \caption{Parameters set used to simulate from the TARMA(1,1) model of Eq.~\ref{tarma.ima}.}\label{SMtab:param}
\end{table}
\par\noindent
The empirical size of the tests at nominal level $\alpha=5\%$, is displayed in Table~\ref{SMtab:size}. As shown previously, the ADF, the KS, the BBC and the EG tests are severely oversized also for $n=500$, as $\theta$ approaches unity. Clearly, the sLMb test is the only test that shows a correct size in all the settings, whereas both the sLM and the M class of tests show some bias, albeit small.

\begin{table}[H]
\spacingset{1}
\centering\small
\caption{Empirical size at $\alpha = 5\%$. Rejection percentages from the TARMA$(1,1)$ model of Eq.~(\ref{tarma.ima}). Sizes over 15\% have been emphasized in bold font.}\label{SMtab:size}
\begin{tabular}{rrrrrrrrrrrrr}
       & \multicolumn{9}{c}{asymptotic} & \multicolumn{2}{c}{bootstrap}\\
 \cmidrule(lr){2-10} \cmidrule(lr){11-12}
$\theta$ & sLM & $\bar{\mathrm{M}}^{\mathrm{g}}$ & $\mathrm{M}^{\mathrm{g}}$ & $\mathrm{MP}_\mathrm{T}$ & ADF & ADF$^{\mathrm{g}}$ & KS & BBC & EG & sLMb & KSb  \\
 \multicolumn{10}{l}{$n=100$} & \multicolumn{2}{c}{}\\
 \cmidrule(lr){1-1} \cmidrule(lr){2-10} \cmidrule(lr){11-12}
  -0.9 &  2.4  & 8.1 &     7.2  & 7.4 &     3.1 &     3.5 &      7.9 &    10.6 &      6.9 & 6.0 &     6.5   \\
  -0.5 &  1.6  & 6.3 &     6.1  & 5.8 &     4.8 &     5.1 &      7.0 &     6.1 &      5.7 & 4.3 &     4.9   \\
   0.0 &  1.7  & 4.9 &     5.0  & 4.6 &     5.5 &     5.4 &      8.3 &     3.1 &      5.0 & 4.8 &     6.0   \\
   0.5 &  1.5  & 5.7 &     6.2  & 5.3 &     6.7 &     7.5 &{\bfseries  64.5} &    10.6 &{\bfseries  57.9} & 5.0 &{\bfseries 58.6}    \\
   0.9 & 12.1  & 6.6 &{\bfseries 17.7}  & 6.6 &{\bfseries 77.2} &{\bfseries 17.1} &{\bfseries 100.0} &{\bfseries 92.7} &{\bfseries 100.0} & 4.6 &{\bfseries 99.4}   \\

\multicolumn{10}{l}{$n=300$} & \multicolumn{2}{c}{}\\
 \cmidrule(lr){1-1} \cmidrule(lr){2-10} \cmidrule(lr){11-12}
 -0.9 & 5.8 & 6.9 & 6.7 & 6.3 &    3.2  &    4.4  &    6.6   &    13.8 &    6.6   & 6.2  &    4.8   \\
 -0.5 & 4.3 & 5.3 & 5.2 & 4.7 &    4.6  &    4.5  &    5.8   &    8.6  &    5.6   & 5.1  &    3.6   \\
  0.0 & 2.9 & 5.0 & 4.9 & 4.6 &    4.9  &    4.8  &    7.1   &    2.9  &    4.9   & 4.2  &    4.8   \\
  0.5 & 2.3 & 5.6 & 5.6 & 5.2 &    5.2  &    6.2  &{\bfseries 74.8}  &{\bfseries 18.9} &{\bfseries 60.6}  & 4.9  &{\bfseries 67.1}     \\
  0.9 & 5.0 & 1.9 & 2.5 & 1.8 &{\bfseries 86.2} &{\bfseries 15.5} &{\bfseries 100.0} &{\bfseries 99.6} &{\bfseries 100.0} & 4.0  &{\bfseries 100.0}  \\

 \multicolumn{10}{l}{$n=500$} & \multicolumn{2}{c}{}\\
 \cmidrule(lr){1-1} \cmidrule(lr){2-10} \cmidrule(lr){11-12}
 -0.9 & 8.3 & 6.0 & 5.9 & 5.4 &    6.8  & 4.2  &    6.4   &{\bfseries 15.3} &    6.1   & 5.8 &  4.2  \\
 -0.5 & 5.2 & 5.1 & 5.1 & 4.8 &    4.8  & 4.5  &    5.6   &    9.1          &    5.4   & 7.0 &  4.9   \\
  0.0 & 3.8 & 4.5 & 4.4 & 4.2 &    5.3  & 4.3  &    7.5   &    3.3          &    5.1   & 5.4 &  5.8   \\
  0.5 & 2.6 & 5.5 & 5.3 & 5.1 &    5.4  & 5.4  &{\bfseries 78.2}  &{\bfseries 23.5} &{\bfseries 61.6}  & 5.1 &{\bfseries 73.2}  \\
  0.9 & 3.7 & 1.3 & 1.3 & 1.2 &{\bfseries 83.3} & 14.1 &{\bfseries 100.0} &{\bfseries 99.9} &{\bfseries 100.0} & 4.9 & {\bfseries 100.0}  \\
 \cmidrule(lr){2-10} \cmidrule(lr){11-12}
\end{tabular}
\end{table}

The size-corrected power of the tests is presented in Tables~\ref{SMtab:1},~\ref{SMtab:2},~\ref{SMtab:3} for $n=100$, 300 and 500, respectively. Note that rows for $\tau=0$ correspond to the size and other rows give size-corrected power. The size correction for bootstrap tests is achieved by calibrating the $p$-values. Notice that, in some cases, the corrected size deviates slightly from the nominal 5\% due to discretization effects on the empirical distribution of bootstrap $p$-values. Clearly, the supLM tests are almost always more powerful than the other tests, especially as $\tau$ increases. The case $\theta=0$ (central panel) corresponds to a TAR model and this is one of two  instances where the KS tests are slightly more powerful than the supLM tests. The power of the bootstrap version of the KS test is zero in three cases, due to its 100\% oversize.

\begin{table}[H]
\spacingset{1}
\centering \small
\caption{n = 100. Size corrected power of the asymptotic and bootstrap tests at $\alpha=5\%$. Simulation from the TARMA(1,1) model of Eq.~(\ref{tarma.ima}).}\label{SMtab:1}
\begin{tabular}{crrrrrrrrrrr}
$n=100$       & \multicolumn{9}{c}{asymptotic} & \multicolumn{2}{c}{bootstrap}\\
 \cmidrule(lr){1-1} \cmidrule(lr){2-10} \cmidrule(lr){11-12}
 $\tau\;;\;\theta$ & sLM & $\bar{\mathrm{M}}^{\mathrm{g}}$ & $\mathrm{M}^{\mathrm{g}}$ & $\mathrm{MP}_\mathrm{T}$ & ADF & ADF$^{\mathrm{g}}$ & KS & BBC & EG & sLMb & KSb \\
\cmidrule(lr){2-10} \cmidrule(lr){11-12}
  0.0;-0.9 & 5.0  & 5.0  & 5.0  & 5.0 & 5.0 & 5.0 & 5.0  & 5.0 & 5.0 &  5.0 &  5.1  \\
  0.5;-0.9 & 8.2  & 6.8  & 7.5  & 7.1 & 5.6 & 8.1 & 4.5  & 6.8 & 3.5 &  8.3 &  4.5  \\
  1.0;-0.9 & 12.6 & 7.4  & 8.0  & 7.6 & 5.8 & 8.9 & 5.6  & 8.1 & 3.8 & 11.2 &  5.7  \\
  1.5;-0.9 & 16.0 & 7.2  & 7.6  & 7.4 & 6.0 & 9.3 & 6.0  & 8.8 & 3.7 & 15.2 &  5.7  \\
  0.0;-0.5 & 5.0  & 5.0  & 5.0  & 5.0 & 5.0 & 5.0 & 5.0  & 5.0 & 5.0 &  5.0 &  5.1  \\
  0.5;-0.5 & 8.3  & 8.5  & 8.9  & 8.6 & 5.6 & 8.9 & 4.9  & 7.5 & 3.8 &  7.3 &  5.2  \\
  1.0;-0.5 & 11.1 & 8.5  & 8.9  & 8.6 & 5.8 & 8.9 & 6.9  & 8.4 & 4.2 & 10.1 &  6.3  \\
  1.5;-0.5 & 14.6 & 9.2  & 9.6  & 9.3 & 5.9 & 9.6 & 9.0  & 8.9 & 4.8 & 13.0 & 10.3  \\
  0.0;0.0  & 5.0  & 5.0  & 5.0  & 5.0 & 5.0 & 5.0 & 5.0  & 5.0 & 5.0 &  5.0 &  5.0  \\
  0.5;0.0  & 7.0  & 9.8  & 10.0 & 10.0& 5.7 & 10.0& 11.0 & 6.3 & 8.8 &  7.3 &  9.7  \\
  1.0;0.0  & 9.3  & 9.8  & 10.4 & 9.8 & 6.0 & 10.1& 15.5 & 7.0 & 9.9 & 11.7 & 16.8  \\
  1.5;0.0  & 11.3 & 10.3 & 10.9 & 10.4& 6.0 & 10.4& 19.5 & 8.2 & 11.3& 12.0 & 18.6  \\
  0.0;0.5  & 5.0  & 5.0  & 5.0  & 5.0 & 5.0 & 5.0 & 5.0  & 5.0 & 5.0 &  5.0 &  0.0  \\
  0.5;0.5  & 6.7  & 8.4  & 8.5  & 8.5 & 5.5 & 8.5 & 12.6 & 6.2 & 9.8 &  5.8 &  0.0  \\
  1.0;0.5  & 7.4  & 9.1  & 9.4  & 9.0 & 5.9 & 9.2 & 13.3 & 7.4 & 10.8&  9.4 &  0.0  \\
  1.5;0.5  & 9.5  & 8.8  & 9.3  & 8.8 & 6.1 & 9.0 & 13.1 & 8.1 & 11.0&  8.5 &  0.0  \\
  0.0;0.9  & 5.0  & 5.0  & 5.0  & 5.0 & 5.0 & 5.0 & 5.0  & 5.0 & 5.0 &  5.1 &  0.0  \\
  0.5;0.9  & 5.2  & 5.1  & 5.0  & 5.2 & 4.9 & 5.0 & 0.5  & 8.0 & 5.9 &  6.6 &  0.0  \\
  1.0;0.9  & 5.7  & 5.3  & 5.6  & 5.4 & 4.7 & 5.1 & 0.1  & 8.2 & 5.4 &  7.8 &  0.0  \\
  1.5;0.9  & 5.5  & 5.1  & 5.1  & 5.2 & 4.7 & 5.1 & 0.1  & 8.0 & 4.7 &  8.0 &  0.0  \\
\cmidrule(lr){2-10} \cmidrule(lr){11-12}
\end{tabular}
\end{table}

\begin{table}[H]
\spacingset{1}
\centering \small
\caption{n = 300.  Size corrected power of the asymptotic and bootstrap tests at $\alpha=5\%$. Simulation from the TARMA(1,1) model of Eq.~(\ref{tarma.ima}).}\label{SMtab:2}
\begin{tabular}{crrrrrrrrrrr}
$n=300$       & \multicolumn{9}{c}{asymptotic} & \multicolumn{2}{c}{bootstrap}\\
\cmidrule(lr){2-10} \cmidrule(lr){11-12}
 $\tau\;;\;\theta$ & sLM & $\bar{\mathrm{M}}^{\mathrm{g}}$ & $\mathrm{M}^{\mathrm{g}}$ & $\mathrm{MP}_\mathrm{T}$ & ADF & ADF$^{\mathrm{g}}$ & KS & BBC & EG & sLMb & KSb  \\
\cmidrule(lr){2-10} \cmidrule(lr){11-12}
  0.0;-0.9 & 5.0  & 5.0  & 5.0  & 5.0  & 5.0 & 5.0  & 5.0  & 5.0  & 5.0  &  4.9 &  5.1   \\
  0.5;-0.9 & 16.1 & 10.7 & 10.6 & 11.1 & 8.0 & 11.7 & 6.4  & 9.9  & 3.6  & 17.3 &  8.3   \\
  1.0;-0.9 & 24.2 & 10.7 & 10.6 & 10.9 & 7.5 & 11.4 & 10.4 & 13.5 & 4.2  & 21.8 & 11.9   \\
  1.5;-0.9 & 31.3 & 10.9 & 11.1 & 11.1 & 7.4 & 12.0 & 14.0 & 17.2 & 4.2  & 25.6 & 13.7   \\
  0.0;-0.5 & 5.0  & 5.0  & 5.0  & 5.0  & 5.0 & 5.0  & 5.0  & 5.0  & 5.0  &  5.1 &  4.8   \\
  0.5;-0.5 & 15.5 & 12.8 & 12.5 & 12.6 & 7.7 & 12.6 & 9.1  & 9.8  & 4.4  & 15.7 &  9.0   \\
  1.0;-0.5 & 22.8 & 13.3 & 13.2 & 13.3 & 8.2 & 13.4 & 14.1 & 13.1 & 5.2  & 22.4 & 15.9   \\
  1.5;-0.5 & 28.2 & 12.8 & 12.8 & 13.1 & 8.7 & 13.1 & 18.3 & 16.3 & 5.8  & 27.9 & 20.2   \\
  0.0;0.0  & 5.0  & 5.0  & 5.0  & 5.0  & 5.0 & 5.0  & 5.0  & 5.0  & 5.0  &  5.0 &  5.0   \\
  0.5;0.0  & 14.0 & 14.2 & 14.5 & 14.1 & 8.0 & 14.2 & 23.2 & 9.8  & 12.6 & 17.0 & 24.7   \\
  1.0;0.0  & 21.7 & 14.6 & 14.8 & 14.4 & 8.0 & 14.5 & 28.8 & 13.3 & 14.0 & 26.2 & 31.1   \\
  1.5;0.0  & 26.2 & 15.0 & 15.3 & 15.0 & 8.1 & 15.1 & 33.1 & 16.1 & 14.5 & 27.1 & 29.3   \\
  0.0;0.5  & 5.0  & 5.0  & 5.0  & 5.0  & 5.0 & 5.0  & 5.0  & 5.0  & 5.0  &  4.9 &  0.0   \\
  0.5;0.5  & 12.5 & 12.1 & 12.0 & 12.0 & 8.3 & 12.1 & 22.1 & 11.1 & 16.3 & 11.2 &  0.0   \\
  1.0;0.5  & 17.3 & 12.3 & 12.4 & 12.5 & 8.8 & 12.5 & 21.6 & 14.8 & 18.1 & 16.7 &  0.0   \\
  1.5;0.5  & 21.0 & 12.0 & 12.0 & 12.1 & 8.6 & 12.2 & 20.9 & 16.9 & 18.4 & 20.8 &  0.0   \\
  0.0;0.9  & 5.0  & 5.0  & 5.0  & 5.0  & 5.0 & 5.0  & 5.0  & 5.0  & 5.0  &  4.9 &  0.0   \\
  0.5;0.9  & 9.1  & 6.2  & 6.2  & 6.0  & 6.7 & 6.2  & 0.1  & 10.5 & 7.0  &  9.2 &  0.0   \\
  1.0;0.9  & 10.1 & 6.3  & 6.6  & 6.1  & 6.3 & 6.8  & 0.0  & 11.6 & 6.8  & 11.0 &  0.0   \\
  1.5;0.9  & 10.9 & 6.3  & 6.5  & 6.2  & 6.2 & 6.4  & 0.0  & 10.9 & 6.0  & 10.6 &  0.0   \\
\cmidrule(lr){2-10} \cmidrule(lr){11-12}
\end{tabular}
\end{table}

\begin{table}[H]
\spacingset{1}
\centering \small
\caption{n = 500.  Size corrected power of the asymptotic and bootstrap tests at $\alpha=5\%$. Simulation from the TARMA(1,1) model of Eq.~(\ref{tarma.ima}).}\label{SMtab:3}
\begin{tabular}{crrrrrrrrrrr}
$n=500$       & \multicolumn{9}{c}{asymptotic} & \multicolumn{2}{c}{bootstrap}\\
\cmidrule(lr){2-10} \cmidrule(lr){11-12}
 $\tau\;;\;\theta$ & sLM & $\bar{\mathrm{M}}^{\mathrm{g}}$ & $\mathrm{M}^{\mathrm{g}}$ & $\mathrm{MP}_\mathrm{T}$ & ADF & ADF$^{\mathrm{g}}$ & KS & BBC & EG & sLMb & KSb  \\
\cmidrule(lr){2-10} \cmidrule(lr){11-12}
  0.0;-0.9 & 5.0  & 5.0  & 5.0  & 5.0  & 5.0 & 5.0  & 5.0  & 5.0  & 5.0 &   5.0 &   5.1 \\
  0.5;-0.9 & 21.4 & 13.1 & 12.8 & 13.2 & 9.6 & 13.4 & 9.7  & 14.8 & 4.3 &   8.3 &   4.5 \\
  1.0;-0.9 & 30.6 & 12.5 & 12.1 & 12.5 & 8.6 & 12.8 & 14.5 & 19.5 & 4.6 &  11.2 &   5.7 \\
  1.5;-0.9 & 35.1 & 13.0 & 12.8 & 13.0 & 9.4 & 13.6 & 17.8 & 23.0 & 4.9 &  15.2 &   5.7 \\
  0.0;-0.5 & 5.0  & 5.0  & 5.0  & 5.0  & 5.0 & 5.0  & 5.0  & 5.0  & 5.0 &   5.0 &   5.1 \\
  0.5;-0.5 & 22.0 & 14.2 & 14.0 & 14.1 & 9.6 & 13.9 & 12.9 & 14.9 & 5.6 &   7.3 &   5.2 \\
  1.0;-0.5 & 30.4 & 14.6 & 14.4 & 14.2 & 9.4 & 14.1 & 18.8 & 19.6 & 5.4 &  10.1 &   6.3 \\
  1.5;-0.5 & 34.9 & 14.1 & 13.9 & 13.8 & 8.9 & 13.7 & 23.1 & 23.3 & 6.1 &  13.0 &  10.3 \\
  0.0;0.0  & 5.0  & 5.0  & 5.0  & 5.0  & 5.0 & 5.0  & 5.0  & 5.0  & 5.0 &   5.0 &   5.0 \\
  0.5;0.0  & 20.2 & 16.7 & 16.8 & 16.4 & 8.5 & 16.4 & 29.2 & 13.0 &14.0 &   7.3 &   9.7 \\
  1.0;0.0  & 28.4 & 16.5 & 16.5 & 16.1 & 8.9 & 16.1 & 34.4 & 18.5 &15.0 &  11.7 &  16.8 \\
  1.5;0.0  & 33.4 & 17.4 & 17.4 & 17.0 & 8.6 & 17.1 & 35.7 & 21.9 &15.0 &  12.0 &  18.6 \\
  0.0;0.5  & 5.0  & 5.0  & 5.0  & 5.0  & 5.0 & 5.0  & 5.0  & 5.0  & 5.0 &   5.0 &   0.0 \\
  0.5;0.5  & 17.3 & 14.6 & 14.6 & 14.5 & 9.1 & 14.6 & 27.7 & 16.4 &19.0 &   5.8 &   0.0 \\
  1.0;0.5  & 23.9 & 14.8 & 15.0 & 14.8 & 9.1 & 15.2 & 25.9 & 20.6 &20.0 &   9.4 &   0.0 \\
  1.5;0.5  & 26.9 & 14.4 & 14.6 & 14.5 & 8.8 & 14.9 & 22.0 & 22.6 &20.0 &   8.5 &   0.0 \\
  0.0;0.9  & 5.0  & 5.0  & 5.0  & 5.0  & 5.0 & 5.0  & 5.0  & 5.0  & 5.0 &   5.1 &   0.0  \\
  0.5;0.9  & 11.0 & 7.3  & 7.7  & 7.3  & 7.6 & 7.9  & 0.1  & 13.4 & 9.4 &   6.6 &   0.0  \\
  1.0;0.9  & 12.9 & 7.8  & 8.2  & 7.6  & 7.0 & 8.2  & 0.0  & 14.8 & 9.5 &   7.8 &   0.0  \\
  1.5;0.9  & 14.0 & 8.0  & 8.4  & 8.0  & 6.6 & 8.2  & 0.0  & 15.1 & 9.6 &   8.0 &   0.0  \\
\cmidrule(lr){2-10} \cmidrule(lr){11-12}
\end{tabular}
\end{table}

\end{document}